# TECHNICAL REPORT:

# Rapid Reviews on Engineering of Internet of Things Software Systems

AUTHORS

**Rebeca Campos Motta**
PhD Student
PESC/COPPE/UFRJ – Brazil
LAMIH UMR 8201 UPHF - France
*rmotta@cos.ufrj.br*

**Káthia Marçal de Oliveira**
Full Professor at UPHF
LAMIH CNRS UMR 8201
Valenciennes, France
*kathia.oliveira@uphf.fr*

**Guilherme Horta Travassos**
Full Professor at UFRJ
PESC/COPPE
Rio de Janeiro, Brazil
*ght@cos.ufrj.br*

# Index

# 1 INTRODUCTION

Interactive systems, ubiquitous systems, Internet of Things compose a new generation of software systems. From our perspective, these new applications require integrating different engineering domains (e.g., software engineering, human-machine interaction) and integrating heterogeneous technologies (e.g., sensors, interactive devices) this, be what we are calling IoT. It is currently possible to develop small devices with embed intelligence, seamless communication, thing-thing interaction, wireless connections, and different technologies. We can observe that the evolution of several areas and the collaboration among them enables the realization of the IoT, the topic of this thesis.

Because of their broad potential, IoT requires not only software solutions but must make use of a variety of technologies (communications, mobile devices, sensors and actuators, big data, cloud computing, artificial intelligence) (Miorandi *et al.*, 2012; Whitmore, Agarwal and Da Xu, 2015). Thus, these systems offer an opportunity for research and development focused on the integration of devices and technologies that motivate to evolve the engineering to a creative vision, innovative and of constant change glimpsing new forms of interaction between the involved actors and new utilities for the interconnected physical objects with embedded software.

New challenges are emerging as a result of these new possibilities, such higher need for the software to be embedded in the product (Miranda *et al.*, 2015; Lu, 2017) and technology diversity and multidisciplinarity to deliver the variety of possible solutions (Chapline and Sullivan, 2010; Gubbi *et al.*, 2013) considering communication and interoperability, essential for the materialization of the paradigm (Things, 2015; Lin *et al.*, 2017). Thus, attention to software development with a holistic vision focusing on integrating with different disciplines can be the excellent differential for developing these systems, since complex systems require systems engineering that integrates across the subsystems to meet requirements (Chapline and Sullivan, 2010). As a consequence, it is necessary multidisciplinarity among the solution and the business needs for IoT. This scenario can motivate the development of new technologies to solve new problems, many of which are unknown and require a high degree of innovation (Atzori, Iera, and Morabito, 2010; Haller, 2011).

This context demands an evolution in knowledge, skills, and technologies distinct from those offered by traditional software engineering (Skiba, 2013; Zambonelli, 2016; Larrucea *et al.*, 2017). For Software Engineering, new research and development challenges emerge in this paradigm, without prejudice to initial concerns of deadlines, costs and quality levels of products and processes (Pfleeger and Atlee, 1998), but involving the intensive internalization of software in the products, high distribution of solutions, diversity and technological multidisciplinarity, communication and systemic interoperability.

# 2 CONTEXT

As a result of our investigation in this direction (Motta, de Oliveira, and Travassos, 2018), we propose six different facets that an IoT software system should take into account. We understand facets as “one side of something many-sided” (Oxford Dictionary), “one part of a subject, a situation that has many parts” (Cambridge Dictionary), representing the multidisciplinarity required in such systems.

**1. Connectivity**

Connectivity is one of the main aspects of future systems. We argue that it is necessary to have available a medium by which things can connect to materialize the IoT paradigm. It is essential some form of connection, a network for the development of solutions. Our idea is not to limit Internet-only connectivity, but to cover other media such as Intranet, Bluetooth, among others, means **the manner by which objects are connected**.

It is important to note that there is no one-fit-for-all solution (Luzuriaga *et al.*, 2015) since it englobes many domains, and each one of them will have particular characteristics and requirements. However, in the literature, we can observe that specific requirements are related to the devices' nature or the application needs, which influence communication directly - such as low latency, bandwidth, and robustness (Poluru and Naseera, 2017). Even though some of the requirements are not directly related to connectivity. Still, they show aspects that will profoundly influence communication; thus, they are requirements that need to be well understood and addressed to make IoT work.

**2. Things**

In this sense, it means the things by themselves in IoT. Tags, sensors, actuators, mobile phones, **all hardware that can traditionally replace the computer,** expanding the connectivity reach.

In our understanding, *things* exist in the physical realm, such as sensors, actuators, or any objects equipped with identifying, sensing, or acting behaviors and processing capabilities that can communicate and cooperate to reach a goal, varying according to the systems requirements (Whitmore, Agarwal and Da Xu, 2015). When an object has enhanced capabilities and uses connectivity to interact with others, it can be considered a *thing* in our context.

**3. Behavior**

The existence of things is not new nor their natural capacities. What IoT provides is the chance of **enhancements in the things, extending their original behaviors**. In the beginning, the things in IoT systems were objects attached to electronic tags, so these systems present Identification's behavior. Subsequently, sensors and actuators composing the software systems enabled the Sensing and Actuation behaviors, respectively. It can be necessary to use software solutions, semantic technologies, data analytics, and other areas to enhance things' behavior.

The idea of the system behavior results from its constituent parts; that is, the behavior is generated by the interaction and collaboration of two or more devices. The combination of simpler behaviors can manage more complex behavior. The behavior of an IoT can be aggregative and emergent, being capable of performing different actions (Jackson, 2013).

**4. Smartness**

Smartness or Intelligence is related to Behavior but as to managing or organizing it. It refers to **orchestration associated with things and to what level of intelligence technology can evolve their initial behavior**.

Artificial intelligence and machine learning techniques can be applied to enhance the intelligence and significant interactions between things to manage smartness. In the development of smart applications, it is important to highlight that having only sensors collecting data does not make it smart. For a system to be *smart*, it needs a set of actions, for example, treating data, making decisions, and acting. The level of smartness depends on the application domain and user needs.

**5. Interactivity**

It refers to actors' involvement **in the interaction** to exchange information with things and the degree to which it happens. The actors engaged with IoT applications are not limited to humans. Therefore, beyond the human actors' sociotechnical concerns, we also have concerns with other actors like animals and thing-thing interactions. The degree to which it happens works together with the medium through which things can connect (connectivity) so that they can understand (interoperability) in addition to being connected.

**6. Environment**

The problem and the solution are embedded in a domain, an environment, or a context. This facet seeks to represent such an environment and how the context information can influence its use. The environment is **the place where things are, actions happen, events occur, and people are**. Smart Environments or Smart Spaces provide intelligent services by acquiring knowledge about itself and its inhabitants to adapt to users' needs and behavior (Aziz, Sheikh, and Felemban, 2016). IoT uses several technologies to meet specific requirements that differ according to the project [13][16]. These systems have a set of *things* which are capable of sensing, reason, collaborate, and act upon ambient. An essential characteristic of this ambient is the user-centric thinking approach in which all of the systems have to be developed to attend to the users in the first place.

The initial conceptual basis for the proposal of the facets presented was based on the research previously performed focused on the Internet of Things (Motta, de Oliveira, and Travassos, 2018). With the progress of the discussions and reservations, we would like to confirm the findings and see if their applicability in this broader IoT context is feasible. The strategy used was to review the technical literature looking for the facets in the context of IoT.

We conducted the review in the Rapid Reviews format, which are adaptations of systematic literature reviews made to fit practitioners' constraints (Tricco *et al.*, 2015) and is beginning to be used in the context of Software Engineering (Cartaxo, Pinto and Soares, 2018). The procedure to be performed is the same. For this, we formatted a generic meta-protocol that was instantiated for each of the six facets presented (**Connectivity**, **Things**, **Behavior**, **Smartness**, **Interactivity,** and **Environment**)and considering the issue of **Security**, one of the most important and frequent challenges in the context of IoT (Motta, de Oliveira, and Travassos, 2018) to give us a transversal view of the challenge. The meta-protocol is detailed in section 3, and the results of each review are presented in the following sections.

# 3 EXECUTION

The revision was carried out in the context of a postgraduate discipline of the program of Systems Engineering and Computing of the Federal University of Rio de Janeiro. The discipline was Special Topics in Software Engineering. The revisions were carried out by the students at the postgraduation (master and Ph.D.) level, being accompanied by a doctoral student and the professor. Follow-up was carried out weekly, and the discussions and doubts were handled individually. The facets were distributed randomly through a lottery, and each student was responsible for instantiating the protocol for their respective facet. All protocols share the same PICOC structure, altering only the intervention for the respective facet. The discussion of the strings and the trials were done together with all participants. The execution occurred in the second half of 2018.

All the results are presented in the protocol, and a summary of the findings are also presented in the format of Evidence Briefings (EB). As described by Cartaxo et al. (2016), an EB is a one-piece document that reports the main findings of empirical research. It combines Information Design and Gestalt Theory elements to enable documents more appealing to the final reader and easier to read.

The main template, available to use under an open-source license (CC-BY) in the link http://cin.ufpe.br/eseg/briefings. We have adapted it to our context, with the main elements, as described below and represented in Figure 1:

- The title of the briefing (1), sometimes simplifying the paper title to make the briefing more appealing to the practitioners;
- The logos and identification of the research group and the university (2).

- A summary (3) to present the objective, motivation, facet definition, and context of the briefing;
- Informative box (4), separated from the main text, to highlight the target audience and the purpose of the briefing and answer the research questions;
- The additional information (5), extracted from the original empirical study;
- The references to the original empirical study (6);

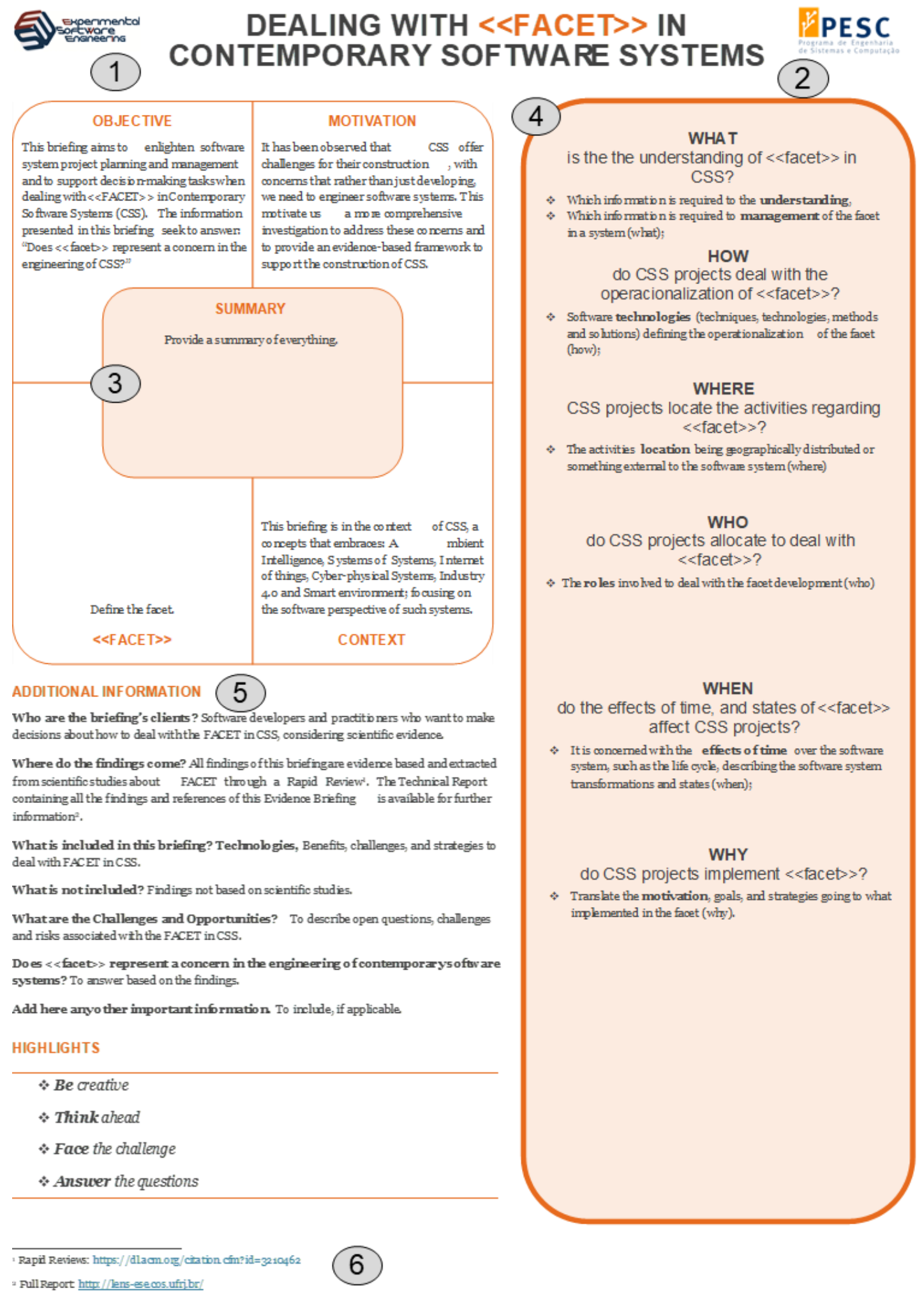

**Figure 1. Overview of the elements of the EB template.**

As we did on the protocol, this is a meta-template that should be instantiated for each facet.

# 4 META-PROTOCOL TEMPLATE

# Rapid Reviews Meta-Protocol:

## Engineering of Internet of Things Software Systems

**Student, Rebeca Motta, Guilherme H. Travassos**

## <<Facet>>

In the investigation regarding the Internet of Things (IoT), it has been observed that these modern software systems offer challenges for their construction since they are calling into question our traditional form of software development. Usually, they rely on different technologies and devices that can interact-capture-exchange information, act, and make decisions. It leads to concerns that, rather than just developing software, we need to engineer software systems embracing multidisciplinarity, integrating different areas. From our initial research, we analyzed the concerns related to this area. We categorized them into a set of facets - Connectivity, Things, Behavior, Smartness, Interactivity, Environment, and Security - representing such projects' multidisciplinarity, in the sense of finding a set of parts composing this engineering challenge.

Since these facets can bring additional perspectives to the software system project planning and management, acquiring evidence regarding such facets is of great importance to providing an evidence-based framework to support software engineers' decision-making tasks. Therefore, the following question should be answered:

***"Does <<facet>> represent a concern in the engineering of Internet of Things software systems?"***

This Rapid Review (RR) [1] aims to analyze <<facet>> to characterize it in the IoT field, regarding *what, how, where, when and why* is used in the context of IoT projects, verifying the existence of published studies supporting the previous results. The 5W1H aims to give the observational perspective on which information is required to the understanding and management of the facet in a system (what); to the software technologies (techniques, technologies, methods, and solutions) defining their operationalization (how); the activities location being geographically distributed or something external to the software system (where); the roles involved to deal with the facet development (who); the effects of time over the facet, describing its transformations and states (when); and to translate the motivation, goals, and strategies going to what is implemented in the facet (why), in respect of IoT projects.

## 4.1 Research Questions

- **RQ1:** What is the understanding and management of <<facet>> in IoT projects?
- **RQ2:** How do IoT projects deal with software technologies (techniques, technologies, methods, and solutions) and their operationalization regarding <<facet>>?
- **RQ3:** Where do IoT projects locate the activities regarding <<facet>>?
- **RQ4:** Whom do IoT projects allocate to deal with <<facet>>?
- **RQ5:** When do the effects of time, transformations, and states of <<facet>> affect IoT projects?
- **RQ6:** Why do IoT projects implement <<facet>>?

## 4.2 Search Strategy

The Scopus[1] search engine and the following search string support this RR, built using PICOC with five levels of filtering:

**P**opulation - Internet of Things software systems

Synonymous:

"ambient intelligence" OR "assisted living" OR "multiagent systems" OR "systems of systems" OR "internet of things" OR "Cyber-Physical Systems" OR "Industry 4" OR "fourth industrial revolution" OR "web of things" OR "Internet of Everything" OR "contemporary software systems" OR "smart manufacturing" OR digitalization OR digitization OR "digital transformation" OR "smart cit*" OR "smart building" OR "smart health" OR "smart environment"

**I**ntervention - <<facet>>

**C**omparison – no

[1] https://www.scopus.com

**O**utcome - Synonymous:

understanding OR management OR technique OR "technolog*" OR method OR location OR place OR setting OR actor OR role OR team OR time OR transformation OR state OR reason OR motivation OR aim OR objective

**C**ontext – (engineering or development or project or planning OR management OR building OR construction OR maintenance)

Limited to articles from 2013 to 2018

Limited to Computer Science and Engineering

LIMIT-TO (SUBJAREA, "COMP" ) OR LIMIT-TO (SUBJAREA, "ENGI" ) ) AND ( LIMIT-TO (PUBYEAR, 2018 ) OR LIMIT-TO (PUBYEAR, 2017 ) OR LIMIT-TO (PUBYEAR, 2016 ) OR LIMIT-TO (PUBYEAR, 2015)

| TITLE-ABS-KEY (<<to include>> ) |
|---|

## 4.3 Selection procedure

One researcher performs the following selection procedure:

1. Run the search string;
2. Apply the inclusion criteria based on the paper Title;
3. Apply the inclusion criteria based on the paper Abstract;
4. Apply the inclusion criteria based on the paper Full Text, and;

After finishing the selection from Scopus, use the included papers set to:

5. Execute snowballing backward (one level) and forward:
    a. Apply the inclusion criteria based on the paper Title;
    b. Apply the inclusion criteria based on the paper Abstract;
    c. Apply the inclusion criteria based on the paper Full Text.

The JabRef Tool[2] must be used to manage and support the selection procedure.

## 4.4 Inclusion criteria

- The paper must be in the context of **software engineering**; and

[2] http://www.jabref.org/

- The paper must be in the context of the **Internet of Things software systems**; and
- The paper must report a **primary or a secondary study**; and
- The paper must report an **evidence-based study** grounded in empirical methods (e.g., interviews, surveys, case studies, formal experiment, etc.); and
- The paper must provide data to **answering** at least one of the RR **research questions**.
- The paper must be written in the **English language**.

## 4.5 Extraction procedure

The extraction procedure is performed by one researcher, using the following form:

| <paper_id>:<paper_reference> | |
|---|---|
| Abstract | <Abstract> |
| Description | <A brief description of the study objectives and personal understanding> |
| Study type | <Identify the type of study reported by paper (e.g., survey, formal experiment)> |
| RQ1: WHAT<br>information required to understand and manage the facet in IoT | - < A1_1><br>- < A1_2><br>- ... |
| RQ2: HOW<br>software technologies (techniques, technologies, methods and solutions) and their operationalization | - < A2_1><br>- < A2_2><br>- ... |
| RQ3: WHERE<br>activities location or something external to the IoT | - < A3_1><br>- < A3_2><br>- ... |
| RQ4: WHO<br>roles involved to deal with the facet development in IoT | -< A4_1><br>-<A4_2><br>- … |
| RQ5: WHEN<br>effects of time over << facet>>, describing its transformations and states in IoT | - < A5_1><br>- < A5_2><br>- ... |
| RQ6: WHY<br>motivation, goals, and strategies regarding <<facet>> in IoT | - < A6_1><br>- < A6_2><br>- ... |

## 4.6 Synthesis Procedure

In this RR, the extraction form provides a synthesized way to represent extracted data. Thus, we do not perform any synthesis procedure.

However, the synthesis is usually performed through a narrative summary or a Thematic Analysis when the number of selected papers is not high [2].

## 4.7 Report

An Evidence Briefing [2] reports the findings to ease the communication with practitioners.

## 4.8 Results

**Execution**

- Execution date:
- Search result:
- Included by Title analysis:
- Included by Abstract analysis:
- Papers not found:
- Included after full read:
- Included on Snowballing:
- Total Included:

**Final Set**

<< to do>>

## 4.9 Summary of the articles

<< to do>>

## 4.10 Tracking Matrix

<< to do>>

## 4.11 Summary of the Findings

<< to do>>

## 4.12 Final Considerations

<< to do>>

## 4.13 References

[1] C. Tricco et al. A scoping review of rapid review methods. BMC Medicine, 2015.

[2] B. Cartaxo et al.: The Role of Rapid Reviews in Supporting Decision -Making in Software Engineering Practice. EASE 2018.

[3] B. Cartaxo et al. Evidence briefings: Towards a medium to transfer knowledge

from systematic reviews to practitioners. ESEM, 2016.

[4] << to complement>>

# DEVELOPING IOT SOFTWARE SYSTEMS? TAKE CONNECTIVITY INTO ACCOUNT

## CONTEXT

IoT allows composing systems from *things* equipped with identifying, sensing, or actuation behaviors and processing capabilities that can communicate and cooperate to reach a goal. For so, different facets should be taken into account, such as connectivity.

## CONNECTIVITY

It is necessary to have available a medium by which *things* can connect to materialize the IoT. The idea is not to limit Internet-only connectivity but to represent different forms of connections, therefore connectivity should be addressed in IoT.

## SUMMARY

As IoT system emerge, the necessity of connection among the components and other systems becomes clear. For this reason, *connectivity* is a fundamental aspect in IoT solutions. The understanding of connectivity, regarding its requirements, limitations and particularities can help develop better IoT solutions.

## OBJECTIVE

This briefing aims to enlighten project planning and to provide guidance when dealing with Connectivity in IoT. The information presented in this briefing seeks to answer: "What taken into account when considering connectivity for IoT?".

## MOTIVATION

IoT offers challenges for their construction, with concerns that rather than just developing, we need to engineer software systems. This motivated us to a more comprehensive investigation to address these concerns and to provide evidence-based support for IoT engineering.

## ADDITIONAL INFORMATION

**Who are the briefing's clients?** Software developers and practitioners who want to make decisions about how to deal with *connectivity* in IoT, considering scientific evidence.

**Where do the findings come?** All findings of this briefing were extracted from scientific studies about *connectivity* through a Rapid Review[1]. The Technical Report[2] containing all the findings is available for further information.

**What is included in this briefing?** Technologies, challenges, and strategies to deal with *connectivity* in IoT projects.

**What are the Challenges and Opportunities?** The articles analyzed indicate many technologies, but they do not show how they converge. There are concepts of topology, transmission, standards, protocols, virtualization, but there is a lack of explanation how these works together and their coverage for connectivity as whole. The main challenge is like a puzzle; there are some pieces already discovered but is not possible to solve it entirely due to the existing gaps and the absence of a blueprint showing how to fit the pieces together.

**Does connectivity represent a concern in the Engineering of Internet of Things Software Systems?** The evidence showed that, though has some characteristics well-defined which helps to understand the connectivity in IoT scenarios, there are still open questions.

## HIGHLIGHTS

❖ *The **Routing Process** became more difficult due to the mobility of the devices and the death of nodes*

❖ *It is difficult to provide **Network Management** due to the IoT specifics requirements*

❖ *To provide seamless connectivity we must pay attention to the **Infrastructure Integration***

## WHAT
### is the the understanding of connectivity in CSS?

- IoT require a **scamless connection** as well as **network traffic control** and **management**, providing low latency even with a limited bandwidth available.
- Connectivity in IoT will be deeply influenced by **devices limitations** and **domain requirements**, system should deal with limited resources (low memory capacity and low processing power), thus, requiring more efficient operations and solutions.

## HOW
### do CSS projects deal with the operacionalization of connectivity?

- Use **specific solutions** according to the application domain and solution requirements. For this reason, it is important to define relevant parameters like mobility, range, precision, availability.
- It is most based on wireless communication technologies such as: **Short-Range**, **Long-Range**, **Cellular-based** or a combination such as **mesh networking.**
- Can use **virtualization techniques** such as SDN and NFV.

## WHERE
### CSS projects locate the activities regarding connectivity?

- Extrinsic Connectivity technologies are scattered outside the system, with the possibility to **re-use legacy cellular** infrastructure or invest on **novel communication solutions.** Intrinsic Connectivity technologies are present in the **Network Architecture** and the **Network layers.**
- Geographically distributed devices, sometimes, in remote and critical regions. For this reason, an extended coverage area is needed no matter what the communication technology.
- The system deployment location directly affects its operation. Therefore, **Mobility** and **constant change** should be in the center of decision-making regarding connectivity.

## WHO
### do CSS projects allocate to deal with connectivity?

- **Engineers**, **architects** and **technicians** are responsible for tailor the model, design and implement proposed solution.
- Definitions in the architecture as well **legal**, **regulatory** and **contractual** issues have significant influence in the connectivity solution.

## WHEN
### do the effects of time, and states of connectivity affect CSS projects?

- The connection must be fluid, and this infrastructure integration must not be noted for the final user. For critical applications **low latency** is required, since a delay could be fatal.

## WHY
### do CSS projects implement connectivity?

- Connectivity is the core of IoT, and by overcoming device's limitations and domain requirements it enables the communication among devices for the solution achieve its goal.

[1] More on Rapid Reviews: https://dl.acm.org/citation.cfm?id=3210462
[2] Full Report: https://goo.gl/no7jtR

# 5 RAPID REVIEW ON CONNECTIVITY

# Rapid Reviews Meta-Protocol:

Engineering of Internet of Things Software Systems

**Andréa Doreste, Rebeca C. Motta, Guilherme H. Travassos**

# Connectivity

In the investigation regarding Internet of Things Software Systems (IoT), it has been observed that these modern software systems offer challenges for their construction since they are calling into question our traditional form of software development. Usually, they rely on different technologies and devices that can interact-capture-exchange information, act, and make decisions. It leads to concerns that, rather than just developing software, we need to engineer software systems embracing multidisciplinarity, integrating different areas. From our initial research, we analyzed the concerns related to this area. We categorized them into a set of facets - Connectivity, Things, Behavior, Smartness, Interactivity, Environment, and Security - representing such projects' multidisciplinarity, in the sense of finding a set of parts composing this engineering challenge.

Since these facets can bring additional perspectives to the software system project planning and management, acquiring evidence regarding such facets is of great importance to providing an evidence-based framework to support software engineers' decision-making tasks. Therefore, the following question should be answered:

***"Does Connectivity represent a concern in the engineering of***

***Internet of Things software systems?"***

This Rapid Review (RR) aims to analyze Connectivity to characterize it in the IoT field, regarding *what, how, where, when, and why* it is used in the context of IoT projects, verifying the existence of published studies supporting the previous results. The 5W1H aims to give the observational perspective on which information is required to the understanding and management of the facet in a system (what); to the software technologies (techniques, technologies, methods, and solutions) defining their operationalization (how); the activities location being geographically distributed or something external to the software system (where); the roles involved to deal with the facet development (who); the effects of time over the facet, describing its transformations and states (when); and to translate the motivation, goals, and strategies going to what is implemented in the facet (why), in respect of IoT projects.

## 5.1 Research Questions

- **RQ1:** What is the understanding and management of Connectivity in IoT projects?

- **RQ2:** How do IoT projects deal with software technologies (techniques, technologies, methods, and solutions) and their operationalization regarding Connectivity?
- **RQ3:** Where do IoT projects locate the activities regarding Connectivity?
- **RQ4:** Whom do IoT projects allocate to deal with Connectivity?
- **RQ5:** When do the effects of time, transformations, and states of Connectivity affect IoT projects?
- **RQ6:** Why do IoT projects implement Connectivity?

## 5.2 Search Strategy

The Scopus[3] search engine and the following search string support this RR:

**P**opulation - Internet of Things software systems

Synonymous:

"ambient intelligence" OR "assisted living" OR "multiagent systems" OR "systems of systems" OR "internet of things" OR "Cyber-Physical Systems" OR "Industry 4" OR "fourth industrial revolution" OR "web of things" OR "Internet of Everything" OR "contemporary software systems" OR "smart manufacturing" OR digitalization OR digitization OR "digital transformation" OR "smart cit*" OR "smart building" OR "smart health" OR "smart environment"

**I**ntervention - Connectivity

Synonymous:

connectivity OR "system connection" OR "software connection" OR "things connection" OR "objects connection"

**C**omparison - no

**O**utcome - answers to RQs

understanding OR management OR technique OR “technolog*” OR method OR location OR place OR setting OR actor OR role OR team OR time OR transformation OR state OR reason OR motivation OR aim OR objective

**C**ontext

engineering or development or project or planning OR management OR building OR construction OR maintenance

Limited to articles from 2013 to 2018

Limited to Computer Science and Engineering

LIMIT-TO (SUBJAREA, "COMP" ) OR LIMIT-TO (SUBJAREA, "ENGI" ) ) AND ( LIMIT-TO (PUBYEAR, 2018 ) OR LIMIT-TO (PUBYEAR, 2017 ) OR LIMIT-TO (PUBYEAR, 2016 ) OR LIMIT-TO (PUBYEAR, 2015)

[3] https://www.scopus.com

| |
|---|
| TITLE-ABS-KEY (( "ambient intelligence" OR "assisted living" OR "multiagent systems" OR "systems of systems" OR "internet of things" OR "Cyber Physical Systems" OR "Industry 4" OR "fourth industrial revolution" OR "web of things" OR "Internet of Everything" OR "contemporary software systems" OR "smart manufacturing" OR digitalization OR digitization OR "digital transformation" OR "smart cit*" OR "smart building" OR "smart health" OR "smart environment") AND (connectivity OR "system connection" OR "software connection" OR "things connection" OR "objects connection") AND (understanding OR management OR technique OR "technolog*" OR method OR location OR place OR setting OR actor OR role OR team OR time OR transformation OR state OR reason OR motivation OR aim OR objective) AND (engineering or development or project or planning OR management OR building OR construction OR maintenance) AND ( LIMIT-TO ( SUBJAREA , "COMP" ) OR LIMIT-TO ( SUBJAREA , "ENGI" ) ) AND ( (LIMIT-TO ( PUBYEAR , 2018 ) OR LIMIT-TO ( PUBYEAR , 2017 ) OR LIMIT-TO ( PUBYEAR , 2016 ) OR LIMIT-TO ( PUBYEAR , 2015)))) |

## 5.3 Selection procedure

One researcher performs the following selection procedure:

1. Run the search string;
2. Apply the inclusion criteria based on the paper Title;
3. Apply the inclusion criteria based on the paper Abstract;
4. Apply the inclusion criteria based on the paper Full Text, and;

After finishing the selection from Scopus, use the included papers set to:

5. Execute snowballing backward (one level) and forward:
    a. Apply the inclusion criteria based on the paper Title;
    b. Apply the inclusion criteria based on the paper Abstract;
    c. Apply the inclusion criteria based on the paper Full Text.

The JabRef Tool[4] must be used to manage and support the selection procedure.

## 5.4 Inclusion criteria

- The paper must be in the context of **software engineering**; and
- The paper must be in the context of the **Internet of Things software systems**; and
- The paper must report a **primary or a secondary study**; and
- The paper must report an **evidence-based study** grounded in empirical methods (e.g., interviews, surveys, case studies, formal experiment, etc.); and
- The paper must provide data to **answering** at least one of the RR **research questions**.
- The paper must be written in the **English language**.

## 5.5 Extraction procedure

The extraction procedure is performed by one researcher, using the following form:

| <paper_id>:<paper_reference> | |
|---|---|
| Abstract | <Abstract> |
| Description | <A brief description of the study objectives and personal understanding> |

[4] http://www.jabref.org/

| | |
|---|---|
| Study type | <Identify the type of study reported by paper (e.g., survey, formal experiment)> |
| RQ1: WHAT information required to understand and manage the Connectivity in IoT | - < A1_1><br>- < A1_2><br>- ... |
| RQ2: HOW software technologies (techniques, technologies, methods and solutions) and their operationalization | - < A2_1><br>- < A2_2><br>- ... |
| RQ3: WHERE activities location or something external to the IoT | - < A3_1><br>- < A3_2><br>- ... |
| RQ4: WHO roles involved to deal with the Connectivity development in IoT | -< A4_1><br>-<A4_2><br>- … |
| RQ5: WHEN effects of time over Connectivity, describing its transformations and states in IoT | - < A5_1><br>- < A5_2><br>- ... |
| RQ6: WHY motivation, goals, and strategies regarding Connectivity in IoT | - < A6_1><br>- < A6_2><br>- ... |
| Limitations | |

## 5.6 Synthesis Procedure

In this RR, the extraction form provides a synthesized way to represent extracted data. Thus, we do not perform any synthesis procedure.

However, the synthesis is usually performed through a narrative summary or a Thematic Analysis when the number of selected papers is not high.

## 5.7 Report

An Evidence Briefing [2] reports the findings to ease the communication with practitioners. It was presented as the cover for this chapter.

## 5.8 Results

### Execution

| Activity | Execution date | Result | Number of papers |
|---|---|---|---|
| First execution | 16/07/2018 | 781 documents added | 781 |
| Remove duplicated | 16/07/2018 | Three documents were withdrawn | 778 |
| Remove conferences/workshops | 16/07/2018 | 26 documents withdrawn | 752 |
| Included by Title analysis | 18/07/2018 | 633 documents withdrawn | 119 |
| Included by Abstract analysis | 19/07/2018 | 88 documents withdrawn | 31 |
| Papers not found | 19/07/2018 | Four documents were withdrawn | 27 |
| Articles for reading | 19/07/2018 | 27 documents | 27 |
| Removed after a full reading | 20/07/2018 – 07/08/2018 | 16 documents withdrawn | 11 |
| Snowballing | 07/08/2018 | 11 documents added | 22 |
| Snowballing after reading | 07/08/2018 – 11/08/2018 | Nine documents were withdrawn | 13 |
| Total included | 11/08/2018 | 13 documents | 13 |
| Papers extracted | 20/07/2018 – 11/08/2018 | 13 documents | 13 |

**Final Set**

| Reference | Author | Title | Year | Source |
|---|---|---|---|---|
| [1] | LUZURIAGA et al. | Handling mobility in IoT applications using the MQTT protocol | 2015 | Regular search |
| [2] | CENTENARO et al. | Long-range communications in unlicensed bands: The rising stars in the IoT and smart city scenarios. | 2016 | Regular search |
| [3] | DHUMANE et al. | Routing issues in the internet of things: a survey | 2016 | Snowballing |
| [4] | BERA et al. | Software-defined networking for the internet of things: A survey | 2016 | Regular search |
| [5] | POLURU et al. | A Literature Review on Routing Strategy in the Internet of Things | 2017 | Regular search |
| [6] | ROJAS et al. | Enabling Connectivity of Cyber-Physical Production Systems: A Conceptual Framework | 2017 | Regular search |
| [7] | YAQOOB et al. | Enabling communication technologies for smart cities. | 2017 | Regular search |
| [8] | AKPAKWU et al. | A survey on 5G networks for the Internet of Things: Communication technologies and challenges | 2018 | Regular search |
| [9] | BAUMANN et al. | Utilizing the Tor Network for IoT Addressing and Connectivity | 2018 | Regular search |
| [10] | CHEN et al. | Optimal routing for multi-hop social-based D2D communications in the Internet of Things | 2018 | Regular search |
| [11] | JIN et al. | Design and Implementation of a Wireless IoT Healthcare System Based on OCF IoTivity | 2018 | Regular search |
| [12] | LI et al. | 5G internet of things: A survey | 2018 | Snowballing |
| [13] | MURAKAMI et al. | Drawing Inspiration from Human Brain Networks: Construction of Interconnected Virtual Networks | 2018 | Regular search |

## 5.9 Summary of the articles

### Handling mobility in IoT applications using the MQTT protocol

Protocols (as known as middleware) are among the elements involved to enable communication among devices in network applications. The most widely adopted protocols in IoT scenarios are MQTT (Message Queuing Telemetry Transport), CoAP (Constrained Application Protocol), and LWM2M (Lightweight Machine-to-Machine), and all of them are based on TCP/IP (Transmission Control Protocol/Internet Protocol) protocol suite. Although TCP/IP can work adequately in wired communication, the same does not occur in the Internet of Things (IoT) scenarios. Devices in IoT systems tend to have an intermittent connection with the Internet, and TCP/IP does not perform well in this case. For example, in a case of failure, TCP/IP will request the lost packet's resend, but it will fail once one node has a different IP address (a common situation in IoT applications).

The MQTT protocol works in a publisher/subscriber architecture. The packets are placing in a queue on the message broker, and all the nodes subscribed to this queue automatically receive the message. The MQTT protocol has a QoS (Quality of Service) parameter set to three levels to achieve good performance. The first discards the message in the queue once sent, but the others are only discarded after messages are achieved. These confirmation messages can cause overhead in the network; thus, this could not be a better solution. To try a better approach, this work proposes a solution to adapt MQTT to mobile scenarios. The main idea is to intermediate buffering the messages and guarantees that the node will receive them in case of a lost connection, whether the node has the same IP address or not.

**Long-range communications in unlicensed bands: The rising stars in the IoT and smart city scenarios**

According to the article, "connectivity is probably the most basic building block of the IoT paradigm." Two main approaches are provided: short-range communication technologies in the unlicensed spectrum (such as ZigBee and Bluetooth) and long-range legacy cellular technologies in licensed frequency bands (2G/GSM/GPRS). Although useful in the IoT scenarios, these two sets of technologies have their limitations. For example, short-range transmission technologies have a limited coverage area, and cellular-based communications cannot deal with a large number of devices present in an IoT system.

Low Power WANS (LPWANs) communication emerges to provide seamless integration in the IoT paradigm as an intermediate solution. LPWANs operate in sub-gigahertz unlicensed frequency bands, characterized by long-range radio links and star topology. The end devices are directly connected to a single node; the collector provides the bridging to the IP world.

In order to understand the issues about connectivity in IoT, the article brings an overview of the LPWANs and shows an experiment where LoRa was used in a big city.

**Routing Issues in the Internet of Things: A Survey**

According to this article, the Internet of Things “promises to build a world where all the objects around us will be connected." These objects will be the primary users of the internet and will 'communicate with each other to gather, share, and forward the information' about the environment. However, differently from the traditional systems, most of the devices in IoT have intermittent connectivity. Besides, one of the main characteristics of IoT systems is an ever-changing network topology due to the mobility and the death of nodes. For these reasons, routing data has become a significant challenge in IoT scenarios.

The present survey explores the routing of the data in IoT in order to analyze, compare and consolidate past research works as well as discuss the applicability of their findings in IoT scenarios

**Software-Defined Networking for Internet of Things: A Survey**

Internet of Things has emerged as one of the most prominent paradigms of the 21st century where devices, also known as "Things," will be connected to the Internet to communicate and exchange data. Although connectivity is the core of this new technology, the traditional network infrastructure is not prepared to support IoT requirements. Traditional devices, such as switches and routers, are usually preprogrammed to do particular tasks and follow precise rules (protocols). This type of behavior does not meet the IoT application-specific requirements because the devices cannot adapt to multiple rules that must be attended to provide optimal network services.

A new concept, known as software-defined networking (SDN), was proposed to address such limitations. SDN is an emerging network architecture where network control can be decoupled from the traditional hardware devices allowing network operators and users to control and access the network devices remotely, providing a global view of the network, allowing network management and dynamic resource allocation. This article provides a brief overview of the existing SDN-based technologies that can be useful for IoT.

**A literature review on routing strategy in the internet of things**

Routing protocols are a meaningful part of network applications as they specify the best route between any two nodes among multiple possibilities. This process provides communication between devices on the Internet and, consequently, is an essential element in order to provide connectivity.

Although routing is a comprehensive process in conventional network applications, some Internet of Things requirements made it difficult to happen in IoT scenarios. Some of the reasons are the limited resource available on the devices (processing power and memory capacity), the topology changes due to the dead of nodes and mobility, and the energy efficiency concern (most devices use a battery and have a limited lifetime). These issues must be solved or managed in IoT systems, and this paper aims to provide a comprehensive discussion and a literature review on routing in IoT.

**Enabling Connectivity of Cyber-Physical Production Systems: A Conceptual Framework**

It is possible to say that Industry 4.0 is a paradigm where Information Technology (IT) and Operational Technology (OT) domains will be joint in order to provide to physical objects in industrial environments cognitive, communication and autonomy capabilities as well as the abilities of perception and actuation in the environment. In order to guarantee that, advanced communication skills are a fundamental element since systems like these are designed to stay connected and are characterized by high-structured information regularly exchanged among the devices.

Implement a complex system like this is not an easy task. To facilitate the process, the present paper aims to develop a conceptual framework capable of exploiting the cognitive and communication capabilities of Industry 4.0 systems. It had a premise that this can be achieved by realizing a typical communication network with a set of standard protocols and was implemented as a proof of concept at the Mini-Factory Laboratory of the Free University of Bolzano.

**Enabling Communication Technologies for Smart Cities**

According to this article, Smart cities are "technology-intensive cities that can offer collection, analyzation, and distribution of information to transform services offered to its citizen, increase operational efficiency, and entail better decisions at the municipal level." Smart home, smart industry, and smart grid are applications that a smart city can involve.

Due to the number of possible applications and their nature, Smart City can be characterized as a particular scenario with some specific requirements and challenges. For example, a transportations management application deals with humans' lives, and network delay can cause serious accidents. Moreover, in the Smart City domain, a considerable number of devices are connected, using different radio technologies and, therefore, interference among them must be managed and avoided. In order to understand these scenarios, this article discusses some enabling communication technologies used in smart cities and exams in some real smart city applications.

**A Survey on 5G Networks for the Internet of Things: Communication Technologies and Challenges**

Internet of Things has emerged as one of the most significant technologies these days. It has the objective of providing communication between devices, allowing them to interact and exchange data, enabling "a conducive environment that will impact and influence several aspects of everyday life and business applications." IoT can be applied in many situations and types, and connectivity is one of the most basic requirements. A variety of communication technologies has emerged with specific characteristics, benefits, and limitations to ensure that.

The primary objective of this study was to review the existing IoT key communication technologies. According to it, there is no one-fits-for all solution, and, because of that, some IoT domains were explored, as Smart Home, Industrial IoT, Smart Healthcare, and so on. Furthermore, the article divides these technologies into three groups: long-range network (Lora, SigFox, Ingenu-RPMA, etc.), Short-Range Network (Bluetooth, ZigBee, WiFi), Cellular-based technologies (2G, 3G,5G, 3GPP LTE) and exposes the main characteristics of each one of them.

**Utilizing the Tor Network for IoT Addressing and Connectivity**

One of the Internet of Things fundamental problems is the devices' addressability. It is essential to address them in every situation, even in scenarios where the network has restrictive settings, like firewalls. In order to present a solution to this problem, this paper proposes a system where Tor (The Onion Routing/Router) network is used.

Tor is an overlay protocol based on the nesting of data packages within each other that are partially unpacked by stations along the communication path. Tor is usually used to allow safe and encrypted data exchanged. Still, its data structure (that can be identified and filtered by firewalls) can provide the technological basis for static addressing in the IoT scenario.

**Optimal Routing for Multi-Hop Social-Based D2D Communications in the Internet of Things**

Internet of Things promises to improve human lives by providing innovative services conceived for a wide range of application domains. A considerable number of communication technologies have been explored to enable the IoT scenario. One of them is Device-to-Device (D2D) communication, a promising technology for IoT due to its high-power efficiency, high spectral efficiency, and low transmission delay. Furthermore, many devices in IoT are based on human behavior and, consequently, the social relationships of people (the owners of these devices) should be considered in the IoT D2D communication.

Most of the models that use this type of communication assume that the nodes are grouped into transmitter/receiver pairs, and each pair trusts each other by default, which is impractical. The present paper aims to derive a routing algorithm for maximizing the trusted connectivity probability (T-CP), a model to select an optimal path for multi-hop social-based D2D communication in the IoT scenario.

**Design and Implementation of a Wireless IoT Healthcare System Based on OCF IoTivity**

Internet of Things (IoT) systems consist of sensing, networking, data processing, and application. There are many possible application domains in IoT, and one of them is the health care system. With IoT technologies, the health care system can easily be operated by users such as doctors, nurses, patients, etc. Moreover, sensing and remote monitoring can improve the quality of healthcare services and help reduce healthcare costs.

In this paper, a wireless IoT healthcare system is proposed and implemented as a proof of concept. It uses OCF IoTivity to implement communication between the devices and the server. Besides, an Intel Edison board is used to implement the E-health device (with sensors) and the server.

**5G Internet of Things: A Survey**

Applications in the Internet of Things (IoT) domain require extensive connectivity, security, trustworthiness, the ultra-reliable connection, among other requirements for a large number of devices and, though used in IoT scenarios, 2G, 3G, and 4G technologies are not fully optimized for IoT applications. To solve this set of specific requirements, 5G technologies emerge. 5G is expected to provide new connectivity interfaces for future IoT applications, solving challenges such as complex communication, low computational devices' capabilities, intermittent connection, and fast communication and capacity.

Although 5G is at its early stage, the main progress has been made in the field. The present paper aims to present the recent research on both 5G technologies and IoT, analyzing the IoT new requirements, the critical communication techniques, and the challenges present in this scope.

**Drawing inspiration from human brain networks: Construction of interconnected virtual networks**

The virtualization of wireless sensor networks (VWSN) is a key technology enabling the IoT scenario. However, no strategies have been proposed to efficiently generate virtualized topologies that can achieve a high level of performance, quick communication between any pair of nodes, high resilience against failures of network components, and low cost.

Moreover, some researches in the complex network field have been using the brain network as an inspiration to understand how efficiently connect two or more nodes as brain networks possess excellent characteristics that have been optimized through the process of evolution. The present study aims to do the same and has the objective of constructing a VWSN Network topology inspired by the human brain. This network has two main layers: physical (where the real network infrastructure is located) and virtual (where the virtual topology is created) and aims to ensure three essential requirements of IoT applications: communication efficiency, robustness, and low construction cost.

## 5.10 Tracking Matrix

| Articles | WHAT | HOW | WHERE | WHO | WHEN | WHY |
|---|---|---|---|---|---|---|
| 1. Handling mobility in IoT applications using the MQTT protocol | X | X | X | | | X |
| 2. Long-range communications in unlicensed bands: The rising stars in the IoT and smart city scenarios. | X | X | X | | | |
| 3. Routing issues in the internet of things: a survey | X | X | X | | | X |
| 4.Software-defined networking for the internet of things: A survey | X | X | X | | | |
| 5. A Literature Review on Routing Strategy in the Internet of Things | X | X | X | | | |
| 6. Enabling Connectivity of Cyber-Physical Production Systems: A Conceptual Framework | X | X | | | | |
| 7. Enabling communication technologies for smart cities. | X | X | | | | X |
| 8. A survey on 5G networks for the Internet of Things: Communication technologies and challenges | X | X | | | | |
| 9. Utilizing the Tor Network for IoT Addressing and Connectivity | X | X | X | | | |
| 10. Optimal routing for multi-hop social-based D2D communications in the Internet of Things | X | X | X | | | |
| 11. Design and Implementation of a Wireless IoT Healthcare System Based on OCF IoTivity | X | X | | | | |
| 12. 5G internet of things: A survey | X | X | X | | | X |
| 13. Drawing Inspiration from Human Brain Networks: Construction of Interconnected Virtual Networks | X | X | X | | | |

## 5.11 Summary of the Findings

**RQ1: WHAT is the understanding of Connectivity in IoT?**

To manage the Connectivity facet in IoT projects, some information and requirements need to be understood.

First, it is important to note that there is no one-fit-for-all solution [1]. Internet of Things englobes many domains, and each one of them will have particular characteristics and requirements.

However, a set of requirements could be captured through the Rapid Review execution. Usually, they are a more intrinsic connection to the devices' nature or application needs, but they will directly influence communication. The main requirements are:

- **Efficient operations** – 10 of the 13 articles in the final set adverts that IoT devices have low power capacity and tends to work with a battery, sometimes in very restricted regions where change the battery is not an option [1][2][3][4][5][6][8][10][12][13]. Besides, four articles advocate that these devices have low memory capacity as well as low processing power [1][4][11][13]. Thus, the communication operations must be done with the maximum efficiency possible to extend the device's lifetime and guarantee proper functioning.
- **A large number of Devices** – 9 of the 13 papers relates that IoT system is composed of a considerable number of nodes [1][2][3][4][5][6][10][12][13].
- **Distributed nodes** – According to 3 articles, the large number of devices above mentioned is geographically distributed [2][4][10] and located, sometimes, in remote and critical regions [1][8][10][13].
- **Extended Coverage area** – 5 of the 13 papers disserts that to attend a large number of devices, geographically distributed, an extended coverage area is needed no matter what the communication technology was chosen for the system [1][3][5][10][12].
- **Mobility** – IoT devices are not static; they tend to move between different coverage areas. According to 6 articles, that is a problem in IoT system, mainly because, in many systems (like Device-to-Device communication), these devices are used to route the packets and, once they left the area, the routing configuration must be redone [1][2][3][4][10][12][13].

- **High Scalability** – According to 6 articles, the IoT system demands high scalability because of the perspective of growing and the need to add new devices as the system evolves [1][3][4][6][12][13].
- **High Availability** – According to 4 papers, high availability is a requirement in the IoT system because it is a way of guarantee the ubiquity of the system [1][2][3][5].
- **Seamless connectivity** – According to 7 articles, seamless connectivity is a fundamental requirement for IoT scenarios [1][2][3][7][9][10][13]. IoT systems involve many devices geographically distributed. The network between them tends to be composed of different technologies that need to be correctly integrated to maintain the service's quality. The connection must be fluid, and this infrastructure integration must not be noted for the final user.
- **Traffic Control and Management** – As mentioned above, the IoT system englobes communication among a vast number of devices that can cause congestion in the network. That is why five papers dissert about the importance of traffic management and control to deal with the enormous data generated by these devices and guarantee the quality of service [1][2][4][10][12].
- **Low latency** – Many IoT applications can be considered critical, for example, transport monitoring, that is why, according to 8 articles, a delay could be fatal and, thus, must be managed and avoided [1][2][3][4][5][6][10][12].
- **Bandwidth** – 7 of the 13 articles cites limited bandwidth as a characteristic of IoT systems. Thus, this must be taken into consideration when managing the system or develop an application [1][2][3][4][5][10][12].
- **Robustness** – Robustness is an essential characteristic of systems that deal with critical applications. That is why four articles emphasize that IoT systems must be robust [3][5][10][12].

As was said before, some of these requirements are not directly related to connectivity (such as the necessity of mobility or efficient operations due to the device's limitation). Still, they show aspects that will profoundly influence communication. Thus, they are requirements that need to be well understood and addressed to make IoT work.

**RQ2: HOW do IoT projects deal with the operationalization of Connectivity?**

As if it was said previously, there is no general solution for connectivity. According to 5 papers, each domain and application will require a specific configuration, but wireless technologies seem to be almost unanimity [1][3][8][9][12]. That said, part of them will work better with short-range network technologies. Another part will prefer long-range technologies, and the rest will opt for cellular-based systems.

In short-range communications, it is possible to cite Bluetooth Low Energy, ZigBee, Z-Wave, NFC (Near Field Communication), RFID (Radio-Frequency Identification), and Wi-Fi as enabling technologies [1][3][4][8][9][10][12].

In long-range communications, Low-Power Wide-Area technologies like LoRa (Long-Range), SigFox, Ingenu-RPMA (Random Phase Multiple Access) can be cited [1][10][12].

In cellular-based communications, the 2G, 3G, 4G appear even with their limitations, as well as Third Generation Partnership Project (3GPP), like LTE (Long Term Evolution) and LTE-Advance [1][3][6][10][12].

Another technique cited in the articles to provide connectivity was virtualization, like Software-Defined Network (SDN) and Network Function Virtualization (NFV). They construct a higher layer where network functionalities can be implemented, removing the responsibility of management from the network infrastructure [1][2][3][5][12].

A common question addressed in the articles is whether IoT should re-use the legacy cellular infrastructure. According to five articles, IoT systems and legacy cellular networks are incompatible because of the specific requirements of the IoT applications [1][2][10][12][13]. Another six articles support the re-use, but, apparently, in this case, some adaptions must be made, and technologies must evolve to support these requirements (like Bluetooth with Bluetooth Low Energy) [1][3][4][6][9][10].

The last technique mentioned to provide connectivity between two devices was protocols, especially routing protocols, cited by 11 articles as an essential aspect of the network and a challenge that must be taken into considerations in IoT systems [1][3][4][6][7][8][9][10][11][12][13].

**RQ3: WHERE IoT projects locate the activities regarding Connectivity?**

Through the Rapid Review execution was possible to note that the activities regarding Connectivity are located on the Network Architecture [1][2][3][4][5][9][10][12][13]; consequently, on some Network Layers [1][3][5][9][10].

**RQ4: WHOM do IoT projects allocate to deal with Facets?**

No evidence was found through the Rapid Review execution to answer this question.

**RQ5: WHEN do the effects of time, transformations, and states of Connectivity affect IoT projects?**

No evidence was found through the Rapid Review execution to answer this question. However, connectivity is a crucial aspect of IoT systems, and the activities regard it as a present from the beginning to the end in IoT projects.

**RQ6: WHY do IoT projects implement Connectivity?**

Connectivity is the core of various IoT systems, like the Internet of Things, Smart City, Smart Home, Industry 4.0, etc. It enables communication among devices and provides a considerable number of possible applications [3] [1]. These applications provide facilities for everyday life [12] or even improve the industrial process [7].

## 5.12 Final Considerations

After the Rapid Review execution, it is possible to say that IoT is yet at its early stage. Connectivity was shown as one of the main aspects of the future systems, being deeply influenced by the device's limitations and domain requirements such as mobility, low power capacity, a large number of devices geographically distributed through an extended area, and so on.

The problem is that many technologies that have been used to develop Contemporary Software Systems cited through the RR cannot attend to the complete set of requirements.

They were created years ago to operate in other domains, such as cellular communication. Furthermore, they need to be adapted to work in IoT scenarios and deal with their limitations and specifications while still compatible with the classic Internet Structure.

Although many people have been searching for this topic, the results do not seem to fit. The articles analyzed through the Rapid Review indicate and explain many technologies, which is already an achievement, but they do not show how they will converge. For example, they talk about network topology, transmission technologies, standards, protocols, virtualization techniques, etc. but they do not show how these technologies will work together to provide connectivity. It is like a puzzle; some pieces are already discovered, but it is not possible to solve it entirely due to the existing gaps and the absence of a blueprint showing how to fit the pieces together.

Perhaps, a suggestion to help solve this problem would be creating a conceptual model about connectivity, representing how the different types of technologies would hierarchically relate with each other.

The OSI (Open Systems Interconnection) model represents how the communication parts fit together at the traditional Internet infrastructure. It is a conceptual model with seven different layers. Each of them makes explicit what types of technologies must operate there, what functions it must be capable of doing, and how it must communicate with the other layer’s technologies. Although not implemented in practice, it helps define the scope of operation, understand how things should work, manage the functionalities, and understand where the technology or technique fits.

However, this type of model does not exist in the IoT domain and, due to the differences between the systems, it is not possible to say whether the OSI model would work. A new conceptual model (or an adapted one) should englobe the specific technologies used in IoT (such as network virtualization techniques) and the various transmission technologies, communication protocols, and so on. It would facilitate understanding the IoT system's connectivity aspect, how each technology should operate, and how to combine them to provide connectivity. In other words, it would work as a blueprint, providing a systematic and useful view of the entire connectivity facet.

## 5.13 References

**Final Set:**

[1] LUZURIAGA, Jorge E; CANO, Juan Carlos; CALAFATE, Carlos; *et al*. Handling mobility in IoT applications using the MQTT protocol. *In*: **Internet Technologies and Applications (ITA), 2015**. [s.l.]: IEEE, 2015, p. 245–250.

[2] CENTENARO, Marco; VANGELISTA, Lorenzo; ZANELLA, Andrea; *et al*. Long-range communications in unlicensed bands: The rising stars in the IoT and smart city scenarios. **IEEE Wireless Communications**, v. 23, n. 5, p. 60–67, 2016.

[3] DHUMANE, Amol; PRASAD, Rajesh; PRASAD, Jayashree. Routing issues in the internet of things: a survey. *In*: **Proceedings of the international multiconference of engineers and computer scientists**. [s.l.: s.n.], 2016, v. 1, p. 16–18.

[4] BERA, Samaresh; MISRA, Sudip; VASILAKOS, Athanasios V. Software-defined networking for the internet of things: A survey. **IEEE Internet of Things Journal**, v. 4, n. 6, p. 1994–2008, 2017.

[5] POLURU, Ravi Kumar; NASEERA, Shaik. A Literature Review on Routing Strategy in the Internet of Things. **Journal of Engineering Science & Technology Review**, v. 10, n. 5, 2017.

[6] ROJAS, Rafael A; RAUCH, Erwin; VIDONI, Renato; *et al*. Enabling Connectivity of Cyber-Physical Production Systems: A Conceptual Framework. **Procedia Manufacturing**, v. 11, p. 822–829, 2017.

[7] YAQOOB, Ibrar; HASHEM, Ibrahim Abaker Targio; MEHMOOD, Yasir; *et al*. Enabling communication technologies for smart cities. **IEEE Communications Magazine**, v. 55, n. 1, p. 112–120, 2017.

[8] AKPAKWU, Godfrey Anuga; SILVA, Bruno J; HANCKE, Gerhard P; *et al*. A survey on 5G networks for the Internet of Things: Communication technologies and challenges. **IEEE Access**, v. 6, p. 3619–3647, 2018.

[9] BAUMANN, Felix W; ODEFEY, Ulrich; HUDERT, Sebastian; *et al*. Utilising the Tor Network for IoT Addressing and Connectivity. *In*: **CLOSER**. [s.l.: s.n.], 2018, p. 27–34.

[10] CHEN, Gaojie; TANG, Jinchuan; COON, Justin P. Optimal routing for multi-hop social-based D2D communications in the Internet of Things. **IEEE Internet of Things Journal**, 2018.

[11] JIN, Wenquan; HONG, Yong-Geun; KIM, Do-Hyeun. Design and Implementation of a Wireless IoT Healthcare System Based on OCF IoTivity. **INTERNATIONAL JOURNAL OF GRID AND DISTRIBUTED COMPUTING**, v. 11, n. 4, p. 87–96, 2018.

[12] LI, Shancang; DA XU, Li; ZHAO, Shanshan. 5G internet of things: A survey. **Journal of Industrial Information Integration**, 2018.

[13] MURAKAMI, Masaya; KOMINAMI, Daichi; LEIBNITZ, Kenji; *et al*. Drawing Inspiration from Human Brain Networks: Construction of Interconnected Virtual Networks. **Sensors**, v. 18, n. 4, p. 1133, 2018.

**Additional References:**

[14] TRICCO C.; ANTONY, J.; ZARIN, W.; STRIFLER, L.; GHASSEMI, M.; IVORY, J.; PERRIER, L.; HUTTON, B.; MOHER, D.; STRAUS, S. A scoping review of rapid review methods. BMC Medicine, 2015.

[15] CARTAXO, B.; PINTO, G.; SOARES, S. The Role of Rapid Reviews in Supporting Decision -Making in Software Engineering Practice. EASE 2018.

[16] CARTAXO, B.; PINTO, G.; SOARES, S. Evidence briefings: Towards a medium to transfer knowledge from systematic reviews to practitioners. ESEM, 2016.

# DEVELOPING IOT SOFTWARE SYSTEMS? TAKE THINGS INTO ACCOUNT

## CONTEXT

IoT allows composing systems from *things* equipped with identifying, sensing, or actuation behaviors and processing capabilities that can communicate and cooperate to reach a goal. For so, different facets should be taken into account, such as *things*.

## THINGS

It exists in the physical realm, such as sensors, actuators or any objects equipped with identifying, sensing or acting behaviors and processing capabilities that can communicate and cooperate to reach a goal, varying according to the systems requirements.

## SUMMARY

Improving quality of life, productivity, reducing costs are some of the advantages provided by IoT. In the core of IoT is the *things*. They make possible to sense, act, identify and interact in new forms with users and environment. Acting as a bridge between the virtual and physical worlds, these smart devices and components should

## OBJECTIVE

This briefing aims to enlighten project planning and to provide guidance when dealing with Things in IoT. The information presented in this briefing seeks to answer: "What taken into account when considering *things* for IoT?".

## MOTIVATION

IoT offers challenges for their construction, with concerns that rather than just developing, we need to engineer software systems. This motivated us to a more comprehensive investigation to address these concerns and to provide evidence-based support for IoT engineering.

## ADDITIONAL INFORMATION

**Who are the briefing's clients?** Software developers and practitioners who want to make decisions about how to deal with *things* in IoT, considering scientific evidence.

**Where do the findings come?** All findings of this briefing were extracted from scientific studies about *things* through a Rapid Review[1]. The Technical Report[2] containing all the findings is available for further information.

**What is included in this briefing?** Technologies, challenges, and strategies to deal with *things* in IoT projects.

**What are the Challenges and Opportunities?** How to define the requirements, the large amount and the heterogeneity of devices, the necessity to deal with limit resources, interoperability, security and privacy are among the main challenges to deal with *things* in IoT.

**Do things represent a concern in the Engineering of Internet of Things Software Systems?** Yes. *Things* are in the core of every IoT solution, constantly evolving in some aspects (such as resources capabilities) but with several open challenges (such as safety and security), it should be carefully addressed in every step of development.

## HIGHLIGHTS

❖ *The **combination of technologies** allows developing solutions for several scenarions, previously not possible.*

❖ ***Microcontrollers and sensors** where widely used in most of the solutions.*

## WHAT
### is the the understanding of things in IoT?

- Everyday devices that enables **identification**, **sensing**, **actuation**, **monitoring**, **interacting** with the user or environment.
- It permits an **evolutionary model of system design and deployment.**

## HOW
### do IoT projects deal with the operacionalization of things?

- Define the components according to their **behavior**, **limitations** and **events**.
- Provide strategies to **adapt**, **build** or **buy** components.
- Establish procedures for **integration** and **protection** in case of external components and do-it-yourself approaches.
- In terms of technologies there are a lots of solutions that were combined to build the devices like sensors, actuators, smartphones, microcontrollers, tags, interactables, cameras, communication and network enablers, among others.

## WHERE
### IoT projects locate the activities regarding things?

- The location is basically the own **environment** where the solution is deployed and depends on the problem domain.
- Consider the environment impact on the components and vice-versa.
- The solutions where in places like houses, shopping, transport systems, smart cities, factories, roads, military, farms, mines, hospitals, offices, airport, among others.

## WHO
### do IoT projects allocate to deal with things?

- **Engineers**, **architects** and **technicians** are responsible for tailor the model, design and implement proposed solution.
- **Users** can actively participate in the process with do-it-yourself approaches and more active roles.
- **Actors** in the system can be other systems and are not limited to humans anymore. Therefore, the interactions, roles and responsibilities should be well defined.

## WHEN
### do the effects of time, and states of things affect IoT projects?

- The context of use can affect the behavior of the device, have consequences on the system. **Repeatable patterns and rules** should be defined.
- It is important to **unify temporality** across different components as well as procedures in case of **real-time operation.**

## WHY
### do IoT projects implement things?

- In general, the use of *things* with part of a solutions in IoT provide a **series of benefits** such as comfort, simplify user life, reduce costs, accessibility, reduce the risk of mistakes, efficiency, enhance productivity, decrease energy consumption, support in decision making, automate a manual process, and remote control and monitoring.

[1] More on Rapid Reviews: https://dl.acm.org/citation.cfm?id=3210462
[2] Full Report: https://goo.gl/no7jtR

# 6 RAPID REVIEW ON THINGS

# Rapid Reviews Meta-Protocol:

## Engineering of Internet of Things Software Systems

**Danyllo V. da Silva, Rebeca C. Motta, Guilherme H. Travassos**

# Things

In the investigation regarding Internet of Things Software Systems (IoT), it has been observed that these modern software systems offer challenges for their construction since they are calling into question our traditional form of software development. Usually, they rely on different technologies and devices that can interact-capture-exchange information, act, and make decisions. It leads to concerns that, rather than just developing software, we need to engineer software systems embracing multidisciplinarity, integrating different areas. From our initial research, we analyzed the concerns related to this area. We categorized them into a set of facets - Connectivity, Things, Behavior, Smartness, Interactivity, Environment, and Security - representing such projects' multidisciplinarity, in the sense of finding a set of parts composing this engineering challenge.

Since these facets can bring additional perspectives to the software system project planning and management, acquiring evidence regarding such facets is of great importance to providing an evidence-based framework to support software engineers' decision-making tasks. Therefore, the following question should be answered:

***"Do Things represent a concern in the engineering of***

***Internet of Things software systems?"***

This Rapid Review (RR) aims to analyze Things to characterize it in the IoT field, regarding *what, how, where, when and why* is used in the context of IoT projects, verifying the existence of published studies supporting the previous results. The 5W1H aims to give the observational perspective on which information is required to the understanding and management of the facet in a system (what); to the software technologies (techniques, technologies, methods, and solutions) defining their operationalization (how); the activities location being geographically distributed or something external to the software system (where); the roles involved to deal with the facet development (who); the effects of time over the facet, describing its transformations and states (when); and to translate the motivation, goals, and strategies going to what is implemented in the facet (why), in respect of IoT projects.

## 6.1 Research Questions

- **RQ1:** What is the understanding and management of Things in IoT projects?
- **RQ2:** How do IoT projects deal with software technologies (techniques, technologies, methods, and solutions) and their operationalization regarding Things?

- **RQ3:** Where do IoT projects locate the activities regarding Things?
- **RQ4:** Whom do IoT projects allocate to deal with Things?
- **RQ5:** When do the effects of time, transformations, and states of Things affect IoT projects?
- **RQ6:** Why do IoT projects implement Things?

## 6.2 Search Strategy

The Scopus[5] search engine and the following search string support this RR:

**P**opulation – Internet of Things software systems

Synonymous:

"ambient intelligence" OR "assisted living" OR "multiagent systems" OR "systems of systems" OR "internet of things" OR "Cyber-Physical Systems" OR "Industry 4" OR "fourth industrial revolution" OR "web of things" OR "Internet of Everything" OR "contemporary software systems" OR "smart manufacturing" OR digitalization OR digitization OR "digital transformation" OR "smart cit*" OR "smart building" OR "smart health" OR "smart environment"

**I**ntervention:

"tag" OR "mobile phone" OR "addressable thing" OR "spime" OR "smart item" OR "virtual thing" OR "identifiable thing" OR "smart object" OR "audio receiver" OR "video receiver"

**C**omparison - no

**O**utcome:

understanding OR management OR technique OR “technolog*” OR method OR location OR place OR setting OR actor OR role OR team OR time OR transformation OR state OR reason OR motivation OR aim OR objective

**C**ontext – (engineering or development or project or planning OR management OR building OR construction OR maintenance)

Limited to articles from 2013 to 2018

Limited to Computer Science and Engineering

LIMIT-TO (SUBJAREA, "COMP" ) OR LIMIT-TO (SUBJAREA, "ENGI" ) ) AND ( LIMIT-TO (PUBYEAR, 2018 ) OR LIMIT-TO (PUBYEAR, 2017 ) OR LIMIT-TO (PUBYEAR, 2016 ) OR LIMIT-TO (PUBYEAR, 2015)

TITLE-ABS-KEY (("ambient intelligence" OR "assisted living" OR "multiagent systems" OR "systems of systems" OR "internet of things" OR "Cyber Physical Systems" OR "Industry 4" OR "fourth industrial revolution" OR "web of things" OR "Internet of Everything" OR

[5] https://www.scopus.com

"contemporary software systems" OR "smart manufacturing" OR digitalization OR digitization OR "digital transformation" OR "smart cit*" OR "smart building" OR "smart health" OR "smart environment") AND ("tag" OR "mobile phone" OR "addressable thing" OR "spime" OR "smart item" OR "virtual thing" OR "identifiable thing" OR "smart object" OR "audio receiver" OR "video receiver") AND (understanding OR management OR technique OR "technolog*" OR method OR location OR place OR setting OR actor OR role OR team OR time OR transformation OR state OR reason OR motivation OR aim OR objective) AND (engineering OR development OR project OR planning OR management OR building OR construction OR maintenance)) AND ( LIMIT-TO ( PUBYEAR , 2018 ) OR LIMIT-TO ( PUBYEAR , 2017 ) OR LIMIT-TO ( PUBYEAR , 2016 ) OR LIMIT-TO ( PUBYEAR , 2015)) AND (LIMIT-TO ( SUBJAREA,"COMP") OR LIMIT-TO (SUBJAREA, "ENGI"))

## 6.3 Selection procedure

One researcher performs the following selection procedure:

1. Run the search string;
2. Apply the inclusion criteria based on the paper Title;
3. Apply the inclusion criteria based on the paper Abstract;
4. Apply the inclusion criteria based on the paper Full Text, and;

After finishing the selection from Scopus, use the included papers set to:

5. Execute snowballing backward (one level) and forward:
    a. Apply the inclusion criteria based on the paper Title;
    b. Apply the inclusion criteria based on the paper Abstract;
    c. Apply the inclusion criteria based on the paper Full Text.

The JabRef Tool[6] must be used to manage and support the selection procedure.

## 1.1 Inclusion criteria

1. The paper must be in the context of **software engineering**; and
2. The paper must be in the context of the **Internet of Things software systems**; and
3. The paper must report a **primary or a secondary study**; and
4. The paper must report an **evidence-based study** grounded in empirical methods (e.g., interviews, surveys, case studies, formal experiment, etc.); and
5. The paper must provide data to **answering** at least one of the RR **research questions**.
6. The paper must be written in the **English language**.

## 6.4 Extraction procedure

The extraction procedure is performed by one researcher, using the following form:

| <paper_id>:<paper_reference> | |
|---|---|
| Abstract | <Abstract> |
| Description | <A brief description of the study objectives and personal understanding> |
| Study type | <Identify the type of study reported by paper (e.g., survey, formal experiment)> |

[6] http://www.jabref.org/

| | |
|---|---|
| RQ1: WHAT information required to understand and manage the << facet>> in IoT | - < A1_1><br>- < A1_2><br>- ... |
| RQ2: HOW software technologies (techniques, technologies, methods and solutions) and their operationalization | - < A2_1><br>- < A2_2><br>- ... |
| RQ3: WHERE activities location or something external to the IoT | - < A3_1><br>- < A3_2><br>- ... |
| RQ4: WHO roles involved to deal with the << facet>> development in IoT | -< A4_1><br>-<A4_2><br>- … |
| RQ5: WHEN effects of time over << facet>>, describing its transformations and states in IoT | - < A5_1><br>- < A5_2><br>- ... |
| RQ6: WHY motivation, goals, and strategies regarding <<facet>> in IoT | - < A6_1><br>- < A6_2><br>- ... |
| Additional Information and Comments | - <Important information not necessarily related to research questions><br>- <Personal comments> |

## 6.5 Synthesis Procedure

In this RR, the extraction form provides a synthesized way to represent extracted data. Thus, we do not perform any synthesis procedure.

However, the synthesis is usually performed through a narrative summary or a Thematic Analysis when the number of selected papers is not high.

## 6.6 Report

An Evidence Briefing [2] reports the findings to ease the communication with practitioners. It was presented as the cover for this chapter.

## 6.7 Results

### Execution

| Activity | Execution date | Result | Number of papers |
|---|---|---|---|
| First execution | 27/09/2018 | 830 documents added | 830 |
| Remove proceedings | 28/09/2018 | 48 documents withdrawn | 782 |
| Included by Title analysis | 29/09/2018 - 30/09/2018 | 622 documents withdrawn | 160 |
| Included by Abstract analysis | 29/09/2018 - 30/09/2018 | 126 documents withdrawn | 33 |
| Papers not found | 02/10/2018 | Four documents were withdrawn | 29 |
| Articles for reading | 02/10/2018 - 07/10/2018 | 29 documents | 29 |
| Removed after a full reading | 02/10/2018 - 07/10/2018 | Eight documents were withdrawn | 21 |
| Snowballing | 08/10/2018 - 10/10/2018 | 47 documents added | 68 |
| Snowballing after reading | 08/10/2018 - 10/10/2018 | 38 documents withdrawn | 30 |
| Total included | 10/10/2018 | 30 documents | 30 |
| Papers extracted | 10/10/2018 - 21/10/2018 | 30 documents | 30 |

**Final Set**

| Reference | Author | Title | Year | Source |
|---|---|---|---|---|
| [1] | Rezvan and Barekatain | The Sensors Are Innovative in the Internet of Things | 2015 | Regular search |
| [2] | Suryavanshi et al. | Integration of Smart Phone and IOT for development of the smart public transportation system | 2016 | Regular search |
| [3] | Ciampolini et al. | MuSA: A Smart Wearable Sensor for Active Assisted Living | 2016 | Regular search |
| [4] | Jha et al. | Internet of things enabled smart switch | 2016 | Regular search |
| [5] | Rathod and Khot | Smart assistance for the public transport system | 2016 | Regular search |
| [6] | Li et al. | A 220-volts power switch controlled through WiFi | 2016 | Regular search |
| [7] | Tew and Ray | ADDSMART: Address digitization and smart mailbox with RFID technology | 2016 | Regular search |
| [8] | Khoury et al. | An IoT approach to vehicle accident detection, reporting, and navigation | 2016 | Regular search |
| [9] | Dalli and Bri | ACQUISITION DEVICES IN INTERNET OF THINGS: RFID AND SENSORS | 2016 | Regular search |
| [10] | Materazzi et al. | RFID temperature sensors for monitoring soil solarization with biodegradable films | 2016 | Regular search |
| [11] | Boonchieng et al. | An integrated system of applying the use of internet of things, RFID and cloud computing: A case study of logistic management of electricity generation authority of Thailand (egat) mae mao lignite coal mining, lampang, Thailand | 2017 | Regular search |
| [12] | Khan, S. F. | Health care monitoring system in the Internet of Things (IoT) by using RFID | 2017 | Regular search |
| [13] | Kumar et al. | A study on the possible application of the RFID system in different real-time environments | 2017 | Regular search |
| [14] | Saraf and Gawali | IoT based smart irrigation monitoring and controlling system | 2017 | Regular search |
| [15] | Juang et al. | Bidirectional smart pillbox monitored through the internet and receiving reminding message from remote relatives | 2017 | Regular search |
| [16] | Pawar and Deosarkar | Health condition monitoring system for a distribution transformer using the Internet of Things (IoT) | 2017 | Regular search |
| [17] | Huertas and Mendez | Biomedical IoT Device for Self-Monitoring Applications | 2017 | Regular search |
| [18] | Meroni et al. | Design and development of a nearable wireless system to control indoor air quality and indoor lighting quality | 2017 | Regular search |
| [19] | Kuo et al. | Development of unmanned surface vehicle for water quality monitoring and measurement | 2018 | Regular search |
| [20] | Las-Heras Andrés et al. | RFID Technology for Management and Tracking: e-Health Applications | 2018 | Regular search |
| [21] | Esteve et al. | A Multimodal Fingerprint-Based Indoor Positioning System for Airports | 2018 | Regular search |
| [22] | Arsenio et al. | Wireless sensor and actuator system for smart irrigation on the cloud | 2015 | Snowballing |
| [23] | Meroni et al. | An open source low-cost wireless control system for a forced circulation solar plant | 2015 | Snowballing |
| [24] | Akkarajitsakul et al. | A control system in intelligent farming by using Arduino technology | 2016 | Snowballing |
| [25] | Shi et al. | Indoor positioning system based on inertial MEMS sensors: Design and realization | 2016 | Snowballing |
| [26] | Ray, P. P. | Internet of things cloud-enabled MISSENARD index measurement for indoor occupants | 2016 | Snowballing |
| [27] | Kamble and Vatti | Bus tracking and monitoring using RFID | 2017 | Snowballing |
| [28] | Huang and Mao | Occupancy estimation in the smart building using hybrid CO2/light wireless sensor network | 2017 | Snowballing |
| [29] | Pitarma et al. | System Based on the Internet of Things for Real-Time Particle Monitoring in Buildings | 2018 | Snowballing |
| [30] | Gong et al. | Active Plant Wall for Green Indoor Climate Based on Cloud and Internet of Things | 2018 | Snowballing |

## 6.8 Summary of the articles

**The Sensors Are Innovative in the Internet of Things [1]**

This paper gives us an extensive discussion of the current scenario where objects and systems are connected to the internet to collect the data (receive and transmit) and share information, such as location. In this context, they discuss RFID tags and NFC sensors since they play a broader role in IoT, and their production, service, and utility cost has decreased. The authors explain the operation and components of these technologies. Also, they examine RFID tags and NFC sensors regarding the protocols used and other attributes for evaluation. At last, they discuss why, among all of the sensors and protocols used in the IoT field, the RFID tags and NFC sensors are the best for different cases.

**Integration of Smart Phone and IOT for development of smart public transportation system [2]**

This paper presented an efficient framework of Intelligent Public Transport Management System that dynamically tracks all the buses' location and estimates the bus terminal's arrival time. The paper is inserted in the area of Smart Cities and gives a general discussion of this context. Their system is updated at regular intervals every time the bus sends an update to the server. In this case, an attractive solution is the use of smartphones since they are ever-present in our daily life, and their ever-increasing power makes them an attractive solution for developing IoT applications. It distributes bus information, on-demand, to passengers who send a request using a smartphone application or through SMS. The traffic issues can be curbed as more and more people will opt for efficient and economical public transportation as a medium of travel on a frequent and regular basis. The user is furnished with explicit information about the nearest buses' current location approaching a mobile application's bus stop. With the information on-demand service, the commuters can plan their journey well in advance, saving much time and making the individual more productive. The next arriving bus terminal's commotion is also solved by the in-bus display module, which will give details of the route at regular intervals. By using readily available Android API’s, 3G network, and SMS based services, it is a useful solution to assist commuters, drivers, and also the administrators of the transport system in a very convenient manner.

**MuSA: A Smart Wearable Sensor for Active Assisted Living [3]**

This paper focuses on features introduced in the wearable sensor MuSA to support behavioral analysis within the context of the HELICOPTER project, funded in the AAL European joint program. In general clinical, wearable, and environmental sensors have been selected and implemented, aiming at early detection of symptoms of age-related diseases. In particular, the wearable device performs two essential functions: on the one hand, it is used as a behavioral data source, continuously monitoring the quantity of user physical activity (through the energy expenditure index evaluation), location, and posture; on the other hand, MuSA enables the fusion of data coming from the environmental sensors, correctly attributing actions on a particular sensor to a specific user in a multi-user environment. These functions are carried out without the need for external devices (RFID tags, etc.), but only relying on sensors embedded on the wearable device and its communication capabilities. Some specific features have been investigated regarding trading off among performance, cost, and system intrusiveness. Some sample results from pilot studies are shown to exploit a living lab environment, which involves nearly 50 elderly users in two different European countries.

**Internet of things enabled smart switch [4]**

This paper has reviewed the currently available home automation solutions focusing on smart switches and has identified a critical issue that has hindered the popularization of such technology. The idea of manually operating a switch is replaced by a smart technology that involves operating the switches using phones, laptops, or other electronic gadgets. However, despite interesting, it presented to be expensive to purchase and implement such a solution. The current work makes use of a Web App and a cloud to control the operation of the switches. A cloud server is created for the environment where the switches are mounted. The switches are interfaced with few electronic components such as logic gates, a 555 timer, flip-flops, and a processor. The user communicates with the processor through the Web App. The processor controls the switches based on the user's commands and updates the user about the switches' status after the control operation is performed to the cloud. The system has been tested, implemented, and found to function as intended, with an overall cost estimated to be approximately 75% less expensive than most current, commercially available solutions.

**Smart assistance for public transport system [5]**

This paper introduces another solution for the Smart Cities domain with a Smart Assistance in Public Transport System. The project is to be implemented for the public bus with an online tracking system by using GPS/GSM system using a smartphone app. Some interesting features are presented concerning public security and safety, safety for women, alcohol detection, driver authentication with RFID tags, and monitoring facilities. Also, an accident detection that includes the accident location to provide immediate assistance. It has an app for users to track the bus on smart by using a smartphone with the possibility for both online (GPS) and offline (GSM) services. Their work also discusses different ideas to make public transport smarter and more accessible so the whole public can take advantage of it. The evaluation was carried out from a prototype-level implementation for the public transport system (bus). The results taken are real-time and help improve a safe and smart system in the field of automation.

**A 220-volts power switch controlled through WiFi [6]**

The materialization of the internet of things is gradually coming into people's lives with communication between things and wireless networking technology. The paper proposes to apply such principles in a household appliance with a power switch controlled through WiFi. This essay's innovation point is that it can control in the distance (so long as places with WiFi), and the loads can be controlled by smartphone in not only on-off but the *watt* level, considering the safety, low voltage control high voltage is another innovation point. WIFI module communicates with MCU by serial ports, which will send the commands from mobile phones to the center proceeding module, then MCU takes the order and controls the appliance (by an android app). The features are included wireless, long-distance, high-speed transporting, seamless integration with the wired network, intelligence monitor, and control system. As the future idea is to apply the solution in a new house, so there is no need to fix up so many electric wires and observe the economic and environmental impact.

**ADDSMART: Address digitization and smart mailbox with RFID technology [7]**

ADDSMART is a research project focused on digitizing addresses of locations and building a smart mailbox by combining wireless sensors, cameras, locks, and RFID readers and tags into a system controlled by an Arduino microcontroller board. The goal was to create a smart mailbox prototype with various services, including digitized building addresses and driveway monitoring. They work with the idea of address digitization incorporated into a mailbox that can communicate wirelessly with the homeowner to provide mail status and home security. The proposal is demonstrated in a proof-of-concept of how these different components and systems can be combined to create a system that may be used for address digitization, mail notification, and home surveillance. They also discussed the challenges and future steps within the research.

**An IoT approach to vehicle accident detection, reporting, and navigation [8]**

This work is concerned with Public Safety Organizations (PSO) regarding vehicle accidents. Their motivation is related to the difficulty in reaching the injured people due to late alerts and insufficient information about the specific accident location. To contribute in this direction, they present an IoT system solution that instantly notifies the PSO headquarter whenever an accident occurs and pinpoints its geographic coordinates on the map. The solution relies on sensors to detects accidents and locations, control central to process information, and find a PSO headquarter, indicating accident occurrence. Some advantages, when compared to traditional systems, are minimizing injured passengers' interaction, providing necessary medical information to rescue teams, recognizing exact and accurate accident locations, and facilitating the routing process. The evaluation that the system is feasible and reliable. As future steps, the data collected can be fed to data mining and generate statistical reports related to the number of accidents, number of injured, a bank of blood donors, and road conditions.

**Acquisition Devices In Internet Of Things: Rfid And Sensors [9]**

This paper discussed the basics concepts of IoT. In particular, the architecture of IoT is introduced. The proposed architecture is based on three layers: the perception layer, a sensor-based technology responsible for collecting real-time data from different sources. The network layer comprises the tools used in the communication of IoT hardware to software applications. The application layer is then responsible for processing the massive amount of data and information using knowledge to achieve companies' objectives, for the perception layer, where RFID and sensors can be placed. The paper presents a review of RFID and sensors technologies, detailing tags and readers, and different categories of sensors (for monitoring space, objects, or monitoring interactions between objects and space). The contribution of the paper is to provide information for choosing the right technology for an IoT system. They also present technical challenges that need to be addressed to realize the IoT mainly related to Dimensions, Energy, Security, QoS, and configuration management.

**RFID temperature sensors for monitoring soil solarization with biodegradable films [10]**

The work can be applied in the field of agriculture and is motivated by the fact that monitoring tools may help in solarization management. Therefore, the paper presents the testing of RFID sensor applications for soil solarization purposes. They used different soils to assess RFID temperature sensors. The soil solarization treatment was carried out as a case of study during a period characterized by changeable weather using a biodegradable film. Their findings show that integrating information technology solutions with new-generation biodegradable films may offer an exciting revaluation of soil solarization in a real farm organization. Also, they conclude that RFID temperature sensors represent easy-to-use and cheap tools to support the decision-making process during long-term treatment such as solarization. A future solution is to integrate the system within smartphone applications and allowed easier real-time monitoring for farmers.

**An integrated system of applying the use of the internet of things, RFID, and cloud computing [11]**

RFID, Internet of things, and Cloud Computing were used to apply for logistic management of lignite coal mine trucks of EGAT. They use RFID tags for lignite coal trucks, and data from RFID proceed through a server and are stored in a private cloud computing. The integrated RFID, IoT, and Cloud system has been used for one year, from 2015 to 2016, and has been operational 24 hours a day, seven days a week. The integrated system was considered stable. Information was considered reliable and accurate (if compared to information done manually). Officers who authorize and are responsible for logistic information were satisfied and reported that the system enhanced logistics' best practices.

**Health care monitoring system in the Internet of Things (IoT) by using RFID [12]**

This paper discusses the use of IoT in health is presenting concepts such as wireless wellbeing monitoring, U-healthcare, E-healthcare, Age-friendly healthcare techniques. Later they propose a monitoring cycle and healthcare system designed by using the IoT and RFID tags. In this proposed system, the configuration comprises the association between microcontroller and actuator using an ECG sensor, Blood Pressure sensor, Temperature sensor, Motion sensor, EEG sensor, and Blood Glucose sensor patient's health is shared with medical professionals via smartphones using loT. The architecture is divided into the sensor, network, internet, and service layers. The study results provided a positive and robust output against various medical emergencies.

**A study on the possible application of the RFID system in different real-time environments [13]**

This paper lists several applications of RFID technology in different sectors. They present a historical presentation and discussion of the area and then examine and classify them by the type (active or passive), frequency range (Low Frequency, High Frequency, and Ultra-High Frequency ranges). It also presents the advantages and disadvantages of deploying technology in the respective fields, and some evaluation in performance is identified by reliability, cost-effectiveness, efficiency, and usage. Some RFID use cases are presented regarding its use in the supply chain, car parking, traffic control, vehicle positioning, and tracking items. A detailed insight is given about RFID in the traffic control system for possible smart city development applications. They conclude that RFID is an upcoming new technology implemented by many companies and organizations to improve their services.

**IoT based smart irrigation monitoring and controlling system [14]**

The proposal presented fits interconnected field studies of the Internet of Things, Machine-to-Machine and Wireless Sensor, and Actuator Networks. The authors begin by presenting general concepts of the area and interconnection. Smart objects embedded with sensors enable interaction with the physical and logical worlds according to the concept of IoT. In this paper, they apply the IoT paradigm in a smart irrigation system for a farm. Some of the technologies used are an android smartphone for controlling the system; Zigbee for communication between sensor and base station; a web-based java graphical user interface for monitoring and data processing and a cloud-based wireless communication system. The evaluation was a proof-of-concept in the form of an automated irrigation system that was liable to be developed and can manage irrigation water supply effectively. The idea is to use the solution for agricultural applications since it helps to optimize water use. The consumption is reduced with the implementation of a soil-moisture-based automated irrigation system.

**Bidirectional smart pillbox monitored through the internet and receiving reminding message from remote relatives [15]**

The papers' motivation is given by the rapid population aging and the need to promote health, well-being, and quality of life for the elderly. The technological and smart solution can contribute to achieving these goals. In related works, they present some assistive devices that integrate sensors as part of a network communicating by wireless communication, transmitting data to a personal computer or mobile phone. In this sense, they propose an interactive smart pillbox that detects if the older person is taking the pills and displaying reminders in the pillbox with text or voice to the users, characterized as bidirectional. This application uses the Webduino module installed in SPB to achieve two-way messaging with remote relatives via the internet. The module first reads the sensing signal in the kit. It uses WiFi to transmit the WiFi Router signal and then sends the medication information to a remote web page or cell phone for monitoring (LCD). Some input care messages can be inserted on the web page or mobile phone, and Webduino will receive the message send it to Arduino to display it in the pillbox. With this solution, the relatives or responsible can efficiently manage the elderly medication and take care of them even in distances.

**Health condition monitoring system for a distribution transformer using the Internet of Things (IoT) [16]**

This paper presents a mobile embedded system's development to monitor and record parameters of a distribution transformer like Current, Temperature, Rise or Fall of Oil level, Vibration, and Humidity. The proposal can be applied in several areas, is useful primarily for the industry. The focus is on the Distribution Transformer since it is a significant component of the power system, and its correct functioning is vital to system operations. They implemented a method to the GSM/GPRS based on monitoring transformer health and a remote terminal unit installed at the distribution transformer site. The system is designed based on a PIC 18F4550 microcontroller, which acts as a data acquisition & transmission system. Suppose any abnormality is observed in the parameters. In that case, the system sends alerts through mobile phones and monitoring units that contain information about abnormality to some predefined instructions programmed in the microcontroller. The primary goal is to optimally utilize the protection of the transformer's power line and identify problems before any catastrophic failure. Some evaluation is presented for the Remote Terminal Unit also for monitoring and communication modules.

**Biomedical IoT Device for Self-Monitoring Applications [17]**

The paper's idea is focused on implementing a local monitoring system for patients to provide a better medical service by keeping track of their vital signals while they stay at home. The motivation is based on a contextual issue in Colombia, where medical services coverage is not enough to fulfill patients' growing demand in medical institutions. Their proposal uses an acquisition system that collects patient data at a scheduled time by integrating different sensors of a physiological monitor. The communication module between the acquisition system and a local mobile device is made with Bluetooth due to the technology's simplicity and availability. This system provides the visualization of data acquired by sensors in a mobile device through an Android application to let the patient study the vital signals' current status. Android was used since it is compatible with as many mobile devices. An evaluation was conducted with volunteers students to observe the solution feasibility, and this preliminary test has shown that all the acquired data from the different sensors can be displayed simultaneously in the mobile phone achieving a real-time local visualization platform, and the users were able to understand their current health status with no complication.

**Design and development of a nearable wireless system to control indoor air quality and indoor lighting quality [18]**

This paper proposes implementing indoor air quality (IAQ), indoor lighting quality (ILQ) manage and control through smart objects using the principles of do-it-yourself. It describes the results of an “open-source smart lamp” able to manage and control the indoor environmental quality (IEQ) of the built environment. They introduce the hardware and software used in their case, relying on nearable wireless for connectivity. An interesting point in the proposal is to be built following a do-it-yourself (DIY) approach. The solution comprises the monitoring and coordination station, actuation, data connection, and control algorithm. They used a microcontroller with temperature and humidity sensors, with a monitor that allows the adjustment of the indoor thermal comfort quality (ICQ) by interacting directly with the air conditioner. The evaluation was conducted in an office environment with mechanical ventilation and an air conditioning system with two different settings by four participants, presenting exciting results at the end of the article.

**Development of unmanned surface vehicle for water quality monitoring and measurement [19]**

The work is inserted in the area of water management. It is of great importance since the rapid growth of populations and high tech industries. Furthermore, recent climate change also plays an essential role in this area. This paper proposes an Unmanned Surface Vehicle (USV) for monitoring water quality. They initially define the hardware requirements - sort of a small boat - and software for sensing and automation, considering water's environmental conditions. After this analysis, the proposed USV is developed and carries a mobile water quality sensor to perform a real-time scan of water qualities; it also includes a communication relay. Some places have no mobile signal coverage. They conduct a field test as proof of concept of the solution where several improvements can be observed for future work.

**RFID Technology for Management and Tracking: e-Health Applications[20]**

The research is inserted in the health domain. The authors introduce radiofrequency technologies used in this context for tracking and monitoring and highlight that this technology has become a key solution in the logistics and management industry, thanks to distinctive features such as the low cost of RFID tags and the easiness of the RFID tags' deployment and integration within the items to be tracked. They also present a brief comparison between different existent technologies that also represents the state of the area. The review focus on RFID for e-Health applications, describing contributions and limitations. Then they introduce their solution for a system that allows the tracking of medical assets that are tagged with RFID tags within a hospital. They explain the solution between hardware, signal, and software. Ultra-high-frequency (UHF) RFID technology is selected over the most extended near-field communication (NFC) and high-frequency (HF) RFID technology to minimize hardware infrastructure. In particular, UHF RFID also makes the coverage/reading area conformation easier by using different antennas. Information is stored in a database, accessed from end-user mobile devices (tablets, smartphones) where the assets' position and status to be tracked are displayed. In the end, they make a proof of concept using the technology in two rooms in a hospital.

**A Multimodal Fingerprint-Based Indoor Positioning System for Airports[21]**

The paper begins by justifying the importance of indoor positioning and then presents different methods, technologies, and indoor positioning techniques. This issue is becoming more popular in order to provide a seamless indoor positioning system applying the concept of the traditional GPS but for indoor environments to be widely deployed in many public and private buildings (e.g., shopping malls, airports, universities, etc.). Later they describe a general overview of the indoor position service process, introducing the Dora algorithm. Their proposal is differentiated since it uses two different technologies in order to provide improved accuracy (multimodal approach with self and remote positioning). Their solution is responsible for estimating the user location depending on the collected real-time measurements by comparing this collection with the ones available in the Fingerprinting database (radio map). In the end, they present the results of the evaluation study, deployed in an airport, showing that their positioning service with a multimodal approach performs better when compared to individual approaches.

**Wireless sensor and actuator system for smart irrigation on the cloud[22]**

The authors introduce the paper with the relevance and applicability of a smart irrigation solution on a smart farm. The idea of  Wireless Sensor Network can perform acquisition, collection, and analysis of data, such as temperature and soil moisture, and be employed to automate the irrigation process in agriculture while decreasing water consumption, resulting in monetary and environmental benefits. However, some of the challenges are the need to acquire data to be further analyzed and compared. If this data is not organized and processed, it will become meaningless. In this sense, they introduce their proposal and present some related work compared to their own. Their solution is implemented using the cloud for data source identification, performing data validation, partitioning, and processing. Also, the weather information is collected from a 3rd party system. The data is fetched by an HTTP Request and is returned in JSON format. The irrigation solution can be visualized to access historical data related to soil moisture and battery measurements for each sensor and precipitation index and predictions through a user dashboard.

**An open source low-cost wireless control system for a forced circulation solar plant[23]**

The system, implemented following the do-it-yourself (DIY) approach and the use of open hardware and low-cost sensors, allows a solar thermal system with forced circulation to be independently managed, using temperature data from a self-powered wireless station. The article details the design phase, development, and practical application of the solution, and the authors initially describe the calibration before the application in a real scenario. The idea is that the system provides an electric pump for the circulation of heat transfer fluid connecting the solar thermal panel to the storage tank. An external control unit is connected with a temperature sensor, and an internal control unit with an algorithm to manage the solution. A calibration activity was conducted to assess the accuracy of the external control unit. This DIY approach influences the selection of devices and solution implementation. After the calibration, they conducted a study to observe the feasibility.

**A control system in intelligent farming by using Arduino technology [24]**

The work is inserted in the agricultural area in a Thai context. The authors propose an intelligent farming system to improve the production process and automate part of the irrigation process. The system is composed of two main parts: a sensor and a control system. The control part is responsible for watering and roofing an outdoor farm based on the statistical data sensed from the sensor systems (including temperature, humidity, moisture, and light intensity sensors). They apply the Kalman filtering to smooth the data to improve its accuracy. An interesting part is that for the decision making, they also count on the weather information alongside the sensed data, based on a decision tree considering the weather condition. A set of decision rules based on both the sensed data and the predicted weather condition is developed to automatically decide whether the watering and roofing system should be on or off. A mobile application is presented to control the watering and roofing systems manually. The proposal is evaluated in an observational study.

**Indoor positioning system based on inertial MEMS sensors: Design and realization [25]**

The authors propose a foot wearable device with a wireless network to transmit movement information to a computer that can calculate the relative position and show the path walked by. It is based on the inertial MEMS sensor (microelectromechanical sensor), including an accelerometer, gyroscope, and magnetometer. The solution is discussed divided into hardware and software, presenting algorithms adjustments for accuracy, and the system can help people get accurate positioning, especially for indoor environments. They propose an alternative solution for the gait phase detection with machine learning and decision tree techniques. It was different from the traditional approaches and is interesting since this detection is the key to the whole positioning accuracy. A study is conducted to observe the feasibility.

**Internet of things cloud-enabled MISSENARD index measurement for indoor occupants [26]**

Thermal comfort is an essential factor in the human body and impacts the livelihood factors of a human, such as lifestyle, productivity, societal activity, and individual health. MISSENARD index is one of the most suitable techniques to calculate the thermal comfort of indoor occupants. It can be determined by parameters like air temperature, air dynamics, relative humidity, age, health status, clothing style (thickness, color), and activity (physical, psychical). This paper has proposed a solution to measure the MISSENARD Index. The author introduces a general presentation of concepts like IoT, Cloud, and IoT in the Cloud - the intersection where their solution fits. He explains the implementation that used a microcontroller system, different sensors, a communication protocol, the cloud, and two services for visualizing the results. The developed system holds a novel way of smarter integration of sensor fueled data analytics with cloud supported visualization at the same time.

**Bus tracking and monitoring using RFID [27]**

The authors introduce the problem of public transportation in India and present some related works that use GPS and RFID technologies for tracking features. In this sense, the authors present an IoT-based bus tracking system that shows the bus's current location and their respective seat availability. They used RFID readers and tags, IR sensors, and an Arduino microcontroller for tracking the bus implemented with a wifi module for connectivity. They used the Thingspeak web server for displaying the location of the bus and seat availability in the android application. A proof of concept is presented to observe the study's feasibility. They discuss some potential benefits of the idea, such as reducing the waiting time and overcrowding at bus stops.

**Occupancy estimation in the smart building using hybrid CO2/light wireless sensor network [28]**

This paper is inserted in the smart building area. They start by pointing out solutions in this area and describing some related works. They propose a solution to monitor CO2 and light with sensors to observe the occupancy in an environment to promote energy-efficient smart building. The concentration level of indoor CO2 is a good indicator of the number of room occupants while protecting building residents' privacy. Once the indoor CO2 level is observed, HVAC equipment is aware of the number of room occupants. HVAC equipment can adjust its operating parameters to fit the demands of these occupants. Thus, the desired quality of service is guaranteed with minimum energy dissipation. The overall architecture consists of sensors, a connection module, and the control system. A simple evaluation is made in an office environment showing the proposal's functionalities and feasibility but is not sufficiently detailed.

**System Based on the Internet of Things for Real-Time Particle Monitoring in Buildings [29]**

The authors discuss the importance of indoor air quality for human health that can impact cardiovascular and respiratory diseases. In this context, the paper presents iDust, a real-time PM exposure monitoring system and decision-making tool for enhanced healthcare based on an IoT architecture. It was developed using open-source technologies and low-cost sensors, mainly the WEMOS D1 mini microcontroller and a PMS5003 PM sensor to measure particles' value for air quality. The data visualization, notification, and control are made through a dashboard. The proposal is evaluated and provided positive results, as the solution could be used to support the building manager for the appropriate operation and maintenance to deliver not just a safe but also a healthful workplace for enhanced occupational health

**Active Plant Wall for Green Indoor Climate Based on Cloud and Internet of Things [30]**

The authors begin the paper by introducing the context of plant walls and some related work with indoor climate and monitoring. The motivation is that the indoor climate is closely related to human health, well-being, and comfort. They propose a remote monitoring and management solution based on IoT and a public cloud platform. The system has been completely developed from hardware to software and from the local control unit to the cloud end. The microprocessor is the control unit connected to the Arduino microcontroller, which controls light, humidity, temperature, CO2, PM, and ultrasonic sensors. The cloud part has the essential role of enabling the remote monitoring and management system for plant walls since the data processing, storage, and interface for administrators are hosted in the cloud. It is a fascinating paper with several details of setup, devices, development, and implementation.

## 6.9 Tracking matrix

| Ref | Paper | WHAT | HOW | WHERE | WHOM | WHEN | WHY |
|---|---|---|---|---|---|---|---|
| [1] | The Sensors Are Innovative in the Internet of Things | X | | X | | | X |
| [2] | Integration of Smart Phone and IOT for development of the smart public transportation system | X | X | X | | | X |
| [3] | MuSA: A Smart Wearable Sensor for Active Assisted Living | X | X | X | | | X |
| [4] | Internet of things enabled smart switch | X | X | X | | | X |
| [5] | Smart assistance for public transport system | X | X | X | | | X |
| [6] | A 220-volts power switch controlled through WiFi | X | X | X | | | X |
| [7] | ADDSMART: Address digitization and smart mailbox with RFID technology | X | X | X | | | X |
| [8] | An IoT approach to vehicle accident detection, reporting, and navigation | X | X | X | | | X |
| [9] | ACQUISITION DEVICES IN INTERNET OF THINGS: RFID AND SENSORS | X | | X | | | X |
| [10] | RFID temperature sensors for monitoring soil solarization with biodegradable films | X | X | X | | | X |
| [11] | An integrated system of applying the use of internet of things, RFID and cloud computing: A case study of logistic management of electricity generation authority of Thailand (egat) mae mao lignite coal mining, lampang, Thailand | X | X | X | | | X |
| [12] | Health care monitoring system in the Internet of Things (IoT) by using RFID | X | X | X | | | X |
| [13] | A study on the possible application of the RFID system in different real-time environments | X | X | X | | | X |
| [14] | IoT based smart irrigation monitoring and controlling system | X | X | X | | | X |
| [15] | Bidirectional smart pillbox monitored through the internet and receiving reminding message from remote relatives | X | X | X | | | X |
| [16] | Health condition monitoring system for a distribution | X | X | X | | | X |
| | transformer using the Internet of Things (IoT) | | | | | | |
| [17] | Biomedical IoT Device for Self-Monitoring Applications | X | X | | | | X |
| [18] | Design and development of a nearable wireless system to control indoor air quality and indoor lighting quality | X | X | | X | | X |
| [19] | Development of unmanned surface vehicle for water quality monitoring and measurement | X | X | | | | X |
| [20] | RFID Technology for Management and Tracking: e-Health Applications | X | X | | | | X |
| [21] | A Multimodal Fingerprint-Based Indoor Positioning System for Airports | X | X | X | X | | X |
| [22] | Wireless sensor and actuator system for smart irrigation on the cloud | X | X | | | | X |
| [23] | An open source low-cost wireless control system for a forced circulation solar plant | X | X | | X | | X |
| [24] | A control system in intelligent farming by using Arduino technology | X | X | | | | X |
| [25] | Indoor positioning system based on inertial MEMS sensors: Design and realization | X | X | | | | X |
| [26] | Internet of things cloud-enabled MISSENARD index measurement for indoor occupants | X | X | | | | X |
| [27] | Bus tracking and monitoring using RFID | X | X | | | | X |
| [28] | Occupancy estimation in the smart building using hybrid CO2/light wireless sensor network | X | X | | | | X |
| [29] | System Based on the Internet of Things for Real-Time Particle Monitoring in Buildings | X | X | | X | | X |
| [30] | Active Plant Wall for Green Indoor Climate Based on Cloud and Internet of Things | X | X | | | | X |

## 6.10 Summary of the Findings

**RQ1: WHAT is the understanding and management of Things in IoT projects?**

Things in the context of IoT software systems are every device that can sense, actuate, or interact with the user or environment. In other words, these devices are all hardware that can traditionally replace the computer, expanding the connectivity reach. There are many examples in the literature like tags [1][9][13], home controller devices [4][6][7], mobile phones [2][5], wearables [3][25], vehicles and transports like buses [5][27], cars [8], and trucks [11], health devices [12][15][17], farm devices [10][14][22][24][30], indoor environment devices [18][26][28][29], water devices [19], indoor location solutions [21], and tracking devices [20].

Sensors can be applied to different solutions and can help sense the ambient collect valuable information to help people make decisions in various scenarios [10][13]. Tags like RFID and NFC [9][13] are most commonly used and optimal solution [1] kind of sensors that actuate on identifying objects used on logistic management [11][20], agriculture [10][13][22], buildings [28], and others [7][27].

Wearable are devices that the end-user wears responsible for collecting information about the environment, physical activity, localization, and inherently carrying user identification information. This equipment is applied in many application scenarios, notably in Ambient and Active Assisted Living contexts [3]. These devices can help prevent and early discover age-related diseases based on effective interaction among different sensing technologies. Because of this, they are of great importance [3].

These devices can be applied to automatize tasks for users on their houses by providing remote control of appliances giving users comfort, security, and simplifying their life's by low-cost solutions [4][6][7]. Other devices can be applied for health monitoring allowing to collecting patient data for prevention and early discovery of diseases or the well-being of the users [12][15][17].

Other expressive applications of things are on vehicles and transports [8][27]. These systems actuate traffic management and help detect accidents, the user’s mobility, and facilitate the rescue process to provide real-time information for the user's comfort.

On agriculture, or more specifically on farms, these things provide essential services by monitoring soil properties supporting the users in the decision-making process [10][22]. These systems are an efficient manner of managing their plantations, enabling them to produce quality products and increasing their productivity [14][22][24].

At last, indoor environment devices are another category mostly applied. These devices enable real-time monitoring and control of the environment properties (CO2 and PM sensor), providing comfort, well-being, and quality of life for the inhabitants of these ambients [18][26][28][29].

**RQ2: HOW do IoT projects deal with software technologies (techniques, technologies, methods, and solutions) and their operationalization regarding Things?**

In terms of technologies there are a lots of solutions that were combined to build the devices like sensors [3][8][[12][30], actuators [5][7][22], smartphones [2][21], microcontrollers [11][6][17], interactables [4][5][15], cameras [7][13], communication and network enablers[7][23], and others. Following, we describe a short aggregation of technologies in eight categories that were found in the literature:

- Sensors: RFID [1][7][20], NFC [1][7][8], accelerometer sensor [3][5][17], angular rate sensor [3], digital magnetic sensor [3], PIR sensor [5], alcohol sensor (MQ3) [5], CO2 sensor [5][18][28][30], LDR (light control) [5], temperature sensor [5][10][12][30], motion sensor [7][12], locking solenoid [7], shock sensor [8], electrocardiogram sensor (ECG) [12][17], electromyography sensor (EMG) [17], blood pressure sensor [12][17], electroencephalography sensor (EEG) [12], blood glucose sensor [12], soil moisture sensor [14][24], humidity sensor [14][16][24][26], water level sensor [14], infrared light emitting diodes (IRLED) [15], photodetectors [15], thermistor [23][26], current sensor [16], oil level sensor [16], vibration sensor [16], pulse and percentage of oxygen in blood sensor [17], glucometer [17], body temperature sensor [17], airflow sensor [17], galvanic skin response (GSR) sensor [17], ARK sensor system [19], light sensor [24][28][30], IR sensor [27], PM (particulate matter) sensor [29][30], gas sensor [30], ultrasonic sensor [30], MEMS Sensors [25];

- Actuators: relay module [14][18][23][30], buzzer [16], speaker [15], voice recorder [15], LED [4][7], DC motor [5];
- Smartphones: Android [2][5][21] and iOS/Apple [21] smartphones;
- Microcontrollers: Arduino [7][11][24][26], Raspberry Pi [4][8], Intel Edison [30], AT89C52 [6], ARM7 (LPC 218) [5], PIC18F4550 [16], SP430F2274 [22], STM32 [25], MSP430TM [28], Atmel's Mega AVR clan [12], WEMOS D1 mini [29];
- Interactables: switch [4][5], button [15];
- Cameras: Logitech c270 [7], ELP USB camera [7], red light camera [13];
- Communication and network enablers: CC2500 RF Transceiver [22], CC2531 SoC [3], Bluetooth module [23][30], WiFi module [7][24][29][30], XBee [18][23], nRF24L01 chip [25];
- Others: LCD [5][15][16], ADC [12][14][16][28], Unmanned Surface Vehicle (USV) [19], Solar Panel [23][24], power supply [14][16][23][25], photoresistor [18], resistor [18], GPS [2][13][19][21];

Besides that, some systems treats Things giving a virtual representation of these devices enabling remote access and control of them [4][5][14][27]. To achieve this is necessary connect the device with the internet. Some technologies were applied to provide communications services to these devices like: WSN [13][22], Wi-Fi APs (Access Point) [21][22], ZigBee [1][14], 4G Network [19], Bluetooth [1][10][12][17], Bluetooth Low Energy (BLE) [8][17][21], Wi-Fi [1][12][15], SMS Gateway [2], GSM/GPRS [5][12][16][22], Cellular IoT [8], and iBeacons [21].

**RQ3: WHRE do IoT projects locate the activities regarding Things?**

There is no general response to this question. The activities' location is the own environment and depends on the domain that is employed.

Based on the literature found the environment can be anywhere. The authors build Internet of Things software systems in places like houses [1][4][6], shopping places [1][9][13], transport [2][5][27], smart cities [5], factory [6], road/streets [8][13][27], military [9][13], industry [13][16], farm [10][24][22], lignite coal mines [11], hospital [12][17][20], office [18][28], water [19], airport [21], and building [28]. Some solutions were generics like: outdoor and indoor locations [23][25][26][29][30].

**RQ4: WHOM do IoT projects allocate to deal with Things?**

Depending on the domain the actors that use them are: common users [2][10][14][28], operators [16], passengers [2][5][8][27], drivers [3][5][11][13], government [2][12], administrators [2][12][16], patient [12][17], medical professionals [12], emergency team [12], doctors/ nurse [12][13][20], family members [4], elderly users [3], caregivers [3][15], professional carers [3], homeowner [7], postal workers [7], rescue teams [8], farmers [20][24], officers [11], police officers [13], managers [13], and inspectors [19].

In software engineering, there is no evidence about who correctly deal with these devices. Some solutions presented the own user construct and programmed the Thing [18][23][29].

**RQ5: WHEN do the effects of time, transformations, and states of Things affect IoT projects?**

No evidence was found.

**RQ6: WHY do IoT projects implement Things?**

In general, apply Things with part of a solution in the Internet of Things software systems provide a series of benefits for users:

- comfort [26][28];
- simplify user life [1][2];
- reduce costs [4][7][13];
- security [5][7][13];
- accessibility [5];
- increase quality of life [15][28][29];
- reduce the risk of mistakes [11][20];
- efficiency [14][22][27];
- improve quality [22][24];
- enhance productivity [14][28];
- decrease energy consumption [18][28];
- support in decision making process [10][13][22];
- automate manual process [11][16][19];
- track objects [13][20];
- indoor environmental quality [18][29];
- remote control and monitoring [14][16][24];
- monitoring of environment and users [9][12][29];

Things can be applied in some scenarios like traffic management and mobility [13][27], chain management [1][13], healthcare [15][17], and agriculture [10][22] to help users on their daily activities. On traffic management, things help with traffic congestion issues [2][13], accident detection [5][8], prevention of accidents [13], and rescuing process [8]. In healthcare they provide health monitoring [3][12][15][17], prevention and early discovery of diseases [3][12], and treatment monitoring [12][15]. On agriculture helps to minimize the time of treatments [10], automatize soil treatments [10], and save water [14][22].

## 6.11 Final Considerations

From the research performed, we could confirm that *things* execute a central role in the IoT scenario. Due to this central role, we can see that things are directly related to the other facets because it is through the *things* that we have the different behaviors and levels of smartness in the system, for example. The different sensors and identifiers extracted give us a broad view of the use of identification and sense behaviors. Already the actuators and controllers give us a vision of autonomy, representing a certain level of smartness that can reduce human interaction and contribute to decision-making.

In addition to behavior and smartness, it is also apparent the need for connectivity and interactivity between things. In the review, it was possible to extract various types of technologies associated with how to do things to communicate and interact. We recovered technologies from the simplest and most common ones like Bluetooth to modern solutions like iBeacon. Among the solutions, we realize that there are concerns beyond the connectivity itself, such as remote access, resource constraints, stable and secure connections, etc. For this reason, the interaction between things, devices and systems, and actors, including humans, is an exciting challenge for the full operationalization of IoT.

Some of the selected papers bring a proposal for DIY - do it yourself - that behind the user to more active participation with the system, making it a hybrid position of use and development of the solution. These initiatives are interesting from various perspectives, for example, the generalization and simplification of solutions to be tailored by users, new construction technologies to include user-in-the-loop, among others. It demonstrates that in addition to the facets, different views such as the user must be taken into account in the construction of things.

The scenario requirements, a considerable number of devices, and the wide variety and heterogeneity solution brings a necessity to deal with small physical size, limited resources, interoperability, integration, data format, security, and privacy that represent a significant challenge to build these things. These demands and concerns demonstrate that building things is not limited to hardware but involves an intertwining of different areas that need to work together to deliver quality and secure solutions. From the perspective of research and engineering, this brings research opportunities related to multidisciplinarity and partnership between teams with different skills to build these things.

Also, the view of the thing itself is expanded to represent a simple object and the bridge between the physical and the virtual world. With this new view of things, it is necessary to question our systems' vision and have a holistic view to represent the properties and limits of these systems present in both worlds (real and virtual).

## 6.12 References

**Final Set:**

[1] Rezvan, M., & Barekatain, M. (2015, May). The Sensors Are Innovative in the Internet of Things. In Wireless Internet: 8th International Conference, WICON 2014, Lisbon, Portugal, November 13-14, 2014, Revised Selected Papers (Vol. 146, p. 253). Springer.
[2] Sutar, S. H., Koul, R., & Suryavanshi, R. (2016, January). Integration of Smart Phone and IOT for development of smart public transportation system. In the Internet of Things and Applications (IOTA), International Conference on (pp. 73-78). IEEE.
[3] Bianchi, V., Losardo, A., Grossi, F., Guerra, C., Mora, N., Matrella, G., ... & Ciampolini, P. (2016, June). MuSA: A Smart Wearable Sensor for Active Assisted Living. In the Italian Forum of Ambient Assisted Living (pp. 197-208). Springer, Cham.
[4] Reddy, V. M., Vinay, N., Pokharna, T., & Jha, S. S. K. (2016, July). Internet of things enabled smart switch. In Wireless and Optical Communications Networks (WOCN), 2016 Thirteenth International Conference on (pp. 1-4). IEEE.
[5] Rathod, R., & Khot, S. T. (2016, August). Smart assistance for the public transport system. In Inventive Computation Technologies (ICICT), International Conference on (Vol. 3, pp. 1-5). IEEE.
[6] Gao, X., Zhang, B., & Li, S. (2016, October). A 220-volts power switch controlled through WiFi. In Computer Communication and the Internet (ICCCI), 2016 IEEE International Conference on (pp. 526-529). IEEE.
[7] Tew, J. R., & Ray, L. (2016, October). ADDSMART: Address digitization and smart mailbox with RFID technology. In Ubiquitous Computing, Electronics & Mobile Communication Conference (UEMCON), IEEE Annual (pp. 1-6). IEEE.
[8] Nasr, E., Kfoury, E., & Khoury, D. (2016, November). An IoT approach to vehicle accident detection, reporting, and navigation. In Multidisciplinary Conference on Engineering Technology (IMCET), IEEE International (pp. 231-236). IEEE.
[9] Dalli, A., & Bri, S. (2016). ACQUISITION DEVICES IN INTERNET OF THINGS: RFID AND SENSORS. Journal of Theoretical & Applied Information Technology, 90(1).
[10] Luvisi, A., Panattoni, A., & Materazzi, A. (2016). RFID temperature sensors for monitoring soil solarization with biodegradable films. Computers and Electronics in Agriculture, 123, 135-141.
[11] Chieochan, O., SaoKaew, A., & Boonchieng, E. (2017, February). An integrated system of applying the use of internet of things, RFID and cloud computing: A case study of logistic management of electricity generation authority of Thailand (egat) mae mao lignite coal mining, lampang, Thailand. In Knowledge and Smart Technology (KST), 2017 9th International Conference on (pp. 156-161). IEEE.
[12] Khan, S. F. (2017, March). Health care monitoring system in the Internet of Things (IoT) by using RFID. In Industrial Technology and Management (ICITM), International Conference on (pp. 198-204). IEEE.
[13] Saravanakumar, K., Deepa, K., & Kumar, N. S. (2017, April). A study on the possible application of the RFID system in different real-time environments. In Circuit, Power and Computing Technologies (ICCPCT), 2017 International Conference on (pp. 1-7). IEEE.
[14] Saraf, S. B., & Gawali, D. H. (2017, May). IoT based smart irrigation monitoring and controlling system. In Recent Trends in Electronics, Information & Communication Technology (RTEICT), 2017 2nd IEEE International Conference on (pp. 815-819). IEEE.

[15] Tsai, H. L., Tseng, C. H., Wang, L. C., & Juang, F. S. (2017, June). Bidirectional smart pillbox monitored through the internet and receiving reminding message from remote relatives. In Consumer Electronics-Taiwan (ICCE-TW), 2017 IEEE International Conference on (pp. 393-394). IEEE.

[16] Pawar, R. R., & Deosarkar, S. B. (2017, July). Health condition monitoring system for a distribution transformer using the Internet of Things (IoT). In Computing Methodologies and Communication (ICCMC), 2017 International Conference on (pp. 117-122). IEEE.

[17] Huertas, T., & Mendez, D. (2017). Biomedical IoT Device for Self-Monitoring Applications. In VII Latin American Congress on Biomedical Engineering CLAIB 2016, Bucaramanga, Santander, Colombia, October 26th-28th, 2016 (pp. 357-360). Springer, Singapore.

[18] Salamone, F., Belussi, L., Danza, L., Galanos, T., Ghellere, M., & Meroni, I. (2017). Design and development of a nearable wireless system to control indoor air quality and indoor lighting quality. Sensors, 17(5), 1021.

[19] Yang, T. H., Hsiung, S. H., Kuo, C. H., Tsai, Y. D., Peng, K. C., Hsieh, Y. C., ... & Kuo, C. (2018, April). Development of unmanned surface vehicle for water quality monitoring and measurement. In 2018 IEEE International Conference on Applied System Invention (ICASI) (pp. 566-569). IEEE.

[20] Álvarez López, Y., Franssen, J., Álvarez Narciandi, G., Pagnozzi, J., González-Pinto Arrillaga, I., & Las-Heras Andrés, F. (2018). RFID Technology for Management and Tracking: e-Health Applications. Sensors, 18(8), 2663.

[21] Molina, B., Olivares, E., Palau, C. E., & Esteve, M. (2018). A Multimodal Fingerprint-Based Indoor Positioning System for Airports. IEEE Access, 6, 10092-10106.

[22] Sales, N., Remédios, O., & Arsenio, A. (2015, December). Wireless sensor and actuator system for smart irrigation on the cloud. In the Internet of Things (WF-IoT), 2015 IEEE 2nd World Forum on (pp. 693-698). IEEE.

[23] Salamone, F., Belussi, L., Danza, L., Ghellere, M., & Meroni, I. (2015). An open source low-cost wireless control system for a forced circulation solar plant. Sensors, 15(11), 27990-28004.

[24] Putjaika, N., Phusae, S., Chen-Im, A., Phunchongharn, P., & Akkarajitsakul, K. (2016, May). A control system in intelligent farming by using Arduino technology. In Student Project Conference (ICT-ISPC), 2016 Fifth ICT International (pp. 53-56). IEEE.

[25] Wu, C., Mu, Q., Zhang, Z., Jin, Y., Wang, Z., & Shi, G. (2016, June). Indoor positioning system based on inertial MEMS sensors: Design and realization. In Cyber Technology in Automation, Control, and Intelligent Systems (CYBER), 2016 IEEE International Conference on (pp. 370-375). IEEE.

[26] Ray, P. P. (2016). Internet of things cloud-enabled MISSENARD index measurement for indoor occupants. Measurement, 92, 157-165.

[27] Kamble, P. A., & Vatti, R. A. (2017, December). Bus tracking and monitoring using RFID. In Image Information Processing (ICIIP), 2017 Fourth International Conference on (pp. 1-6). IEEE.

[28] Huang, Q., & Mao, C. (2017). Occupancy estimation in the smart building using hybrid CO2/light wireless sensor network. Journal of Applied Sciences and Arts, 1(2), 5.

[29] Marques, G., Roque Ferreira, C., & Pitarma, R. (2018). A System Based on the Internet of Things for Real-Time Particle Monitoring in Buildings. International journal of environmental research and public health, 15(4), 821.
[30] Liu, Y., Hassan, K. A., Karlsson, M., Weister, O., & Gong, S. (2018). Active Plant Wall for Green Indoor Climate Based on Cloud and Internet of Things. IEEE Access, 6, 33631-33644.

**Additional References:**

C. Tricco et al. A scoping review of rapid review methods. BMC Medicine, 2015.
B. Cartaxo et al.: The Role of Rapid Reviews in Supporting Decision -Making in Software Engineering Practice. EASE 2018.
B. Cartaxo et al. Evidence briefings: Towards a medium to transfer knowledge from systematic reviews to practitioners. ESEM, 2016.

# DEVELOPING IOT SOFTWARE SYSTEMS?
# TAKE BEHAVIOR INTO ACCOUNT

## CONTEXT

IoT allows composing systems from *things* equipped with identifying, sensing, or actuation behaviors and processing capabilities that can communicate and cooperate to reach a goal. For so, different facets should be taken into account, such as behavior.

## BEHAVIOR

IoT provides the chance of enhancements in the *things*, extending their original behaviors. It relates to data processing and functions that enable Identification, Sensing and Actuation behaviors, for example.

## SUMMARY

All the actions in IoT are triggered by some event or rule. But also, combining different components can lead to emergent behaviors in the system. The behavior of a system is paramount for its goals; therefore, it needs to be well defined and understood.

## OBJECTIVE

This briefing aims to enlighten project planning and to provide guidance when dealing with Behavior in IoT. The information presented in this briefing seeks to answer: "What taken into account when considering behavior for IoT?".

## MOTIVATION

IoT offers challenges for their construction, with concerns that rather than just developing, we need to engineer software systems. This motivated us to a more comprehensive investigation to address these concerns and to provide evidence-based support for IoT engineering.

## ADDITIONAL INFORMATION

**Who are the briefing's clients?** Software developers and practitioners who want to make decisions about how to deal with *behavior* in IoT, considering scientific evidence.

**Where do the findings come?** All findings of this briefing were extracted from scientific studies about *behavior* through a Rapid Review[1]. The Technical Report[2] containing all the findings is available for further information.

**What is included in this briefing?** Technologies, challenges, and strategies to deal with *behavior* in IoT projects.

**What are the Challenges and Opportunities?** Is it possible to identify emerging behavior in a computable way? Is it possible to make behaviors decentralized? These are some of the open questions that make the concept of behavior so challenging.

**Does behavior represent a concern in the Engineering of Internet of Things Software Systems?** Yes. When it comes to IoT is necessary to define the behaviors and predict how well system will respond to them. There should be guaranteed safety and security procedures for each implemented behavior, considering the users and the system itself.

## HIGHLIGHTS

❖ ***Analyze the behaviors as:*** *Expected and beneficial; Unexpected and beneficial; Expected and harmful; and Unexpected and harmful.*

❖ ***Define procedures and strategies*** *for each case.*

## WHAT
### is the the understanding of behavior in CSS?

- Define the **events** and **rules** that will trigger the behaviors, as well as the consequences for the performed behavior.
- An IoT system may have **Identification**, **Sensing**, **Monitoring**, **Actuation** behaviors, or a combination of them.

## HOW
### do CSS projects deal with the operacionalization of behavior?

- Take care of the growing complexity by **breaking bigger behaviors** in phases or smaller behaviors.
- Each behavior will be addressed by a **virtual or physical component** that should be defined regarding specific requirements and properties.
- It is necessary to be able to **visualize and represent** the components, how they behave, with whom they interact. Therefore, system dynamics, agent-based modeling, discrete event modeling and simulation approaches can be used.
- Behaviors can be **aggregative, expected, beneficial.** But since the system can interact with other in different ways, unexpected, harmful, emergent behaviors can rise, that should be addressed.

## WHERE
### CSS projects locate the activities regarding behavior?

- Throughout the study, it was not possible to observe any evidence that answered the question.

## WHO
### do CSS projects allocate to deal with behavior?

- Should consider the **components** that are part of the system and the **people** that will interact with the system in general, how they are related to each defined behavior.
- The roles identified in the development were software **engineers**, **programmers**, **software** and **system architects.** Active users were also present.

## WHEN
### do the effects of time, and states of behavior affect CSS projects?

- To manage system behaviors should be of great concern in the **initial phases** of the project. It is also necessary to create activities to take care of each behavior throughout the project since it is costly to change known behaviors and address new behaviors that have not yet been treated.

## WHY
### do CSS projects implement behavior?

- The behavior of the system is should be treated carefully through the project phases, since the goal is to obtain the desired behavior in the physical world, outside the product.
- The system behavior is often seen as the final act of a specific solution. When a user or another system interacts with it, it is concerned with the final result of it, that is, with the behavior that the system will generate.

[1] More on Rapid Reviews: https://dl.acm.org/citation.cfm?id=3210462
[2] Full Report: https://goo.gl/no7jtR

# 7 RAPID REVIEW ON BEHAVIOR

# Rapid Reviews Meta-Protocol:

## Engineering of Internet of Things Software Systems

**Diego C. B. Castro, Rebeca C. Motta, Guilherme H. Travassos**

# Behavior

In the investigation regarding Internet of Things Software Systems (IoT), it has been observed that these modern software systems offer challenges for their construction since they are calling into question our traditional form of software development. Usually, they rely on different technologies and devices that can interact-capture-exchange information, act, and make decisions. It leads to concerns that, rather than just developing software, we need to engineer software systems embracing multidisciplinarity, integrating different areas. From our initial research, we analyzed the concerns related to this area. We categorized them into a set of facets - Connectivity, Things, Behavior, Smartness, Interactivity, Environment, and Security - representing such projects' multidisciplinarity, in the sense of finding a set of parts composing this engineering challenge.

Since these facets can bring additional perspectives to the software system project planning and management, acquiring evidence regarding such facets is of great importance to provide an evidence-based framework to support software engineers' decision-making tasks. Therefore, the following question should be answered:

***"Does behavior represent a concern in the engineering of***

***Internet of Things software systems?"***

This Rapid Review (RR) aims to analyze behavior to characterize it in the IoT field, regarding *what, how, where, when and why* it is used in the context of IoT projects, verifying the existence of published studies supporting the previous results. The 5W1H aims to give the observational perspective on which information is required to the understanding and management of the facet in a system (what); to the software technologies (techniques, technologies, methods, and solutions) defining their operationalization (how); the activities location being geographically distributed or something external to the software system (where); the roles involved to deal with the facet development (who); the effects of time over the facet, describing its transformations and states (when); and to translate the motivation, goals, and strategies going to what is implemented in the facet (why), in respect of IoT projects.

## 7.1 Research Questions

- **RQ1:** What is the understanding and management of behavior in IoT projects?
- **RQ2:** How do IoT projects deal with software technologies (techniques, technologies, methods, and solutions) and their operationalization regarding behavior?
- **RQ3:** Where do IoT projects locate the activities regarding behavior?
- **RQ4:** Whom do IoT projects allocate to deal with behavior?
- **RQ5:** When do the effects of time, transformations, and states of behavior affect IoT projects?
- **RQ6:** Why do IoT projects implement behavior?

## 7.2 Search Strategy

The Scopus[7] search engine and the following search string support this RR, built using PICOC with five levels of filtering:

**P**opulation - Internet of Things software systems
Synonymous:
"ambient intelligence" OR "assisted living" OR "multiagent systems" OR "systems of systems" OR "internet of things" OR "Cyber-Physical Systems" OR "Industry 4" OR "fourth industrial revolution" OR "web of things" OR "Internet of Everything" OR "contemporary software systems" OR "smart manufacturing" OR digitalization OR digitization OR "digital transformation" OR "smart cit*" OR "smart building" OR "smart health" OR "smart environment"

**I**ntervention - behavior
Synonymous:
"system service" OR "software service" OR "system behavior" OR "software behavior" OR "system function*" OR "software function*" OR "application service" OR "application function*" OR "application behavior" OR "solution behavior" OR "solution service" OR "solution function*" OR "program behavior" OR "program function*" Or "program service" Or "product behavior" OR "product function*" OR "product service" OR “emergent behavior.”

**C**omparison - no

**O**utcome -
Synonymous:
understanding OR management OR technique OR “technolog*” OR method OR location OR place OR setting OR actor OR role OR team OR time OR transformation OR state OR reason OR motivation OR aim OR objective

**C**ontext -
Synonymous:

[7] https://www.scopus.com

engineering OR development OR project OR planning OR management OR building OR construction OR maintenance

Limited to articles from 2013 to 2018
Limited to Computer Science and Engineering
LIMIT-TO (SUBJAREA, "COMP" ) OR LIMIT-TO (SUBJAREA, "ENGI" ) ) AND ( LIMIT-TO (PUBYEAR, 2018 ) OR LIMIT-TO (PUBYEAR, 2017 ) OR LIMIT-TO (PUBYEAR, 2016 ) OR LIMIT-TO (PUBYEAR, 2015)

```
TITLE-ABS-KEY (( "ambient intelligence" OR "assisted living" OR "multiagent systems" OR "systems of systems" OR "internet of things" OR "Cyber Physical Systems" OR "Industry 4" OR "fourth industrial revolution" OR "web of things" OR "Internet of Everything" OR "contemporary software systems" OR "smart manufacturing" OR digitalization OR digitization OR "digital transformation" OR "smart cit*" OR "smart building" OR "smart health" OR "smart environment" ) AND ("system service" OR "software service" OR "system behavior" OR "software behavior" OR "system function*" OR "software function*" OR "application service" OR "application function*" OR "application behavior" OR "solution behavior" OR "solution service" OR "solution function*" OR "program behavior" OR "program function*" Or "program service" Or "product behavior" OR "product function*" OR "product service" OR "emergent behavior") AND (engineering or development or project or planning OR management OR building OR construction OR maintenance) AND ( LIMIT-TO ( SUBJAREA , "COMP" ) OR LIMIT-TO ( SUBJAREA , "ENGI" ) ) AND ( LIMIT-TO ( PUBYEAR , 2018 ) OR LIMIT-TO ( PUBYEAR , 2017 ) OR LIMIT-TO ( PUBYEAR , 2016 ) OR LIMIT-TO ( PUBYEAR , 2015)))
```

## 7.3 Selection procedure

One researcher performs the following selection procedure:

1. Run the search string;
2. Apply the inclusion criteria based on the paper Title;
3. Apply the inclusion criteria based on the paper Abstract;
4. Apply the inclusion criteria based on the paper Full Text, and;

After finishing the selection from Scopus, use the included papers set to:

5. Execute snowballing backward (one level) and forward:
    a. Apply the inclusion criteria based on the paper Title;
    b. Apply the inclusion criteria based on the paper Abstract;
    c. Apply the inclusion criteria based on the paper Full Text.

The JabRef Tool[8] must be used to manage and support the selection procedure.

## 7.4 Inclusion criteria

- The paper must be in the context of **software engineering**; and
- The paper must be in the context of the **Internet of Things software systems**; and
- The paper must report a **primary or a secondary study**; and
- The paper must report an **evidence-based study** grounded in empirical methods (e.g., interviews, surveys, case studies, formal experiment, etc.); and

[8] http://www.jabref.org/

- The paper must provide data to **answering** at least one of the RR **research questions**.
- The paper must be written in the **English language**.

## 7.5 Extraction procedure

The extraction procedure is performed by one researcher, using the following form:

| <paper_id>:<paper_reference> | |
|---|---|
| Abstract | <Abstract> |
| Description | <A brief description of the study objectives and personal understanding> |
| Study type | <Identify the type of study reported by paper (e.g., survey, formal experiment)> |
| RQ1: WHAT<br>information required to understand<br>and manage the behavior in IoT | - < A1_1><br>- < A2_2><br>- ... |
| RQ2: HOW<br>software technologies (techniques, technologies, methods and solutions) and their operationalization | - < A2_1><br>- < A2_2><br>- ... |
| RQ3: WHERE<br>activities location or something external to the IoT | - < A3_1><br>- < A3_2><br>- ... |
| RQ4: WHO<br>roles involved to deal with the behavior development in IoT | -< A4_1><br>-<A4_2><br>- … |
| RQ5: WHEN<br>effects of time over behavior, describing its transformations and states in IoT | - < A5_1><br>- < A5_2><br>- ... |
| RQ6: WHY<br>motivation, goals, and strategies regarding behavior in IoT | - < A6_1><br>- < A6_2><br>- ... |

## 7.6 Synthesis Procedure

In this RR, the extraction form provides a synthesized way to represent extracted data. Thus, we do not perform any synthesis procedure.

However, the synthesis is usually performed through a narrative summary or a Thematic Analysis when the number of selected papers is not high.

## 7.7 Report

An Evidence Briefing [2] reports the findings to ease the communication with practitioners. It was presented as the cover for this chapter.

## 7.8 Results

### Execution

| Activity | Execution date | Result | Number of papers |
|---|---|---|---|
| First execution | 13/07/2018 – 21h:44min | 592 documents added | 592 |
| Remove conferences/workshops | 14/07/2018 | 21 documents withdrawn | 571 |
| Remove books | 14/07/2018 | Eight documents were withdrawn | 563 |
| Included by Title analysis | 14/07/2018 | 460 documents withdrawn | 103 |
| Included by Abstract analysis | 15/07/2018 | 75 documents withdrawn | 28 |
| Papers not found | 15/07/2018 | One documents were withdrawn | 27 |
| Articles for reading | 15/07/2018 | 27 documents | 27 |
| Removed after a full reading | 17/07/2018 – 20/07/2018 | Ten documents were withdrawn | 17 |
| Snowballing | 21/07/2018 | 9 documents added | 26 |
| Snowballing not found | 21/07/2018 | Two documents were withdrawn | 24 |
| Snowballing after reading | 21/07/2018 – 23/07/2018 | Five documents were withdrawn | 19 |
| Total included | 23/07/2018 | 19 documents | 19 |
| Papers extracted | 23/07/2018 – 07/08/2018 | 19 documents | 19 |

### Final Set

| Reference | Author | Title | Year | Source |
|---|---|---|---|---|
| [1] | JACKSON, M. | Behaviors as design components of cyber-physical systems. Lecture Notes in Computer Science | 2015 | Regular search |
| [2] | KOPETZ, H. et al. | Towards an understanding of emergence in systems-of-systems. | 2015 | Regular search |
| [3] | WACHHOLDER, D.; STARY, C | Enabling emergent behavior in systems-of-systems through bigraph-based modeling | 2015 | Regular search |
| [4] | MITTAL, S.; RAINEY, L. | Harnessing emergence: The control and design of emergent behavior in a system of systems engineering | 2015 | Regular search |
| [5] | LUO et al. | Emergent properties and requirements evolution in engineering systems and a roadmap. | 2015 | Regular search |
| [6] | ZEIGLER, B. P.; MUZY, A. | Some modeling & simulation perspectives on emergence in system-of-systems. | 2016 | Regular search |
| [7] | RUPPEL, S. R. | System behavior models: a survey of approaches. | 2016 | Snowballing |
| [8] | GABOR, T. et al. | A simulation-based architecture for smart cyber-physical systems. | 2016 | Regular search |
| [9] | GRACIANO NETO, V. V. | Validating emergent behaviors in systems-of-systems through model transformations. | 2016 | Regular search |
| [10] | ROCA, D. et al. | Emergent Behaviors in the Internet of Things: The Ultimate Ultra-Large-Scale System. | 2016 | Regular search |
| [11] | GARCÉS, L.; NAKAGAWA, E. Y. | A process to establish, model and validate missions of systems-of-systems in reference architectures. | 2017 | Regular search |
| [12] | GIAMMARCO, K. | Practical modeling concepts for engineering emergence in systems of systems. | 2017 | Regular search |
| [13] | GIAMMARCO, K. | Comprehensive use case scenario generation: An approach for modeling system of systems behaviors. | 2017 | Regular search |
| [14] | OQUENDO, F. | Architecturally describing the emergent behavior of software-intensive system-of-systems with SosADL. | 2017 | Regular search |
| [15] | ZURITA, N. F. S.; TUMER, I. Y. | A survey: Towards understanding emergent behavior in complex engineered systems. | 2017 | Regular search |
| [16] | HAYNES, C. et al. | Engineering the emergence of norms: A review. | 2017 | Regular search |
| [17] | BRINGS, J. | Verifying Cyber-Physical System Behavior in the Context of Cyber-Physical System-Networks | 2017 | Regular search |
| [18] | OQUENDO, F. | Formally Describing the Architectural Behavior of Software-Intensive Systems-of-Systems with SosADL. | 2017 | Regular search |
| [19] | BOSMANS, S. et al. | Towards evaluating the emergent behavior of the internet of things using large-scale simulation techniques. | 2018 | Snowballing |

## 7.9 Summary of the articles

To obtain a better understanding of the behavior within an IoT, the procedure that has already been described in previous sections has been executed.

In this procedure, a search string was executed that returned a total of 592 papers. All of these attitudes were evaluated according to the criteria that were described in section 4. In the first instance, all the findings that had information about conferences and workshops were removed; in this stage, 21 papers were excluded. Then, all the books were removed from the research base, thus removing eight papers. Then, all the titles of the papers were read to know if the same sent the idea researched, with that, 460 papers were removed, leaving a base with 103 papers. All the abstracts of these papers have been read to know the general idea of the same and to verify if it should be included or excluded from the base. At the end of this stage, 75 papers were removed, leaving a base of 28 papers. Of these papers, 1 of them cannot get the full version, so it has been removed from the base. All 27 remaining papers were read in full, 10 of them did not respond to inclusion criterion 5 of section 4.1 (one of the questions answered should be answered), for this reason, they were removed, leaving a base with 17 papers. In all 17 papers, a snowballing procedure was performed, which returned a total of 9 papers, 7 of them forward, and 2 of them backward. According to the same inclusion criteria, these nine papers were analyzed, and 2 of them were included in the final database. Section 8.1 has shown a table summarizing the procedure for analyzing these papers.

At the end of this procedure, 19 articles have been selected to be studied. The following is a brief description of each of them.

Jackson, M. [1] is concerned with a system's behavior, taking into account its chief characteristic. The behavior can be understood as the desired end effect after running a program. There are several different behaviors; it will depend very much on which objects are interacting, what environment the object is immersed in, and what stimulus has been received. Some examples of behaviors that can be given are when a car detects that it will have a collision and reacts to this stimulus, when an elevator discovers that the suspension cable has ruptured and reacted to this stimulus, among others.

Throughout his article, Jackson, M. [1] explains how this behavior can be understood through several different definitions. After these definitions, it is described how it can be treated, for example, using transition states, using tree structure or through an abstraction, where a greater behavior is constituted of several smaller behaviors (behaviors where nothing goes wrong, and no interaction interferes with other behaviors) which are independent and work together to achieve a goal.

Due to the creation of SoS, the idea of an emergency is increasingly addressed. For this reason, Kopetz, H. et al. [2] briefly introduce the concept of emergence in philosophy and computer science and then defines emergent behavior within an SoS. According to them, there are four cases of emergent behavior that must be distinguished in an SoS: expected moreover, beneficial emergent behavior is the usual case; unexpected and useful emergent behavior is a positive surprise; the expected harmful emergent behavior can be avoided by adhering to the appropriate design rules, and the problematic case is an unexpected harmful emergent behavior. Finally, it still draws a parallel between SoS and monolithic systems, showing some of their differences.

Wachholder and Stary [3], argues that the main characteristics of behavior in an IoT are the action, the reaction, the evolutionary character, and the emergency. One of the examples given that addresses these characteristics a configuration change in a system, where it undergoes a stimulus that is the change of configuration. The system reacts to that stimulus, which can emerge at an unexpected moment and maybe changing due to the character evolutionary. They state that for these smaller systems to work together, they need to be well connected. Aiming at this better relationship, they proposed a bigraph modeling to explore the distinctive features of SoS (Autonomy, Belonging, Connectivity, Diversity, and Emergency).

Mittal, S. and Rainey, L. [4] claim that an emergent behavior occurs because of the lack of understanding of any system's complete behavior and define behavior as an observed phenomenon that results from the interactions: agent-agent, agent-environment, and environment-environment. When one possesses a good knowledge of a particular system, its behaviors are no longer emerging and are known. However, it is a challenging task to understand and see the timing of such behaviors. They found four different emergencies (simple, weak, intense, and frightening) presented to characterize such behaviors. It is worth remembering that these categories consider how difficult it is to simulate such an event.

Current software systems are built from the union of several smaller components that operate in collaboration to achieve a goal. Each of these smaller systems behaves differently and acts independently; however, with the union of these minor behaviors, a new, more massive behavior is created, something that smaller components would not be able to achieve alone [4].

Luo et al. [5] define behavior as an interaction between the components of an application and their environment and state that in some contexts, these behaviors can be beneficial, harmful, or unexpected, and that users adapt products to support tasks that designers have never planned. They still append and explain an IoT property called "emergency." It is tough to understand, predict, or model. However, it seeks to find solutions to the problems that are called undesirable emerging properties. In their work, they demonstrate the impact of requirements throughout the project's evolution, showing how creating requirements for IoT is a difficult task. Finally, it is shown how difficult it is to model the interfaces between humans and systems because the emergent behavior can hardly be predicted and, therefore, mostly unknown.

Zeigler and Muzy [6] present some features of the DEVS model to better model emerging behaviors. Among the mentioned characteristics, it is worth mentioning that this model provides components and couplings for models of complex systems; supports a dynamic framework for adapting and evolving systems; has a temporary structure that supports the observation of emergent behavior; supports emergency forecasting models, and finally allows for structural emergency modeling.

Nowadays, systems are getting bigger and more complex, with more things to be managed, like the behavior of those systems that are often unpredictable and can arise at any time. To improve the understanding of these systems, Ruppel, S. R. [7] reviewed to obtain a better conceptualization of these systems' behaviors. From this research, it was possible to get different definitions of the meaning of the emergent behavior, understand when these behaviors can arise, and, finally, know how these behaviors can be modeled.

Gabor, T. et al. [8] state that some event triggers all behavior exerted by the cyber-physical system. Mostly, a behavior is nothing more than a translation from observations to actions.

They introduced the digital twin idea so that it was possible to create an architectural structure centered on the flow of information that is exchanged within a system[8]. An analysis of this flow was elaborated within the scope of a cyber-physical system's components to define a classification of the methods of controlling a system's behavior concerning its mechanisms. Finally, to predict the possible states of a cyber-physical system, simulations were developed to improve the understanding of the constraints of a project, taking into account the potential consequences of external influences during the operation of the system [8].

SoS has a property called emergent behavior that arises from the interaction of one constituent system and another. Modeling for these systems is generally static; however, how can someone shape something that does not know how or when it will happen? It is the main point raised by Graciano Neto, V. V. [9]. He introduced the simulation's idea so that a system's behaviors were dynamically modeled and then proposed a modeling approach to describe an SoS according to a simulation to support emerging behaviors' validation.

Graciano Neto, V. V. [9] states that behavior is triggered by the reception of stimuli and data exchanged between the constituents and between the SoS and the environment. Such behaviors are holistic phenomena that manifest themselves after a certain number of interactions between the components. An example of behavior given is home security that can emerge from a set of individual systems installed in a smart home [9].

Roca, D. et al. [10] contextualizes a general idea about emerging computations. They define behavior as a collection of actions and patterns that result from local interactions between elements and their environment and introduce a model used to understand the coordination of birds' flight in the flock, where three main ideas are presented: alignment, separation, and cohesion. Then this model is introduced to coordinate smart cars, dividing them into hierarchies, where they begin to observe the behaviors of the parts (of one car to another, the speed of the other car, etc.) so that the behavior of the whole (Automatic Coordination of cars) is observed.

Garcés, L. et al. [11] states that the most relevant characteristics for behavior in an SoS are: the action, the reaction, the evolutionary character, and the emergency and seek to improve the viability of SoS architecture. For this, he presented a systematic process to establish, model, and validate the missions of SoS that occur through the behaviors of the constituent systems. This process explains how to identify these systems' tasks so that it is possible to assign responsibilities, allocate operational and communication resources to abstract the idea behind the constituent systems to determine the entities involved in the systems in which behaviors emerge.

Giammarco, K. [12] defines behavior as the way in which a system or one of its components acts on its own accord and in response to stimuli via interaction with other systems or components.

They present a system definition focused on emerging behaviors of the same, differentiating positive and negative behaviors. Aiming at this definition, it illustrates five models for modeling SoS [12]. After defining some concepts and demonstrations of the models, one can understand that some practices must be followed when dealing with SoS; Among them, behaviors and interactions must be separated, model system behaviors and environmental behaviors are required (The system should always consider the behaviors that the environment and other objects can create), formalization of models for automatic execution is also required, as well as the proper allocation of each task to a human or machine, using abstraction to manage large models [12].

Giammarco, K. [13] still presents an idea of an automatic generation of scenarios that an SoS can present. It uses a language called Monterey Phoenix (MP) to generate these possible scenarios through 5 necessary steps, beginning with system behavior modeling, which starts to model the interactions between these behaviors and objects. Next, the results are checked and validated to verify that no scenario ended up being left out, and there is still another that was not addressed. Finally, documentation is generated to demonstrate cases that violated expectations.

Oquendo, F. [14][18] attributes the aggregative, new, and unpredictable features as the main ones of an SoS behavior and, for this reason, aims to improve its modeling to be more evident to the user. They present the concept of SosADL, a description language for an architecture of emerging behaviors within a system of systems. Finally, he demonstrates an example of birds flying in flocks using SosADL.

Software systems can be very complicated due to several factors; however, one of the main factors that can increase this complexity is the emergence of behaviors where these behaviors are difficult to predict and, once predicted, are challenging to manage. Zurita and Tumer [15] affirm that these behaviors are influenced by several factors that involve the system, including the inputs of the information, the environment, taking into account the actions and reactions, etc. Having a better understanding of these behaviors is critical to achieving the creation of a better system.

Haynes, C. et al. [16] present a new approach to understanding the emergence of behaviors within a system of systems. A literature review was conducted to answer three central questions: How can an emerging standard or behavior pattern be detected in a multi-agent system? How can the emerging norm be evaluated about individual agents' needs and the multi-agent system as a whole? How can valuable standards be encouraged, and can dangerous ones be discouraged? The following is a general model of how behavior can emerge and how it can be treated.

Brings, J. [17] states that cybernetic systems are connected through a networked form of objects where behavior emerges from connected systems' interaction. Checking this behavior is a challenging task as there are several different systems within the same network, and each of them behaves differently. Throughout the research, we sought to understand how a cyber-physical system's behavior can be detected considering the various systems networks of which they can be part.

With the increase of IoT architectures, the concept of behavior is gaining more and more importance. With this, you can see new types of interactions between different types of objects, and because of the massive scale of existing objects, this is not an easy task to execute. Therefore, Bosmans, S. et al. [19] seek to explore techniques to evaluate IoT applications' emerging behavior. Its primary focus is to find a simulation and modeling approach to provide an overview of possible methods that can optimize the overall performance of a simulation.

Throughout the research, many definitions were attributed to behavior. In general, behavior can be understood as the materialization of an action of a system in the physical world in which a system or one of its components acts on its own and in response to stimuli via interaction with other systems or components. However, it is worth remembering that these behaviors often occur unexpectedly; they are not known when executed. For this reason, this is one of the main problems behind the treatment of these behaviors.

In general, this behavior has some significant characteristics that influence when it is treated. The main features that were found were: action, reaction, emergence, evolutionism, unpredictability, being aggregative, etc.

The Internet of Things software systems are made up of many related objects at all times; each of these objects may be performing different behaviors, either concurrently or alone. Bearing in mind these ideas presented, it is possible to understand that many systems are related at all times and that it is necessary to know what these systems are so that it is possible to know how to treat each of the particularities of each object that is related within the system.

For the system can run smoothly, the emergent behavior should be decentralized and robust about the replacement of individuals; in other words, the behavior must be stable even if some individuals are removed or replaced. It is another area of research that is still open.

The behavior can arise in 3 ways. The first is through a stimulus that the object feels and to which it reacts; the second is through a collaboration/interaction of the constituent systems in an IoT and, finally, through a reaction to another behavior emitted by another system.

In general, four distinct behavior cases must be analyzed in the Internet of Things system. The first of these is expected and beneficial behavior, which is the most usual case. The second is known as unexpected and useful behavior that can be understood as a positive surprise. Then there is expected harmful behavior that can be avoided by adhering to appropriate design rules. Moreover, finally, there is the problematic case, which is detrimental and unexpected behavior.

The location of the activities in an IoT is a problem to be solved because the behavior generated by an object in one place in a given condition may be otherwise complementary when those conditions change or when that place changes. However, the place where the behavior will occur is directly related to the system's scope to be developed.

When a sensor can, for example, only provide a certain quality of data or just measure specific inputs, which can result in a severe constraint on the behavior of the cyber-physical system as it can no longer discern all different states of the physical world and is thus forced to treat situations similarity when they yield the same sensor data.

Taking into consideration all the information that has been presented, it is possible to understand that the behavior is an essential and delicate part of an IoT and, for this reason, must be treated very carefully.

## 7.10 Tracking matrix

| Paper | WHAT | HOW | WHERE | WHO | WHEN | WHY |
|---|---|---|---|---|---|---|
| A Process to Establish, Model and Validate Missions of Systems-of-Systems in Reference Architectures | X | X | | X | X | X |
| A Simulation-Based Architecture for Smart Cyber-Physical Systems | X | | | X | X | X |
| A survey: towards understanding emergent behavior in complex engineered systems | X | | | | X | X |
| Architecturally Describing the Emergent Behavior of Software-intensive System-of-Systems with SosADL | X | X | | | | X |
| Behaviors as Design Components of Cyber-Physical Systems | X | X | | X | X | X |
| Comprehensive Use Case Scenario Generation | | X | | | | X |
| Emergent Behaviors in the Internet of Things: The Ultimate Ultra-Large-Scale System | X | X | | X | X | X |
| Emergent Properties and Requirements Evolution in Engineering Systems and a Roadmap | X | X | | X | X | X |
| Enabling Emergent Behavior in Systems-of-Systems Through Bigraph-based Modeling | X | X | | | X | |
| Engineering the emergence of norms: a review | X | X | | | | X |
| Formally Describing the Architectural Behavior of Software-intensive Systems-of-Systems with SosADL | X | X | | X | | X |
| Harnessing Emergence The Control and Design of Emergent Behavior in System of Systems Engineering | X | | | X | X | X |
| Practical Modeling Concepts for Engineering Emergence in Systems of Systems | X | | | X | | |
| Some Modeling Simulation Perspectives on Emergence in System-of-Systems | X | X | | | | |
| System behavior models: a survey of approaches | X | X | | | X | X |
| Towards an Understanding of Emergence in Systems-of-Systems | X | | | | | X |
| Towards evaluating the emergent behavior of the Internet of Things using large-scale simulation techniques | X | X | | X | | |
| Validating Emergent Behaviors in Systems-of-Systems through Model Transformations | X | | | X | X | X |
| Verifying Cyber-Physical System Behavior in the Context of Cyber-Physical System-Networks | X | X | | | | X |

## 7.11 Summary of the Findings

For a better understanding of how behavioral characteristics in an IoT, the procedure described in the previous sections has been performed. At the end of this procedure, it was intended to answer the six questions described in section 5 of this paper.

**RQ1: WHAT is the understanding of behavior in IoT?**

Many authors relate the idea of behavior with the result of constituent systems; that is, the behavior is generated by the interaction and collaboration of two or more devices. Garcés and Nakagawa [11] state that an SoS behavior arises from the component systems' synergistic cooperation. As for Oquendo F. [18], new behaviors result from interactions between component systems. Gabor, T. et al. [8] states that all behavior is exercised by some event generated by a cyber-physical system. Jackson M. [1] states that complex behavior is created and understood by combining simpler behaviors.

An IoT's behavior is aggregative and emergent; that is, a set of smaller behaviors form a greater behavior. However, this greater behavior can perform different actions, which would be impossible if these behavior combinations were not possible. Oquendo [14] states that the behavior of the whole SoS is more than the sum of the behaviors of its constituent systems. For Roca, D. et al. [10], the whole's behavior is occasioned by the parts' joining, but when these parts are together, they can create actions that could not be separated.

Graciano Neto, V. V. [9] observed that IoT is formed by independent systems denominated as constituents. They display dynamic properties called emergent behaviors, a global functionality resulting from interoperability between constituent systems. Zurita and Tumer [15] argue that emergent behavior is a distinct aspect of operations. The system's displayed behavior is more complicated than the behavior of the individual components shaping the system. On the other hand, the System of Systems (SoS) field provides a different approach defining emergence as the development of patterns, structures, and properties within an SoS where the final behavior cannot be analyzed as a manifestation of the parts and cannot be predicted.

One of the main factors that make understanding behavior so tricky is the ability to behave at any moment, not knowing when it will happen. The characteristic that makes this happen is called "emergence" and is one of the primary studies that involve the behavior of an IoT.

Luo et al. [5] highlight that emerging properties represent one of the most significant challenges for complex systems engineering. Thus, having a good understanding of this emergence is one of the principal needs to have a good knowledge of the behavior facet. Jackson, M. [1] states that the behavior of the system of any practical system is inevitably complicated, and the non-formal nature of the problematic world adds significantly to this complexity. Therefore, the behavior is a challenging task to be understood and dealt with.

**RQ2: HOW do IoT projects deal with the operationalization of behavior?**

For this reason, some researchers have developed some ideas to try to improve the way they manage these behaviors. The first and most common are through an abstraction where the behaviors are divided into stages. According to this idea, the greater behaviors are constituted by smaller ones. As a result, the complex difficulty of dealing with huge behaviors decreases, making it possible to treat these behaviors in lower parts. Haynes,

C. et al. [16] states that behavior patterns generated at a macro level are caused by objects' interactions at the system's micro-level. Ruppel, S. R. [7] states that separation of concerns is a conceptual approach to dealing with complexity; problems can be "separated" (or modularized) and treated individually, reducing the perceived complexity. Therefore, this idea was also applied to the treatment of IoT behaviors.

Another way to treat this behavior was observed by Haynes et al. [16] was through a conceptual overview of some steps needed to harness and control emerging behavior in a system. Assuming a behavior has arisen in the system, the first step is to detect this behavior. Once this behavior is detected, some assessments are necessary to know what behavior this is. Then Haynes divides behavior into two types: beneficial and maleficent. Once you have analyzed the type of behavior, the next stage begins to encourage beneficial behaviors or discourage harmful behaviors. This idea intends to make the object itself understand what behaviors were good or bad for it through this encouragement or discouragement of actions.

Another way to manage behavior is by using a state machine, where each state will show what behavior was executed by a given object. For example, an object in an X state after performing a Y behavior was directed to the Z state. Gabor, T. et al. [8] uses this idea of states to control the behaviors generated by objects in a system. A central controller always observes the behaviors that were executed and then did the necessary state transitions. Jackson, M. [1] also uses a state machine to coordinate a given system's different behaviors.

When dealing with a particular behavior, it is also necessary to visualize/represent the same quickly and directly, being able to see the objects of a system, how they behave, with whom they interact, etc. For this task, several SoS behavior modeling approaches include system dynamics (SD), agent-based modeling (ABM), and discrete event modeling & simulation approaches using languages such as Systems Modeling Language (SysML), Enhanced Functional Flow Block Diagram (EFFBD), and Lifecycle Modeling Language (LML). SD involves control and feedback modeling in system processes. SosADL and Monterey Phoenix (MP) are the most commonly used behavioral modeling frameworks to describe the IoT architecture regarding abstract specifications of possible constituent systems, mediators, and behaviors [13].

SosADL was created to overcome the limitations of existing ADLs, providing the expressive power to describe the architectural concerns of the five essential features of SoS and describe emerging behaviors in evolutionary architectures. Oquendo, F [14] discusses the idea of SosADL, which is an SoS architecture described regarding abstract specifications of possible constituent systems, mediators, and their coalitions. Therefore, the central concepts are the system to represent the constituents, the mediator to describe the possible connectors between the components, and the coalition to represent their composition to form an SoS.

To model these behaviors Giammarco, K. [12] [13] works with another idea, making use of the Monterey Phoenix (MP) that uses the principle of "separation of interests" to model the behaviors of the system separately and unite these separate system models with the interactions and restrictions at the SoS level.

According to Ruppel et al., another way of modeling these behaviors is by using Petri nets that are well established as a behavior modeling approach, facilitating graphical representation based on formal semantics. According to it, a static model describes only the states of a system, while a dynamic model also describes system transitions. Considering that Petri nets are active, they can be used to model the behavior of an IoT.

**RQ3: WHERE do IoT projects locate the activities regarding behavior?**

Throughout the study that was executed, it was not possible to observe any characteristic and no pattern that answered the question in this section, which sought to know where this behavior is treated.

However, in section 8.5.5, it is possible to observe a significant concern in the initial part of the project that was when the problem of behavior should be treated. For this reason, the question of where to manage this behavior should also be of great concern in the initial parts of the project, taking into consideration that where to treat and when to treat may be very close. Therefore, activities must be created within these phases of the project to address this treatment of behavior.

In another perspective, it is also necessary to create activities to take care of this behavior throughout the execution and implementation of the system, taking into account that many times this behavior is costly to change and throughout the execution and implementation can be discovered new behaviors that have not yet were treated.

**RQ4: WHOM do IoT projects allocate to deal with behavior?**

Once explained how and where these behaviors could be treated, you need to know what roles are involved in an IoT project and what roles are in place after building the system.

In general, the stakeholders of a system are those people and organizations with a legitimate claim to influence the design of the system behavior [1]. When talking about the roles involved in building the systems, two roles are present: the objects that are part of the system and the people who will interact with the system in general (these are the primary stakeholders in the system). People participate in systems in many different roles, such as a vending machine's casual user, as a plant operator, as the driver of a car or train or the pilot of an aircraft, etc. [1].

On the other hand, managing a project requires different professionals' profiles, each with a different specialty. In an IoT project, the most critical roles identified were software engineers, programmers, software architects, and system architects [8][9]. The other roles found were the system users, who are the people involved in the system, and the role of each of the objects within a system [12].

**RQ5: WHEN do the effects of time and states of behavior affect IoT projects?**

Once you have made the above explanations, it becomes necessary to understand when these behaviors can and should be addressed.

The main phases of the lifecycle of an IoT project that were identified were initialization, development, validation, implementation, and verification [5].

There is a significant concern in the early stages of the project, Mittal and Rainey [4] claim that emerging behavior arises from a lack of understanding of the system. They also recognized that, as systems grow in complexity, the work required to understand the system's parts increases significantly. In conclusion, it is difficult to capture the whole behavior of a complex system [4]. For this reason, the initial phases of the project are very relevant for good project progress. A project's requirements are one of the main points of any project since it contains all the information needed to build the system. To have a good understanding of a system's behavior, it is necessary to have a good understanding of the requirements of the system; for this reason, the IoT treatment main emphasis is attributed to the initial phase of requirements engineering [4].

Another phase of extreme relevance for the construction of an IoT is the development phase. In this phase, the system's structuring is done, where the system will be constructed, and where all the behaviors are structured. Jackson, M. [1] states that the development problem also involves the process of designing a behavior that satisfies the requirements, with software specifications that can ensure good behavior. Considering that the development procedure is one of the main tasks by which behavior appears, Luo et al. [5] states that great care must be taken at this stage. The developer must conduct a risk analysis to develop risk management strategies to support management and decision making.

As already seen, many factors influence an IoT project's excellent progress, for example, the emerging behavior that is changing at every moment, the requirements in those systems that also change frequently, etc. For these reasons, many aspects are often changing in IoT. With this, it is expected that frequent updates will occur in the life cycles of projects involving IoT.

Taking these frequent updates into account, a critical phase of a project that works with IoT is the process of maintaining it, where new changes will happen in the system. Luo et al. [5] present a maintenance process consisting of four main phases: the first is the initialization phase of the change request, which is the initial phase of the change process, in which any project engineer or development team member can send a proposed change and insert the change request into the project database. The validation and evaluation phase of the change request begins to validate the change request form. In the third phase, the implementation of the change in the system starts. And finally, the verification to analyze if the change was made correctly comes.

**RQ6: WHY do IoT projects implement behavior?**

Finally, it is worth remembering some reasons why understanding, studying, and knowing how to deal with IoT behavior is such a relevant task.

First, the behavior of the system is considered as the central object of software development. For a cyber-physical system, the software's execution is only a means to obtain the desired behavior in the physical world outside the machine [1].

Today's systems are becoming increasingly sophisticated, with more component numbers and behavioral combinations, resulting in more emergent and unintended behaviors. Understanding these behaviors is critical to an accurate assessment of the system being designed [15]. When it comes to the Internet of Things systems, it is often difficult to predict how correctly the system will behave in advance [8]. However, there must be specific assurances about the system's behavior for mainly all practical applications since it would not be safe to implement it otherwise. It is worth remembering that the early identification of these behaviors can reduce the cost of the program and its risk [13].

## 7.12 Final Considerations

Based on the study that was executed, it was possible to observe that behavior is still a weak point to be treated when the subject is IoT and that there are still many gaps to be filled.

Looking at the definitions presented, it was possible to observe that the behavior is seen as the final act of a specific system, that is, the result of it. When a user or another system interacts with a particular system, it is only concerned with the final result of it, with the behavior that the system will generate.

Still looking at the definitions that have been presented, the behavior is seen as something aggregative, emergent, and unpredictable, and that is interacting with other objects at all times in different ways.

The aggregative characteristic causes a behavior to be created by several minor behaviors. Developers significantly exploit this feature; they take advantage of this feature to be able to manage the more substantial behaviors, they try to control the smaller behavior so that the more important behavior that is the junction of these is also achieved.

Usually, this management attempt is sufficient, but sometimes it is not so relevant. As stated by several authors, this behavior is created by small parties that can have different actions that the smaller parties could not execute, bearing an idea that the sum of the behaviors is higher than the union of the parties' behaviors. At that moment, there is still a gap to be solved; how can one treat this greater behavior, which is not known for sure what this behavior is?

The second characteristic presented was the emergency that causes the behavior to come to arise at unexpected moments. This is the second gap found in this study. How is it possible to find emergent behavior computationally?

The third characteristic was that the behavior is interacting with several other components at all times. Considering one of the shortcomings encountered was trying to understand how to work to make this behavior as decentralized as possible so that greater behavior could occur even if some of the constituent components of that system were inaccessible.

Looking a little at how these behaviors are treated and represented, one can observe a great enthusiasm in this area, mainly in the field of representation. Some ideas and tools that are used to generate this representation have been presented throughout the paper. However, many of them are not yet entirely adequate to be used to represent a behavior.

Many of these tools (except some) still use static modeling to represent behaviors and their interactions. However, the behavior is more dynamic, and a tool that captures this idea is necessary.

Regarding the stages of construction of a system, it was possible to observe a significant concern in the initial parts of the projects that treat IoT. However, it is also necessary to be very careful in the other stages, such as development and execution.

In the developmental stage, all behavioral treatments are done; for this reason, it is necessary to take great care so that nothing is forgotten. Already in the project's execution is the phase that the system is working in a real environment with several other variables that can influence its behavior. Therefore, it is necessary to have a more significant concern at this stage to deal with possible changes, considering that depending on the environment that an object was placed, it can react differently.

Finally, it was not possible to find any activity that dealt with this behavior. However, it was believed that it would be necessary to create specific tasks to handle this behavior in the project phases that were already listed (initial stages, development, and execution).

## 7.13 References

**Final set**

[1] JACKSON, M. Behaviours as design components of cyber-physical systems. Lecture Notes in Computer Science (including subseries Lecture Notes in Artificial Intelligence and Lecture Notes in Bioinformatics), v. 8987, p. 43–62, 2015.

[2] KOPETZ, H. et al. Towards an understanding of emergence in systems-of-systems. 2015 10th System of Systems Engineering Conference, SoSE 2015. Anais...Institute of Electrical and Electronics Engineers Inc., 2015

[3] WACHHOLDER, D.; STARY, C. Enabling emergent behavior in systems-of-systems through bigraph-based modeling. 2015 10th System of Systems Engineering Conference, SoSE 2015. Anais...Institute of Electrical and Electronics Engineers Inc., 2015

[4] MITTAL, S.; RAINEY, L. Harnessing emergence: The control and design of emergent behavior in a system of systems engineering. (M. S. Moon I.-C. Syriani E., Ed.)Simulation Series. Anais...The Society for Modeling and Simulation International, 2015

[5] LUO, J.; SAHRAOUI, A.-E.-K.; HESSAMI, A. G. Emergent properties and requirements evolution in engineering systems and a roadmap. (E. M. Nemiche M., Ed.)Proceedings of 2015 IEEE World Conference on Complex Systems, WCCS 2015. Anais...Institute of Electrical and Electronics Engineers Inc., 2015

[6] ZEIGLER, B. P.; MUZY, A. Some modeling & simulation perspectives on emergence in system-of-systems. (P. S. Mittal S. Michael P. .. Severinghaus R. .. Martin J. L. R. .. Cetinkaya D. .. Zapater M. .. Elfrey P., Ed.)Simulation Series. Anais...The Society for Modeling and Simulation International, 2016

[7] RUPPEL, S. R. System behavior models: a survey of approaches. Ph.D. Thesis—[s.l.] Monterey, California: Naval Postgraduate School, 2016.

[8] GABOR, T. et al. A simulation-based architecture for smart cyber-physical systems. (L. J. Giese H. Kounev S., Ed.)Proceedings - 2016 IEEE International Conference on Autonomic Computing, ICAC 2016. Anais...Institute of Electrical and Electronics Engineers Inc., 2016

[9] GRACIANO NETO, V. V. Validating emergent behaviors in systems-of-systems through model transformations. (B. R. Gray J., Ed.)CEUR Workshop Proceedings. Anais...CEUR-WS, 2016.

[10] ROCA, D. et al. Emergent Behaviors in the Internet of Things: The Ultimate Ultra-Large-Scale System. IEEE Micro, v. 36, n. 6, p. 36–44, 2016

[11] GARCÉS, L.; NAKAGAWA, E. Y. A process to establish, model and validate missions of systems-of-systems in reference architectures. Proceedings of the ACM Symposium on Applied Computing. Anais...Association for Computing Machinery, 2017

[12] GIAMMARCO, K. Practical modeling concepts for engineering emergence in systems of systems. 2017 12th System of Systems Engineering Conference, SoSE 2017. Anais...Institute of Electrical and Electronics Engineers Inc., 2017

[13] GIAMMARCO, K.; GILES, K.; WHITCOMB, C. A. Comprehensive use case scenario generation: An approach for modeling system of systems behaviors. 2017 12th System of Systems Engineering Conference, SoSE 2017. Anais...Institute of Electrical and Electronics Engineers Inc., 2017

[14] OQUENDO, F. Architecturally describing the emergent behavior of software-intensive system-of-systems with SosADL. 2017 12th System of Systems Engineering Conference, SoSE 2017. Anais...Institute of Electrical and Electronics Engineers Inc., 2017a

[15] ZURITA, N. F. S.; TUMER, I. Y. A survey: Towards understanding emergent behavior in complex engineered systems. Proceedings of the ASME Design Engineering Technical Conference. Anais...American Society of Mechanical Engineers (ASME), 2017

[16] HAYNES, C. et al. Engineering the emergence of norms: A review. Knowledge Engineering Review, v. 32, p. 1–31, 2017.

[17] BRINGS, J. Verifying Cyber-Physical System Behavior in the Context of Cyber-Physical System-Networks. Requirements Engineering Conference (RE), 2017 IEEE 25th International. Anais...IEEE, 2017

[18] OQUENDO, F. Formally Describing the Architectural Behavior of Software-Intensive Systems-of-Systems with SosADL. (M. M. Mokhtari M. Wang H. H., Ed.)Proceedings of the IEEE International Conference on Engineering of Complex Computer Systems, ICECCS. Anais...Institute of Electrical and Electronics Engineers Inc., 2017b

[19] BOSMANS, S. et al. Towards evaluating the emergent behavior of the internet of things using large-scale simulation techniques (WIP). Proceedings of the Theory of Modeling and Simulation Symposium. Anais...Society for Computer Simulation International, 2018

**Additional References:**

[20] TRICCO et al. A scoping review of rapid review methods. BMC Medicine, 2015.

[21] CARTAXO et al.: The Role of Rapid Reviews in Supporting Decision -Making in Software Engineering Practice. EASE 2018.

[22] CARTAXO et al. Evidence briefings: Towards a medium to transfer knowledge from systematic reviews to practitioners. ESEM, 2016.

# DEVELOPING IOT SOFTWARE SYSTEMS? TAKE SMARTNESS INTO ACCOUNT

## CONTEXT

IoT allows composing systems from *things* equipped with identifying, sensing, or actuation behaviors and processing capabilities that can communicate and cooperate to reach a goal. For so, different facets should be taken into account, such as smartness.

## SMARTNESS

It refers to orchestration associated with *things* and to what level of intelligence technology it evolves, allowing *things* to acquire a higher or lower degree of smartness. A smart system needs a set of actions, for example treating data, making decisions and acting through software.

## SUMMARY

It focuses on capturing and representing smartness, an essential non-functional requirement for IoT. It represents the extension of things' original capabilities, allowing things to acquire a higher or lower degree of smartness through software. It is related to things behavior, orchestration, and to whatever level of intelligence such a thing achieves from its initial capacity, without human intervention.

## OBJECTIVE

This briefing aims to enlighten project planning and to provide guidance when dealing with Smartness in IoT. The information presented in this briefing seeks to answer: "What taken into account when considering smartness for IoT?".

## MOTIVATION

IoT offers challenges for their construction, with concerns that rather than just developing, we need to engineer software systems. This motivated us to a more comprehensive investigation to address these concerns and to provide evidence-based support for IoT engineering.

## ADDITIONAL INFORMATION

**Who are the briefing's clients?** Software developers and practitioners who want to make decisions about how to deal with *smartness* in IoT, considering scientific evidence.

**Where do the findings come?** All findings of this briefing were extracted from scientific studies about *smartness* through a Rapid Review[1]. The Technical Report[2] containing all the findings is available for further information.

**What is included in this briefing?** Technologies, challenges, and strategies to deal with *smartness* in IoT projects.

**What are the Challenges and Opportunities?** One of the challenges is the lack of standardization or understanding of what smartness is. Having only sensors and collecting data, does not make a solution "smart", therefore more research should be performed. Another challenge is that most devices cannot alone: process data satisfactory; storage space to deal with AI algorithms; high energy consumption; lack of memory. So, limitations on the components is a limitation for smartness as well.

**Does smartness represent a concern in the Engineering of Internet of Things Software Systems?** Yes. The objective found about smartness in the context of the systems are to become them more invisible, with less user interaction, and autonomous, performing a set of tasks in the environments. Thus facilitating the life of their users.

## HIGHLIGHTS

- ***Limitations*** *on the components is a limitation for smartness as well.*
- *It is necessary to define which virtual or physical component will* ***encapsulate the intelligence*** *of the system, for smartness.*

## WHAT
### is the the understanding of smartness in CSS?

- It deals with how data is **collected, analyzed, treated, and transmitted to manage, decision making, and actions**. The combination of these characteristics let the system to be semi or entirely autonomous for performing any action in the environment, reducing human intervention.
- The system behavior is associated with **different levels of smartness**, depending on the application domain and the user`s needs.

## HOW
### do CSS projects deal with the operacionalization of smartness?

- It uses a combination of technologies such as: **Sensors** (since they are responsible for gathering data from the environment), **Wearables** (due to their capacity of embedding several sensors and interaction to human users), **Decision makers** (since they deal with data from the environment and data processing), **Actuators** (all hardware performing any action that can affect the environment, causing a change of state within the ambient), and solutions on **Artificial Intelligence**, machine learning, neural networking, fuzzy logic and others.

## WHERE
### CSS projects locate the activities regarding smartness?

- Throughout the study, it was not possible to observe any evidence that answered the question.

## WHO
### do CSS projects allocate to deal with smartness?

- It is necessary to define which virtual or physical component will **encapsulate the intelligence** of the system.
- The **stakeholders** should decide the requirements and smartness levels from the beginning of the project, since it is fundamental for dealing with smartness.

## WHEN
### do the effects of time, and states of smartness affect CSS projects?

- The system should be defined considering real-time operation that can vary **Real-time decision, real-time monitoring and or real-time visualization.**
- The response time and autonomy of the system should reflect this operation.

## WHY
### do CSS projects implement smartness?

- The goal to make the systems smarter is to improve their **intelligence** and **autonomy**.
- The purpose of making smarter solution is to make them less dependent of users. To improve the quality of life of end users, and, management of the environment, such as save energy, sustainable building, healthcare, and others.

[1] More on Rapid Reviews: https://dl.acm.org/citation.cfm?id=3210462
[2] Full Report: https://goo.gl/no7jtR

# 8 RAPID REVIEW ON SMARTNESS

# Rapid Reviews Meta-Protocol:

## Engineering of Internet of Things Software Systems

**Bruno P. de Souza, Rebeca C. Motta, Guilherme H. Travassos**

# Smartness

In the investigation regarding Internet of Things Software Systems (IoT), it has been observed that these modern software systems offer challenges for their construction since they are calling into question our traditional form of software development. Usually, they rely on different technologies and devices that can interact-capture-exchange information, act, and make decisions. It concerns that rather than just developing software, we need to engineer software systems embracing multidisciplinary, integrating different areas. From our initial research, we analyzed the concerns related to this area. We categorized them into a set of facets - Connectivity, Things, Behavior, Smartness, Interactivity, Environment, and Security - representing such projects' multidisciplinary, in the sense of finding a set of parts composing this engineering challenge.

Since these facets can bring additional perspectives to the software system project planning and management, acquiring evidence regarding such facets is of great importance to provide an evidence-based framework to support software engineers' decision-making tasks. Therefore, the following question should be answered:

***"Does Smartness represent a concern in the engineering of***

***Internet of Things software systems?"***

This Rapid Review (RR) aims to analyze Smartness to characterize it in the IoT field, regarding *what, how, where, when and why* it is used in the context of IoT projects, verifying the existence of published studies supporting the previous results. The 5W1H aims to give the observational perspective on which information is required to the understanding and management of the facet in a system (what); to the software technologies (techniques, technologies, methods, and solutions) defining their operationalization (how); the activities location being geographically distributed or something external to the software system (where); the roles involved to deal with the facet development (who); the effects of time over the facet, describing its transformations and states (when); and to translate the motivation, goals, and strategies going to what is implemented in the facet (why), in respect of IoT projects.

## 8.1 Research Questions

- **RQ1:** What is the understanding and management of Smartness in IoT projects?
- **RQ2:** How do IoT projects deal with software technologies (techniques, technologies, methods, and solutions) and their operationalization regarding Smartness?
- **RQ3:** Where do IoT projects locate the activities regarding Smartness?

- **RQ4:** Whom do IoT projects allocate to deal with Smartness?
- **RQ5:** When do the effects of time, transformations, and states of Smartness affect IoT projects?
- **RQ6:** Why do IoT projects implement Smartness?

## 8.2 Search Strategy

The Scopus[9] search engine and the following search string support this RR:

**P**opulation - Internet of Things software systems

Synonymous:

"ambient intelligence" OR "assisted living" OR "multiagent systems" OR "systems of systems" OR "internet of things" OR "Cyber-Physical Systems" OR "Industry 4" OR "fourth industrial revolution" OR "web of things" OR "Internet of Everything" OR "contemporary software systems" OR "smart manufacturing" OR digitalization OR digitization OR "digital transformation" OR "smart cit*" OR "smart building" OR "smart health" OR "smart environment"

**I**ntervention – (smartness OR intelligence OR "autonomous reaction" OR "learning capability")

**C**omparison – no

**O**utcome – (understanding OR management OR technique OR “technolog*” OR method OR location OR place OR setting OR actor OR role OR team OR time OR transformation OR state OR reason OR motivation OR aim OR objective)

**C**ontext – (engineering or development or project or planning OR management OR building OR construction OR maintenance)

Limited to articles from 2015 to 2018

Limited to Computer Science and Engineering

LIMIT-TO (SUBJAREA, "COMP" ) OR LIMIT-TO (SUBJAREA, "ENGI" ) ) AND ( LIMIT-TO (PUBYEAR, 2018 ) OR LIMIT-TO (PUBYEAR, 2017 ) OR LIMIT-TO (PUBYEAR, 2016 ) OR LIMIT-TO (PUBYEAR, 2015)

| |
|---|
| TITLE-ABS-KEY (("ambient intelligence" OR "assisted living" OR "multiagent systems" OR "systems of systems" OR "internet of things" OR "Cyber Physical Systems" OR "Industry 4" OR "fourth industrial revolution" OR "web of things" OR "Internet of Everything" OR "contemporary software systems" OR "smart manufacturing" OR digitalization OR digitization OR "digital transformation" OR "smart cit*" OR "smart building" OR "smart health" OR "smart environment") AND (smartness OR intelligence OR "autonomous reaction" OR "learning capability") AND (understanding OR management OR technique OR “technolog*” OR method OR location OR place OR setting OR actor OR role OR team OR time OR transformation OR state OR reason OR motivation OR aim OR objective) AND (engineering or development or project or planning OR management OR building OR construction OR maintenance) LIMIT-TO ( SUBJAREA , "COMP" ) OR LIMIT-TO ( SUBJAREA , "ENGI" ) ) AND ( LIMIT-TO ( |

[9] https://www.scopus.com

PUBYEAR , 2018 ) OR LIMIT-TO ( PUBYEAR , 2017 ) OR LIMIT-TO ( PUBYEAR , 2016 ) OR LIMIT-TO ( PUBYEAR , 2015))

## 8.3 Selection procedure

One researcher performs the following selection procedure:

1. Run the search string;
2. Apply the inclusion criteria based on the paper Title;
3. Apply the inclusion criteria based on the paper Abstract;
4. Apply the inclusion criteria based on the paper Full Text, and;

After finishing the selection from Scopus, use the included papers set to:

5. Execute snowballing backward (one level) and forward:
    a. Apply the inclusion criteria based on the paper Title;
    b. Apply the inclusion criteria based on the paper Abstract;
    c. Apply the inclusion criteria based on the paper Full Text.

The JabRef Tool[10] must be used to manage and support the selection procedure.

## 8.4 Inclusion criteria

- The paper must be in the context of **software engineering**; and
- The paper must be in the context of the **Internet of Things software systems**; and
- The paper must report a **primary or a secondary study**; and
- The paper must report an **evidence-based study** grounded in empirical methods (e.g., interviews, surveys, case studies, formal experiment, etc.); and
- The paper must provide data to **answering** at least one of the RR **research questions**.
- The paper must be written in the **English language**.

## 8.5 Extraction procedure

The extraction procedure is performed by one researcher, using the following form:

| <paper_id>:<paper_reference> | |
|---|---|
| Abstract | <Abstract> |
| Description | <A brief description of the study objectives and personal understanding> |
| Study type | <Identify the type of study reported by paper (e.g., survey, formal experiment)> |
| RQ1: WHAT information required to understand and manage the << facet>> in IoT | - < A1_1><br>- < A1_2><br>- ... |
| RQ2: HOW software technologies (techniques, technologies, | - < A2_1><br>- < A2_2><br>- ... |

10 http://www.jabref.org/

| | |
|---|---|
| methods and solutions) and their operationalization | |
| RQ3: WHERE<br>activities location or something external to the IoT | - < A3_1><br>- < A3_2><br>- ... |
| RQ4: WHO<br>roles involved to deal with the << facet>> development in IoT | -< A4_1><br>-<A4_2><br>- … |
| RQ5: WHEN<br>effects of time over << facet>>, describing its transformations and states in IoT | - < A5_1><br>- < A5_2><br>- ... |
| RQ6: WHY<br>motivation, goals, and strategies regarding Smartness in IoT | - < A6_1><br>- < A6_2><br>- ... |

## 8.6 Synthesis Procedure

In this RR, the extraction form provides a synthesized way to represent extracted data. Thus, we do not perform any synthesis procedure.

However, the synthesis is usually performed through a narrative summary or a Thematic Analysis when the number of selected papers is not high.

## 8.7 Report

An Evidence Briefing [2] reports the findings to ease the communication with practitioners. It was presented as the cover for this chapter.

## 8.8 Results

**Execution**

| Activity | Execution date | Result | Number of papers |
|---|---|---|---|
| First execution | 12/07/2018 | 2070 documents added | 2070 |
| Included by Title analysis (v1) | 14/07/2018 | 1474 documents withdrawn | 596 |
| Included by Title analysis (v2) | 19/07/2018 | 243 documents withdrawn | 353 |
| Included by Abstract analysis (v1) | 24/07/2018 | 231 documents withdrawn | 122 |
| Included by Abstract (v2) | 28/07/2018 | 31 documents withdrawn | 91 |
| Papers not found | 29/07/2018 | One document withdrawn | 28 |
| Articles for reading | 29/07/2018 | 91 documents | 17 |
| Removed after a full reading | 01/08/2018 | 74 documents withdrawn | 17 |
| Snowballing | 05/08/2018 - 13/08/2018 | Seven documents added | 24 |
| Snowballing after reading | 05/08/2018 - 13/08/2018 | None paper was excluded | 24 |
| Total included | 13/08/2018 | 24 documents | 24 |
| Papers extracted | 13/08/2018 | 24 documents | 24 |

**Final Set**

| Reference | Author | Title | Year | Source |
|---|---|---|---|---|
| [1] | Bartolozzi et al. | A Smart Decision Support System for Smart City | 2015 | Regular search |
| [2] | Gutierreza et al. | Smart Waste Collection System Based on Location Intelligence | 2015 | Regular search |
| [3] | Paola et al. | SmartBuildings: an AmI system for energy efficiency | 2015 | Regular search |
| [4] | Ghaffarianhoseini et al., | What is an intelligent building? Analysis of recent interpretations from an international perspective | 2015 | Snowballing |
| [5] | Atabekov et al. | Internet of Things-Based Temperature Tracking System | 2015 | Snowballing |
| [6] | Neuhofer et al. | Smart technologies for personalized experiences: a case study in the hospitality domain | 2015 | Snowballing |
| [7] | Korzun et al. | Performance evaluation of smart-M3 applications: A SmartRoom case study | 2016 | Regular search |
| [8] | He et al. | Internet-of-Things Based Smart Resource Management System: A Case Study Intelligent Chair System | 2016 | Regular search |
| [9] | Babli et al. | An Intelligent System for Smart Tourism Simulation in a Dynamic Environment | 2016 | Regular search |
| [10] | Costa et al. | NuSense: A Sensor-Based Framework for Ambient Awareness applied in Game Therapy Monitoring | 2016 | Regular search |
| [11] | El-Faouri et al. | A smart street lighting system using solar energy | 2016 | Snowballing |
| [12] | Saifuzzaman et al. | IoT based street lighting and traffic management system | 2017 | Regular search |
| [13] | Corno et al. | On the design of energy and user aware study room | 2017 | Regular search |
| [14] | Rekha et al. | High yield groundnut agronomy: An IoT based precision farming framework | 2017 | Regular search |
| [15] | Chen et al. | Smart Home 2.0: Innovative Smart Home System Powered by Botanical IoT and Emotion Detection | 2017 | Regular search |
| [16] | Oliveira et al. | SmartCoM: Smart Consumption Management Architecture for Providing a User-Friendly Smart Home based on Metering and Computational Intelligence | 2017 | Regular search |
| [17] | Medina et al. | Retrofit of air conditioning systems through a Wireless Sensor and Actuator Network: An IoT-based application for smart buildings | 2017 | Regular search |
| [18] | Shyam et al. | Smart waste management using Internet-of-Things (IoT) | 2017 | Regular search |
| [19] | Paola et al. | An ambient intelligence system for assisted living | 2017 | Snowballing |
| [20] | Yong et al. | IoT-based intelligent fitness system | 2018 | Regular search |
| [21] | Espinilla et al. | The Experience of Developing the UJAmI Smart Lab | 2018 | Regular search |
| [22] | Bisio et al. | Exploiting Context-Aware Capabilities over the Internet of Things for Industry 4.0 Applications | 2018 | Regular search |
| [23] | Medina et al. | Intelligent multi-dose medication controller for fever: From wearable devices to remote dispensers | 2018 | Snowballing |
| [24] | Palacios et al. | Approximation and Temperature Control System via an Actuator and a Cloud: An Application Based on the IoT for Smart Houses | 2018 | Snowballing |

## 8.9 Summary of the articles

In this section, we presented some of the key concepts about smartness to improve this research's understanding. Also, a summary was performed about the works founds. According to Liao et al. [30], we are experiencing a new technological era, in which the characteristics of the system are changing, such as context awareness, autonomy, omnipresence, and smartness [31].

The fourth industrial revolution has emerged as a way to automate systems that use many resources and be in combination with new systems technologies such as the Internet of Things (IoT) and Cyber-Physical Systems (CPS). These systems mainly influenced the manufacturing industry. Within the new wave of the industrial revolution, new types of systems mentioned above have emerged to facilitate the end user's lives. These systems incorporate new features that the conventional systems did not have [30].

Some of the main challenges of the fourth industrial revolution are the lack of computational power, the large volume of data, the necessity of maintaining connectivity; the emergence of analytical and business intelligence capabilities; the emergence of new forms of human-computer interactions; improvements in transferring digital instructions from the virtual world to the physical world, interoperability between SoS and so on [30] [31]. All these challenges focus on efficient production. However, this research is focused only on the "smartness" of the Internet of Things software systems (IoT).

Nowadays, several IoTs have been proposed in the literature and industry [1] [2] [3] [30]. These systems are capable of collecting data from the environment, making decisions, and acting. For that to happen, a series of devices, sensors, actuators, and things are used to reach a goal. One of the types of systems included in the 4th industrial revolution is IoT systems [30].

IoT is defined as “a set of heterogeneous technologies and devices available that interact through a network, being able to capture data and information, exchange data, information, commands and make decisions and act.” IoT systems have many characteristics, such as addressability, autonomy, context-awareness, heterogeneity, and interoperability [31].

In this research, we are evaluating how smartness is applied in the project of IoT. For this, smartness is defined according to Motta et al. [31] in: "smartness or intelligence is related to behavior, but also as managing or organizing it." It refers to orchestration associated with things and to what level of intelligence technology can evolve from their initial behavior. Also, the system can collect data/information from the ambient, then use this data/information to make a decision and acting". The purpose of this research is based on the GQM [29] with the objective to **analyze** the facet of smartness **with the purpose of** characterizing **with respect to** its concerns in IoT projects **from the point of view of** the researchers in software engineering and **in the context of** the knowledge that is available in the literature. Some works that were found in this RR and used smartness are described below.

Bartolozzi et al. [1] carried out a proof of concept to adapt technologies (of communication, data) to build a framework that would support decision-making in a smart city. The framework works with communication, data, and knowledge-driven trailers simultaneously. The proposed approach is the Km4city (knowledge model for the city). This approach is an intelligent city environment (also called intelligent city ontology) for data aggregation and interoperability semantics and an API for mobile and web applications. This framework was built using the paradigm of thinking systems focusing on a hierarchical analytical process. Initial results showed promise.

On the other hand, in Gutierrez et al. [2], a system was proposed that use IoT to integrate data access network, geographic information system, optimization, and electronic engineering. The proposed system is an intelligent location-based waste collection solution. The simulation where the system is based is in the city of Copenhagen. The study was carried out throughout one month; also, a comparison between the proposed system and how the information was collected previously was made. The system uses sensors to collect and monitor data, internet, microcontroller, server, and optimization algorithms. Regarding smartness, it is the set of equipment (sensors, microcontrollers) used to obtain data and send information. Regarding the results, the authors obtained promising findings from the previous collection; however, the solution's implementation is relatively high in financial resources.

In the research done by Paola et al. [3], an intelligent system was proposed to improve a smart building's energy efficiency through a pervasive infrastructure of monitoring and artificial intelligence. In the authors' perspective, a "smart system" is a set of hardware and software that work together to reach a level of intelligence "almost or totally" autonomous. The intelligent system's implementation is based on a multilayer architecture (which has three layers: the physical layer, middleware layer, and application layer); about data collection, the system has a series of sensors and actuators (specified in the extraction form - appendix A). An algorithm was also used, where a predictive controller is trained using past and forecast information from an external weather forecasting service. As a final result, a smart system was developed to improve a smart building's energy efficiency and was also designed to make the decision. It is worth mentioning that the system is still in development.

Another work-related to Smart buildings (IBs) is from Ghaffarianhoseini et al. [4]. In this research, the authors conducted a literature review on IBs and defined "smart" as Smart can be described by various capacities such as reasoning, problem-solving, acquisition of knowledge, memory, speed of operation, creativity, general knowledge, and motivation. In summary, the authors assumed that smartness applied in smart buildings should carry some characteristics: 1 - incorporation of intelligent technologies and economic principles. 2 - Interwoven with advanced sensors and artificial intelligence—3 - alignment of innovative future technologies and upgrades.

Atabekov et al. [5] constructed a temperature monitoring system based on IoT technologies. The purpose of systems is to help solve overheating problems in a home environment or a server depot. What makes the system intelligent in this case are the sensors used, decision making, and a client-server architecture also used in the system. The main result reported by the authors was the knowledge acquired in the process of implementing this solution. The authors did not define smart in their text. However, they do not mention the system efficiency. Korzun, Marchenkov, and Balandin [7] have built a system that helps humans in collaborative work activities such as conferences, meetings, and seminars. It uses the Smart-M3-based implementation of existing software. The system uses an architecture called M3 (multi-device, multi-vendor, multi-domain) for intelligent spaces.

He, Atabekov, and Haddad [8] proposed an intelligent resource management system. To implement the systems, they used a REST architecture, an RFID Reader, Pressure Sensor, Miscellaneous (Three LEDs are used to indicate the data transmission status.) Three 180-ohm resistors are used for LEDs, and a 10 Ω ohm resistor is used for the pressure sensor), and the way the sensor collects the data. The authors had good results about the study, although the system is still in its initial state.

Babli et al. [9] constructed a simulator decision-making system for tourists; this system aims to draw up an agenda of tasks according to its users' preferences. The system uses the Planning Domain Definition Language (PLDD) and an architecture that was created by a project called GLASS (Goal-management for Long-term Autonomy in Smart citieS). This architecture consists of a two-process loop: planning module and simulation+monitoring that share common information. As a result, the authors report that the system generally worked in the proposed case study.

In the work of Saifuzzaman et al. [12], they proposed an intelligent system based on IoT technology for street traffic and light management. This system can make decisions and act according to the information received from sensors for luminosity control. The intelligent system is still in its initial phase, but it shows very promisingly.

There are a number of other works that use smartness in the IoT development process [6] [10] [11] [13] [15] [18] [20] [23] [24].

## 8.10 Tracking Matrix

| Ref | Paper | WHAT | HOW | WHERE | WHO | WHEN | WHY |
|---|---|---|---|---|---|---|---|
| [1] | A Smart Decision Support System for Smart City | | | | | | X |
| [2] | Smart Waste Collection System Based on Location Intelligence | X | X | X | | | X |
| [3] | SmartBuildings: an AmI system for energy efficiency | X | X | X | | | X |
| [4] | What is an intelligent building? Analysis of recent interpretations from an international perspective | X | X | X | | | X |
| [5] | Internet of Things-Based Temperature Tracking System | X | X | X | | | X |
| [6] | Smart technologies for personalized experiences: a case study in the hospitality domain | X | X | X | | | X |
| [7] | Performance evaluation of smart-M3 applications: A SmartRoom case study | | | X | | | X |
| [8] | Internet-of-Things Based Smart Resource Management System: A Case Study Intelligent Chair System | X | X | X | | | |
| [9] | An Intelligent System for Smart Tourism Simulation in a Dynamic Environment | X | X | X | | | |
| [10] | NuSense: A Sensor-Based Framework for Ambient Awareness applied in Game Therapy Monitoring | X | X | X | | | X |
| [11] | A smart street lighting system using solar energy | X | X | X | | | X |
| [12] | IoT based street lighting and traffic management system | X | X | X | | X | X |
| [13] | On the design of energy and user aware study room | X | X | X | | X | |
| [14] | High yield groundnut agronomy: An IoT based precision farming framework | X | X | X | | X | X |
| [15] | Smart Home 2.0: Innovative Smart Home System Powered by Botanical IoT and Emotion Detection | X | X | X | | | |
| [16] | SmartCoM: Smart Consumption Management Architecture for Providing a User-Friendly Smart Home based on Metering and Computational Intelligence | X | X | X | | | X |
| [17] | Retrofit of air conditioning systems through a Wireless Sensor and Actuator Network: An IoT-based application for smart buildings | X | X | X | | | X |
| [18] | Smart waste management using Internet-of-Things (IoT) | X | X | X | | | X |
| [19] | An ambient intelligence system for assisted living | X | X | | | | X |
| [20] | IoT-based intelligent fitness system | X | X | X | | X | X |
| [21] | The Experience of Developing the UJAmI Smart Lab | X | X | X | | | X |
| [22] | Exploiting Context-Aware Capabilities over the Internet of Things for Industry 4.0 Applications | | X | | | | X |
| [23] | Intelligent multi-dose medication controller for fever: From wearable devices to remote dispensers | X | X | X | | X | X |
| [24] | Approximation and Temperature Control System via an Actuator and a Cloud: An Application Based on the IoT for Smart Houses | X | X | X | | | X |

## 8.11 Summary of the Findings

### RQ1: WHAT is the understanding of Smartness in IoT projects?

This question was answered considering the understanding and managing used to construct the Internet of Things software systems.

In general, the papers report that **data acquisition** or **data collect**ion is a more important characteristic to understand and manage smartness. With the data collected from the environment, the system may send the data to the cloud or another server. Hence, it makes a decision and acts autonomously, in other words, without user interaction. The level of smartness depends on the application domain and user needs. For a system to become smart, it is not necessary to follow an order such as: collect data, analyze data, make decisions, and act, in other words, be autonomous. It is enough that it can make decisions (becoming semi-autonomous). For example, send an e-mail to the user to inform about an event that is going on in the environment.

As we may understand, the way that system will collect the data from the environment is directly associated with the requirements and systems' specification phase in the development cycle. In this step, it was mentioned by the authors which things, sensors, and actuators were needed in the system development. After defining the systems' requirements, the next step was designing the project, which consisted of constructing diagrams.

After finishing the **data collection**, the **data are analyzed** to process, manage, and utilize the data flow that is treated and sent to other applications or services. The vital part of the information is captured about treatments and transmissions of data, then sent to the next application, service, or node.

The other fact to understand or manage smartness in IoT systems is **artificial intelligence** (AI) techniques and **machine learning** that can be applied to enhance intelligence and significant interactions with the end-user. About the development of smart applications, it is essential to highlight that having only sensors collecting data does not make it smart. For a system to be or become smart, it needs a set of actions, for example, treating data, making decisions, and acting. A recommendation system was also mentioned. Figure 1 shows the main characteristics that are cited in the papers that smart systems ought to have.

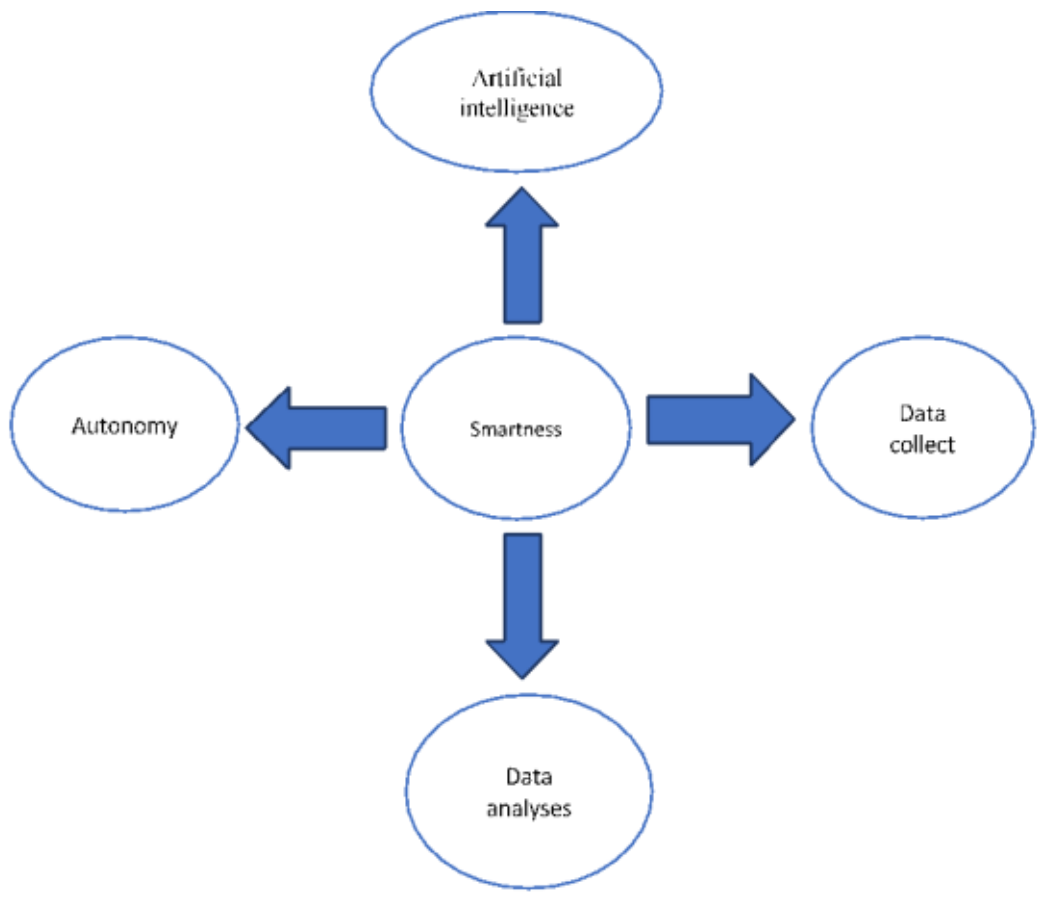

Figure 1. Main characteristics of smartness cited in the papers.

In summary, the papers did not deal with the software development process or mentioned it what made the process of data synthesis flawed and speculative during the data analyses. However, we can consider that these data and analysis collection parts were performed in the software development cycle's initial steps.

**RQ2: HOW do IoT projects deal with software technologies (techniques, technologies, methods, and solutions) and their operationalization regarding Smartness?**

This question was answered considering the technologies used to construct the smart systems. Besides, this question is a compliment from question one. Therefore the systems used many devices, sensors, wearables, and so on to build them.

Another case that was not mentioned in the context of the process of software construction is how the technologies apply to the system project. The papers suggest the algorithms, hardware, or architectures were used. Nevertheless, the authors did not describe software buildings' steps what speculative some information about the technologies used.

***Sensors*** are the essential requirement to construct the smart systems because they get data/information from the ambient and send the data to the system and server—for example, humidity, light, motion sensors. The objective of the sensor in this facet is to capture data from the environment. Thus the systems in the future can make decisions and act. The sensors comprise one of the parts of smartness of the systems, however, on their own they are not "smart" [2] [3] [4] [5] [6] [8] [9] [10] [11] [12] [13] [14] [15] [16] [17] [18] [19] [20] [21] [23] [24].

***Wearables*** are other requirements which consist of technological devices that can be used by users as garments. They are like sensors and serve to get data from the user. The smartphone is another device that is widely used in smart systems. (Wearable was not classified as a sensor because many authors differentiated wearable from the sensor).

***Algorithms*** are a set of logical rules and procedures defined that guide a problem's solution in a finite number of steps. The main algorithms used in these systems are coming from machine learning or artificial intelligence. Some authors described in their research that they used Artificial Intelligent to build smartness in systems, such as Fuzzy logic and neural network [1] [2] [3] [8] [13] [18] [20] [21] [24].

In another case, if there is not **WNS** and **gateways** (optional), there is no smartness or something else because the network is a fundamental characteristic of the Internet of Things software systems. **WNS** and **gateways** are directly linked to the facets of behavior, connectivity, and things. However, they were also intelligently managed, due to sending data, operating as a type of "sensor," and that the data/information browses within them moving from application to application.

**RQ3: WHERE do IoT projects locate the activities regarding Smartness?**

This question was not responded to in most articles in our representation. For example, they mention some artificial intelligence or learning machine algorithms and techniques. Some articles mention that smart is in the cloud, Client-server architecture, SOA, GLASS, and REST APIs. However, they do not mention which part of the software project the smartness is treated. Nevertheless, it was perceived that many studies worked with smartness as a set of operations such as sensing, decision making, data processing, and acting.

In summary, we consider that there is not enough evidence to respond to this question because several papers do not spell out the software project's details. It may become moderately intuitive that smartness can be addressed early in the requirements specification step, or maybe in the smart system's coding step. Therefore, with the evidence we have momentarily, it becomes difficult to have a conclusive result.

**RQ4: WHOM do IoT projects allocate to deal with Smartness?**

In almost all papers, the smart systems were proposed to end-users. It may be because they are the main character in that scenario.

According to [17], a framework is proposed to build smart systems without compromising interoperability between hardware to hardware, hardware to software, and software to software. We can observe that the concern is the integration of hardware and software.

We can consider that many stakeholders evolved in the software development process, such as software engineers, architects, domain analysts, etc.

**RQ5: WHEN do the effects of time, transformations, and states of Smartness affect IoT projects?**

Concerning this question, four results were identified: real-time decision, real-time information, real-time monitoring, and real-time visualization. A real-time decision, this effect is concerned with how to deal with information or data received from the environment and how to make a decision and act. Another feature is real-time information that can be described how the information is sent to the real-time decision. Real-time monitoring only “see” the data or information and show off in and dashboard, and the last one is real-time visualization.

It is important to say that this question was not answered in many articles, due that the same situation happened as question 03 (where). Many authors did not specify which part of the IoT development project. The evidence collected in this RR was not sufficient to state anything regarding the "when."

**RQ6: WHY do IoT projects implement Smartness?**

The Internet of Things systems is playing a vital and essential role in our everyday life. There are several applications that IoT technology can be applied, such as healthcare, smart buildings, the automotive industry, smart farming, and so on. To turn the system more autonomous.

For these reasons, one of them is to improve the lives of end-users. Several intelligent systems have been proposed to deal with these end users, such as the intelligent system of Tourism to recommend a tour of the city being visited by visitors. According to the tourist profile and preference, the smart system suggests tourist attraction be visited [9].

Another example can be saving energy or energy efficiency [3] [11] [13]. Electric energy has become one of the main problems to be solved since the time of its origin. One of the reasons is that there is no communication without it; everything that depends on it would stop working. Since that energy always has an impact on nature. Thus creating smart systems that make it easier to save energy is a challenge.

The paper that reports the development of a smart environment is from Espinilla et al. [21]; this environment aims to improve people's lives with some disease. The intelligent environment has numerous actuators, sensors of different types to capture the environment data, make decisions, and act. Smartness is the set of these actions to make the environment more autonomous and omnipresent.

There are several other reasons, objectives described in the RR for the development of smartness in IoT project, such as smart waste collection [2] [18], body temperature control to indicate if the person has a fever [23], health improvement (healthcare) [6] [8] [10] [20] [21].

The objective found about smartness in the systems' context is to become them more invisible, with less user interaction, and autonomous, performing a set of tasks in the environments and thus facilitating their users' lives.

## 8.12 Final Considerations

In this research, an RR was carried out to understand if smartness is a concern in the IoT software systems. Respond to the research question, smartness is a concern in IoT, due to it is an essential characteristic in these new categories of systems.

One of the reasons for this smartness concern in IoTs may be the lack of standardization or understanding of smart systems. According to the research, to understand the smartness in IoT, the system should have some autonomy level and is a set of operations such as sensing, data collection, data processing, decision-making, and acting to orchestrate things in the environment that are immersed. To the system become "smart," it is not necessary to follow all these operations [8] [15] [18].

Another challenge in our representation is that most devices cannot: processing of data and information satisfactory; storage space to deal with AI algorithms; high energy consumption; lack of memory. Moreover, there are several problems with interoperability between software systems due to the lack of standardization among existing technologies.

## 8.13 References

**Final Set:**

[17] Bartolozzi, M.; Bellini, B.; Nesi, P.; Pantaleo, G.; Santi, L. A Smart Decision Support System For Smart City. 2015 Ieee International Conference On Smart City/Socialcom/Sustaincom (Smartcity). Anais... In: 2015 Ieee International Conference On Smart City/Socialcom/Sustaincom (Smartcity). Dez. 2015

[18] Gutierrez, J. M.; Jensen, M.; Henius, M.; Riaz, T. Smart Waste Collection System Based On Location Intelligence. Procedia Computer Science, Complex Adaptive Systems San Jose, Ca November 2-4, 2015. V. 61, P. 120–127, 1 Jan. 2015.

[19] Paola, A. D.; Re, G.; Morana, M.; Ortolani, M. Smart buildings: An Ami System For Energy Efficiency. 2015 Sustainable Internet And It For Sustainability (Sustained). Anais... In: 2015 Sustainable Internet And Ict For Sustainability (Sustained). Apr. 2015.

[20] Ghaffarianhoseini, A.; Berardi, U.; Alwaer, H.; Chang, S.; Halawa, E.; Ghaffarianhoseini, A.; Croome, D. What Is An Intelligent Building? Analysis Of Recent Interpretations From An International Perspective. Architectural Science Review, V. 59, N. 5, P. 338–357, 2 Set. 2016.

[21] Atabekov, A.; Starosielsky, M.; Lo, D. C.; He, J. Internet Of Things-Based Temperature Tracking System. 2015 Ieee 39th Annual Computer Software And Applications Conference. Anais... In: 2015 Ieee 39th Annual Computer Software And Applications Conference. Jul. 2015

[22] Neuhofer, B.; Buhalis, D.; Ladkin, A. Smart Technologies For Personalized Experiences: A Case Study In The Hospitality Domain. Electronic Markets, V. 25, N. 3, P. 243–254, 1 Set. 2015.
[23] Korzun, D. G.; Marchenkov, S.; Balandin, S. Performance Evaluation Of Smart-M3 Applications: A Smartroom Case Study. 2016 18th Conference Of Open Innovations Association And Seminar On Information Security And Protection Of Information Technology (Fruct-Ispit). Anais... In: 2016 18th Conference Of Open Innovations Association And Seminar On Information Security And Protection Of Information Technology (Fruct-Ispit). Apr. 2016
[24] He, J.; Atabekov, A.; Haddad, H. M. Internet-Of-Things Based Smart Resource Management System: A Case Study Intelligent Chair System. 2016 25th International Conference On Computer Communication And Networks (Icccn). Anais... In: 2016 25th International Conference On Computer Communication And Networks (Icccn). Ago. 2016
[25] Babli, M.; Ibáñez, J.; Sebastiá, L.; Garrido, A.; Onaindia, E. An Intelligent System For Smart Tourism Simulation In A Dynamic Environment.
[26] Costa, T. H.; Paiva, B. L. A.; Menezes, R. V.; Ramos, L. T.; Lopes, A. G.; Bublitz, F. M. Nusense: A Sensor-Based Framework For Ambient Awareness Applied In Game Therapy Monitoring - Pdf.
[27] El-Faouri, F. S.; Sharaiha, M.; Bargouth, D.; Faza, A. A Smart Street Lighting System Using Solar Energy. 2016 Ieee Pes Innovative Smart Grid Technologies Conference Europe (Isgt-Europe). Anais... In: 2016 Ieee Pes Innovative Smart Grid Technologies Conference Europe (Isgt-Europe). Out. 2016
[28] Saifuzzaman, M.; Moon, N. N.; Nur, F. N. IoT Based Street Lighting And Traffic Management System. 2017 Ieee Region 10 Humanitarian Technology Conference (R10-Htc). Anais... In: 2017 Ieee Region 10 Humanitarian Technology Conference (R10-Htc). Dez. 2017
[29] Corno, F.; Russis, L. D.; Sáenz, J. P. On The Design Of An Energy And User Aware Study Room. 2017 Ieee Pes Innovative Smart Grid Technologies Conference Europe (Isgt-Europe). Anais... In: 2017 Ieee Pes Innovative Smart Grid Technologies Conference Europe (Isgt-Europe). Set. 2017
[30] Rekha, P.; Ramesh, V.; Rangan, V. P.; Nibi, K. V. High Yield Groundnut Agronomy: An IoT Based Precision Farming Framework. 2017 Ieee Global Humanitarian Technology Conference (Ghtc). Anais... In: 2017 Ieee Global Humanitarian Technology Conference (Ghtc). Out. 2017
[31] Chen, M.; Yang, J.; Zhu, X.; Wang, X.; Liu, M.; Song, J. Smart Home 2.0: Innovative Smart Home System Powered By Botanical Iot And Emotion Detection. Mobile Networks And Applications, V. 22, N. 6, P. 1159–1169, 1 Dez. 2017.
[32] Oliveira, E. Da L.; Alfaia, R. D.; Souto, A. V. F.; Silva, M. S.; Francês, C. R. L. Smartcom: Smart Consumption Management Architecture For Providing A User-Friendly Smart Home Based On Metering And Computational Intelligence. Journal Of Microwaves, Optoelectronics And Electromagnetic Applications, V. 16, N. 3, P. 736–755, Set. 2017.
[33] Medina, B. E.; Manera, L. T. Retrofit Of Air Conditioning Systems Through An Wireless Sensor And Actuator Network: An IoT-Based Application For Smart Buildings. 2017 Ieee 14th International Conference On Networking, Sensing And Control (Icnsc). Anais... In: 2017 Ieee 14th International Conference On Networking, Sensing And Control (Icnsc). Maio 2017

[34] Shyam, G. K.; Manvi, S. S.; Bharti, P. Smart Waste Management Using Internet-Of-Things (IoT). 2017 2nd International Conference On Computing And Communications Technologies (Iccct). Anais... In: 2017 2nd International Conference On Computing And Communications Technologies (Iccct). Fev. 2017
[35] Paola, A. D; Ferraro, P.; Gaglio, S.; Re, G.; Morana, M.; Ortolani, M.; Peri, D. An Ambient Intelligence System For Assisted Living. 2017 Aeit International Annual Conference. Anais... In: 2017 Aeit International Annual Conference. Set. 2017
[36] Yong, B.; Xu, Z.; Wang, X.; Cheng, L.; Li, X.; Wu, X.; Zhou, Q. IoT-Based Intelligent Fitness System. Journal Of Parallel And Distributed Computing, V. 118, P. 14–21, 1 Ago. 2018.
[37] Espinilla, M.; Martínez, L.; Medina, L.; Nugent, C. The Experience Of Developing The Ujami Smart Lab. Ieee Access, V. 6, P. 34631–34642, 2018.
[38] Bisio, I.; Garibotto, C.; Grattarola, A.; Lavagetto, F.; Sciarrone, A. Exploiting Context-Aware Capabilities Over The Internet Of Things For Industry 4.0 Applications. Ieee Network, V. 32, N. 3, P. 101–107, Maio 2018.
[39] Medina, J. Espinilla, M.; Martínez, L.; Fernández, G. Intelligent Multi-Dose Medication Controller For Fever: From Wearable Devices To Remote Dispensers. Computers & Electrical Engineering, V. 65, P. 400–412, 1 Jan. 2018.
[40] Palacios, P.; Córdova, A. Approximation And Temperature Control System Via An Actuator And A Cloud: An Application Based On The IoT For Smart Houses. 2018 International Conference On E-democracy eGovernment (Icedeg). Anais... In: 2018 International Conference On E-democracy eGovernment (Icedeg). Apr. 2018

Additional References:
[41] Tricco C.; Antony, J.; Zarin, W.; Strifler, L.; Ghassemi, M.; Ivory, J.; Perrier, L.; Hutton, B.; Moher, D.; Straus, S. A Scoping Review Of Rapid Review Methods. Bmc Medicine, 2015.
[42] Cartaxo, B.; Pinto, G.; Soares, S. The Role Of Rapid Reviews In Supporting Decision -Making In Software Engineering Practice. Ease 2018.
[43] Cartaxo, B.; Pinto, G.; Soares, S. Evidence Briefings: Towards A Medium To Transfer Knowledge From Systematic Reviews To Practitioners. Esem, 2016.
[44] Bahill, T.; Botta, R.; Daniels, J, The Zachman Framework Populated With Baseball Models.
[45] Basili, V. R.;Caldeira, G.; Rombach, H. D. Goal Question Metric Paradigm, (1994).
[46] Liao, Y.; Deschamps, F.; Loures, E. F. R.; Ramos, L. F. P. Past, Present And Future Of Industry 4.0 - A Systematic Literature Review And Research Agenda Proposal. International Journal Of Production Research, Vol. 55, No. 12, 3609–3629, 2017
[47] Motta, R.; Marçal, K.; Travassos, G. H. Challenges In Engineering Contemporary Software Systems. Technical Report, 2018

# DEVELOPING IOT SOFTWARE SYSTEMS? TAKE INTERACTIVITY INTO ACCOUNT

## CONTEXT

IoT allows composing systems from *things* equipped with identifying, sensing, or actuation behaviors and processing capabilities that can communicate and cooperate to reach a goal. For so, different facets should be taken into account, such as smartness.

## INTERACTIVITY

It refers to the involvement of actors in the exchange of information with things and the degree to which it happens. Beyond the sociotechnical concerns surrounding the human actors (human-thing interaction), we also have concerns with other actors (thing-thing interaction).

## SUMMARY

The most important of them is the fact that there is no more unique system, but an enormous quantity of systems, components, things, and applications, which has to work together. For this reason, understanding, defining and dealing with interactivity is so important for IoT solutions.

## OBJECTIVE

This briefing aims to enlighten project planning and to provide guidance when dealing with Interactivity in IoT. The information presented in this briefing seeks to answer: "What taken into account when considering interactivity for IoT?".

## MOTIVATION

IoT offers challenges for their construction, with concerns that rather than just developing, we need to engineer software systems. This motivated us to a more comprehensive investigation to address these concerns and to provide evidence-based support for IoT engineering.

## ADDITIONAL INFORMATION

**Who are the briefing's clients?** Software developers and practitioners who want to make decisions about how to deal with *interactivity* in IoT, considering scientific evidence.

**Where do the findings come?** All findings of this briefing were extracted from scientific studies about *interactivity* through a Rapid Review[1]. The Technical Report[2] containing all the findings is available for further information.

**What is included in this briefing?** Technologies, challenges, and strategies to deal with *interactivity* in IoT projects.

**What are the Challenges and Opportunities?** IoT generally involve devices developed and implemented by a diverse range of vendors, using different languages, architectures, and technologies, without a pattern, causing a high level of heterogeneity. The worst case is heterogeneity in data, which affects the semantic (the meaning) and the syntax (the format) and harms interactivity because even if the systems can communicate, they cannot understand each other. Everyone who wants to propose a solution to CSS context must know that if the "conventional" systems are not made homogeneously, the level of heterogeneity increases a lot as well as the level of interaction and autonomy.

**Does interactivity represent a concern in the Engineering of Internet of Things Software Systems?** Yes, and interactivity challenges should be overcame to achieve interoperable, seamlessly and collaborative solutions.

## HIGHLIGHTS

- ***Heterogeneity*** *is an intrinsic characteristic of IoT.*
- *It can affect the semantic (the meaning) and the syntax (the format) and harms interactivity because even if the systems can communicate, they cannot understand each other. That is a great challenge in Interactivity.*

## WHAT
### is the the understanding of interactivity in CSS?

- Interactivity is characterized by the interaction involving things, systems, and humans where interaction is characterized by the ability to communicate, exchange information and control actions.
- It is important to **define actors' control and mechanisms** for the interaction methods (Voice, gesture, actions...).

## HOW
### do CSS projects deal with the operacionalization of interactivity?

- **To guarantee connectivity:** Zig-Bee, Bluetooth, Radio Frequency, RFID, 6LowPAN, WSN, WiFi, IPv6 and others.
- **To guarantee communication:** HTTP, XMPP, TCP, UDP, CoAP, MQTT and others.
- **To guarantee understanding:** JSON, XML, OWL, SSN Ontology, COCI and others.
- Real-world objects are virtualized and represented as Web Resources and accessed through Web Interfaces based on REST principles and Producer and Consumers methods.
- Alternative solutions should also be considering when dealing with human in the loop, to address human-thing interaction – such as voice and gestures commands.

## WHERE
### CSS projects locate the activities regarding interactivity?

- Most of the times, the activities regarding interactivity are located in a system's architecture, on frameworks, middleware, and platform.

## WHO
### do CSS projects allocate to deal with interactivity?

- **Designers, architects, developers, managers, and engineers** deal with interactivity in different phases of IoT projects.
- **Changing the scenario:** "Engineering is no more a set of vertical activities developed by different engineers but a collaborative process in which people and technology is completely involved in the engineering process".

## WHEN
### do the effects of time, and states of interactivity affect CSS projects?

- Activities regarding Interactivity are present all over a system or application **lifecycle**. Especially, they affect the design, development, integration, deployment, and operation phases.
- Interactivity is achieved by the exchange of information among "Things" and the orchestration of these operations occur during **runtime**.

## WHY
### do CSS projects implement interactivity?

- To **bridge the gap** between the massive heterogeneity present in IoT in order to create an interoperable systems, that can overcome different standards, protocols and technologies to perform more efficiently than isolated ones.
- **Interactivity** is one of the main characteristics of IoT projects, making new types of application possible (such as smart environments), facilitating everyday life, enhancing products competitivity, and sustainability.

[1] More on Rapid Reviews: https://dl.acm.org/citation.cfm?id=3210462
[2] Full Report: https://goo.gl/no7jtR

# 9 RAPID REVIEW ON INTERACTIVITY

# Rapid Reviews Protocol:
## Engineering of Contemporary Software Systems

**Vitor Maia, Rebeca Motta, Káthia de Oliveira, Guilherme Travassos**

# Interactivity

In the investigation regarding Internet of Things Software Systems (IoT), it has been observed that these modern software systems offer challenges for their construction since they are calling into question our traditional form of software development. Usually, they rely on different technologies and devices that can interact-capture-exchange information, act, and make decisions. It leads to concerns that, rather than just developing software, we need to engineer software systems embracing multidisciplinarity, integrating different areas. From our initial research, we analyzed the concerns related to this area. We categorized them into a set of facets - Connectivity, Things, Behavior, Smartness, Interactivity, Environment, and Security - representing such projects' multidisciplinarity, in the sense of finding a set of parts composing this engineering challenge.

Since these facets can bring additional perspectives to the software system project planning and management, acquiring evidence regarding such facets is of great importance to provide an evidence-based framework to support software engineers' decision-making tasks. Therefore, the following question should be answered:

***"Does interactivity represent a concern in the engineering of***

***Internet of things software systems?"***

This Rapid Review (RR) aims to analyze interactivity to characterize it in the IoT field, regarding *what, how, where, when, and why* it is used in the context of IoT projects, verifying the existence of published studies. The 5W1H aims to give the observational perspective on which information is required to the understanding and management of the facet in a system (what); to the software technologies (techniques, technologies, methods, and solutions) defining their operationalization (how); the activities location being geographically distributed or something external to the software system (where); the roles involved to deal with the facet development (who); the effects of time over the facet, describing its transformations and states (when); and to translate the motivation, goals, and strategies going to what is implemented in the facet (why), in respect of IoT projects.

## 9.1 Research Questions

- **RQ1:** What is the understanding and management of interactivity in IoT projects?

- **RQ2:** How do IoT projects deal with software technologies (techniques, technologies, methods, and solutions) and their operationalization regarding interactivity?
- **RQ3:** Where do IoT projects locate the activities regarding interactivity?
- **RQ4:** Whom do IoT projects allocate to deal with interactivity?
- **RQ5:** When do the effects of time, transformations, and states of interactivity affect IoT projects?
- **RQ6:** Why do IoT projects motivate the implementation of interactivity?

## 9.2 Search Strategy

The Scopus search engine and the following search string support this RR, built using PICOC with five levels of filtering:

**P**opulation - Internet of Things software systems
Synonymous:
"ambient intelligence" OR "assisted living" OR "multiagent systems" OR "systems of systems" OR "internet of things" OR "Cyber-Physical Systems" OR "Industry 4" OR "fourth industrial revolution" OR "web of things" OR "Internet of Everything" OR "contemporary software systems" OR "smart manufacturing" OR digitalization OR digitization OR "digital transformation" OR "smart cit*" OR "smart building" OR "smart health" OR "smart environment"

**I**ntervention - ("human thing interaction" OR "thing thing interaction" OR "user interaction" OR interactivity)

**C**omparison – no

**O**utcome - Synonymous:
understanding OR management OR technique OR “technolog*” OR method OR location OR place OR setting OR actor OR role OR team OR time OR transformation OR state OR reason OR motivation OR aim OR objective

**C**ontext – (engineering or development or project or planning OR management OR building OR construction OR maintenance)

Limited to articles from 2013 to 2018

Limited to Computer Science and Engineering

LIMIT-TO (SUBJAREA, "COMP" ) OR LIMIT-TO (SUBJAREA, "ENGI" ) ) AND ( LIMIT-TO (PUBYEAR, 2018 ) OR LIMIT-TO (PUBYEAR, 2017 ) OR LIMIT-TO (PUBYEAR, 2016 ) OR LIMIT-TO (PUBYEAR, 2015)

| |
|---|
| TITLE-ABS-KEY ( ( “ambient intelligence” OR “assisted living” OR “multiagent systems” OR “systems of systems” OR “internet of things” OR “Cyber Physical Systems” OR “Industry 4” OR “fourth industrial revolution” OR “web of things” OR “Internet of Everything” OR “contemporary software systems” OR “smart manufacturing” OR digitalization OR digitization OR “digital transformation” OR “smart cit*” OR “smart building” OR “smart health” OR “smart environment” |
| ) AND ( “human thing interaction” OR “thing thing interaction” OR “user interaction” OR Interactivity ) AND ( understanding OR management OR technique OR “technolog*” OR method OR location OR place OR setting OR actor OR role OR team OR time OR transformation OR state OR reason OR motivation OR aim OR objective ) AND ( engineering OR development OR project OR planning OR management OR building OR construction OR maintenance ) ) AND ( LIMIT-TO ( PUBYEAR , 2019 ) OR LIMIT-TO ( PUBYEAR , 2018 ) OR LIMIT-TO ( PUBYEAR , 2017 ) OR LIMIT-TO ( PUBYEAR , 2016 ) OR LIMIT-TO ( PUBYEAR , 2015 ) ) AND ( LIMIT-TO ( SUBJAREA , “COMP” ) OR LIMIT-TO ( SUBJAREA , “ENGI” ) ) ) |

## 9.3 Selection procedure

One researcher performs the following selection procedure:

1. Run the search string;
2. Apply the inclusion criteria based on the paper Title;
3. Apply the inclusion criteria based on the paper Abstract;
4. Apply the inclusion criteria based on the paper Full Text, and;

After finishing the selection from Scopus, use the included papers set to:

1. Execute snowballing backward (one level) and forward:

a. Apply the inclusion criteria based on the paper Title;

b. Apply the inclusion criteria based on the paper Abstract;

c. Apply the inclusion criteria based on the paper Full Text.

The JabRef Tool must be used to manage and support the selection procedure.

## 9.4 Inclusion criteria

1. The paper must be in the context of **software engineering**; and
2. The paper must be in the context of the **Internet of Things software systems**; and
3. The paper must report a **primary or a secondary study**; and
4. The paper must report an **evidence-based study** grounded in empirical methods (e.g., interviews, surveys, case studies, formal experiment, etc.); and
5. The paper must provide data to **answering** at least one of the RR **research questions**.
6. The paper must be written in the **English language**.

## 9.5 Extraction procedure

The extraction procedure is performed by one researcher, using the following form:

| <paper_id>:<paper_reference> | |
|---|---|
| Abstract | <Abstract> |
| Description | <A brief description of the study objectives and personal understanding> |
| Study type | <Identify the type of study reported by paper (e.g., survey, formal experiment)> |
| RQ1: WHAT information required to understand and manage the << facet>> in IoT | - < A1_1><br>- < A1_2><br>- ... |
| RQ2: HOW | - < A2_1> |
| software technologies (techniques, technologies, methods and solutions) and their operationalization | - < A2_2><br>- ... |
| RQ3: WHERE<br>activities location or something external to the IoT | - < A3_1><br>- < A3_2><br>- ... |
| RQ4: WHO<br>roles involved to deal with the << facet>> development in IoT | -< A4_1><br>-<A4_2><br>- … |
| RQ5: WHEN<br>effects of time over << facet>>, describing its transformations and states in IoT | - < A5_1><br>- < A5_2><br>... |
| RQ6: WHY<br>motivation, goals, and strategies regarding <<facet>> in IoT | - < A6_1><br>- < A6_2><br>... |

## 9.6 Synthesis Procedure

In this RR, the extraction form provides a synthesized way to represent extracted data. Thus, we do not perform any synthesis procedure.

However, the synthesis is usually performed through a narrative summary or a Thematic Analysis when the number of selected papers is not high.

## 9.7 Report

An Evidence Briefing [2] reports the findings to ease the communication with practitioners. It was presented as the cover for this chapter.

## 9.8 Results

To obtain a better understanding of Interactivity within IoT projects, we performed the previous sections' procedure. Throughout the revision process, some papers were removed based on the selection criteria defined. Firstly, all the findings containing information about conferences, workshops, and books were removed, totaling 25 exclusions. Then, titles were read to filter researches possibly not related to the subject of Interactivity, totaling 1487 exclusions, and leaving a base with 538 papers. All the remaining papers' abstracts have been read and interpreted to know if they contain the general idea of the research and decide if they should be included or excluded, resulting in 1378 exclusions, leaving a base with 109 papers. Besides, 12 papers were not found, leaving a base with 97 papers. All the remaining papers were fully read, and 50 of them did not fulfill criterion 5 of section 4.1, “The paper must provide data to answering at least one of the research questions,” leaving a base with 47 papers. Then we performed forward and backward snowballing procedures, which resulted in 40 possible papers to be included. Following the selection criteria, 8 of the were added, leaving a base with 55 papers.

Two reviewers performed an analysis of these 55 paper extractions, and 16 papers were removed due to their smaller coverage of the review topic referring to the 5W1H answers. At last, 39 papers compose the final set.

## Execution

| Activity | Execution date | Result | Number of papers |
|---|---|---|---|
| First execution | 08/05/2020 | 2050 documents added | 2050 |
| Remove conferences/workshops/books | 08/05/2020 | 25 documents withdrawn | 2025 |
| Included by Title analysis | 09-10/05/2020 | 1487 documents withdrawn | 538 |
| Included by Abstract analysis | 13/05/2020 – 04/06/2020 | 429 documents withdrawn | 109 |
| Papers not found | 04/06/2020 | 12 papers not found | 97 |
| Articles for reading | 05/06/2020 | - | 97 |
| Removed after a full reading | 05/06 - 04/07/2020 | 50 papers withdrawn | 47 |
| Snowballing | 09 - 12/07/2020 | 40 papers to read | - |
| Snowballing after reading | 09 - 12/07/2020 | 8 papers added | 55 |
| Total included | 08/2020 | - | 55 |
| Papers extracted | 08/2020 | 16 removed | 39 final set |

## Final Set

| Reference | Author | Title | Year | Source |
|---|---|---|---|---|
| 1 | Xian Wang, Ana M. Bernardos, Juan A. Besada, Eduardo Metola, José R. Casar | A gesture-based method for natural interaction in smart spaces | 2015 | Regular search |
| 2 | Borja Gamecho, Raúl Miñón, Amaia Aizpurua, Idoia Cearreta, Myriam Arrue, Nestor Garay-Vitoria, Julio Abascal | Automatic Generation of Tailored Accessible User Interfaces for Ubiquitous Services | 2015 | Regular search |
| 3 | Börge Kordts, Bashar Altakrouri, Andreas Schrader | Capturing and Analysing Movement using Depth Sensors and Labanotation | 2015 | Regular search |
| 4 | Byron Lowens, Vivian Motti, Kelly Caine | Design Recommendations to Improve User Interaction with Wrist-Worn Devices | 2015 | Regular search |
| 5 | Fulvio Corno, Elena Guercio, Luigi De Russis, Eleonora Gargiulo | Designing for user confidence in intelligent environments | 2015 | Regular search |
| 6 | Ian Oakley, Doyoung Lee, Islam Md Rasel, Augusto Esteves | Beats: Tapping Gestures for Smart Watches | 2015 | Snowballing |
| 7 | Michael Vacher, Sybille Caffiau, François Portet, Brigitte Meillon, Camille Roux, Elena Elias, Benjamin Lecouteux, Pedro Chahuara | Evaluation of a Context-Aware Voice Interface for Ambient Assisted Living: Qualitative User Study vs. Quantitative System Evaluation | 2015 | Snowballing |
| 8 | Silvia Lovato, Anne Marie Piper | "Siri, is this you?": Understanding young children's interactions with voice input systems | 2015 | Snowballing |
| 9 | Paul Whittington, Huseyin Dogan | A SmartDisability Framework: enhancing user interaction | 2016 | Regular search |
| 10 | Andrej Grguric, Alejandro M. Medrano Gil, Darko Huljenic, Zeljka Car, Vedran Podobnik | A survey on user interaction mechanisms for enhanced living environments | 2016 | Regular search |
| 11 | Mohamad A. Eid, Hussein Al Osman | Affective Haptics - Current Research and Future Directions | 2016 | Regular search |
| 12 | Alessandro Di Nuovo, Ning Wang, Frank Broz, Tony Belpaeme, Ray Jones, Angelo Cangelosi | Experimental evaluation of a multi-modal user interface for a robotic service | 2016 | Regular search |
| 13 | Ali Asghar Nazari Shirehjini, Azin Semsar | Human Interaction with IoT-based smart environments | 2016 | Regular search |
| 14 | Christopher C. Mayer, Gottfried Zimmermann, Andrej Grguric, Jan Alexandersson, Miroslav Sili, Christophe Strobbe | A comparative study of systems for the design of flexible user interfaces | 2016 | Regular search |
| 15 | Seong M. Kim, Eui S. Jung, Jaekyu Park | Effective quality factors of multi-modal interaction in simple and complex tasks of using a smart television | 2016 | Regular search |
| 16 | Matteo Zallio, Niccolò Casiddu | Lifelong Housing Design: User Feedback Evaluation of smart objects and accessible houses for healthy aging | 2016 | Regular search |
| 17 | André Mewes, Bennet Hensen, Frank Wacker, Christian Hansen | Touchless interaction with the software in interventional radiology and surgery: a systematic literature review | 2016 | Snowballing |
| 18 | Micael Carreira, Karine Lan Hing Ting, Petra Csobanka, Daniel Gonçalves | Evaluation of in-air hand gestures interaction for older people | 2016 | Snowballing |
| 19 | Shaikh Shawon Arefin Shimon, Courtney Lutton, Zichun Xu, Sarah Morrison-Smith, Christina Boucher, Jaime Ruiz | Exploring Non-touchscreen Gestures for Smartwatches | 2016 | Snowballing |
| 20 | Cheng Zhang, Abdelkareem Bedri, Gabriel Reyes, Bailey Bercik Omer T. Inan, Thad E. Starner, Gregory D. Abowd | TapSkin: Recognizing On-Skin Input for Smartwatches | 2016 | Snowballing |
| 21 | Hyoseok Yoon, Se-Ho Park, Kyung-Taek Lee, Jung Wook Park, Anind K. Dey, SeungJun Kim | A case study on iteratively assessing and enhancing wearable user interface prototypes | 2017 | Regular search |
| 22 | David Ledo, Fraser Anderson, Ryan Schmidt, Lora Oehlberg, Saul Greenberg, Tovi Grossman | Pineal - Bringing passive objects to life with embedded mobile devices | 2017 | Regular search |
| 23 | Rossana M.C. Andrade, Rainara M. Carvalho, Italo Linhares de Araújo, Káthia M. Oliveira, Marcio E.F. Maia | What changes from ubiquitous computing to internet of things in interaction evaluation | 2017 | Regular search |
| 24 | Aida Mostafazadeh Davani, Ali Asghar Nazari Shirehjini, Sara Daraei | Towards interacting with smarter systems | 2017 | Snowballing |
| 25 | Rainara Maia Carvalho, Rossana M. C. Andrade, Káthia Marçal de Oliveira | AQUArIUM - A suite of software measures for HCI quality evaluation of ubiquitous mobile applications | 2017 | Regular search |
| 26 | Thomas van de Werff, Karin Niemantsverdriet, Harm van Essen, Berry Eggen | Evaluating interface characteristics for shared lighting systems in the office environment | 2017 | Regular search |
| 27 | F. Marulli, L. Vallifuoco | Internet of Things for Driving Human-Like Interactions: A Case Study for Cultural Smart Environment | 2017 | Regular search |
| 28 | Amr Alanwar, Moustafa Alzantot, Bo-Jhang Ho, Paul Martin, Mani Srivastava | SeleCon: Scalable IoT Device Selection and Control Using Hand Gestures | 2017 | Regular search |
| 29 | Guiying Du, Auriol Degbelo, Christian Kray, Marco Painho | Gestural interaction with 3D objects shown on public displays - An elicitation study | 2018 | Regular search |
| 30 | Stephanie Van Hove, Jolien De Letter, Olivia De Ruyck, Peter Conradie, Anissa All, Jelle Saldien, and Lieven De Marez | Human-Computer Interaction to Human-Computer-Context Interaction: Towards a conceptual framework for conducting user studies for shifting interfaces | 2018 | Regular search |
| 31 | Ricardo Rosales, Manuel Castañón-Puga, Felipe Lara-Rosano, Josue Miguel Flores-Parra, Richard Evans, Nora Osuna-Millan, Carelia Gaxiola-Pacheco | Modelling the interaction levels in HCI using an intelligent hybrid system with interactive agents A case study of an interactive museum exhibition module in Mexico | 2018 | Regular search |
| 32 | Vijay Rajanna, Tracy Hammond | A gaze gesture-based paradigm for situational impairments, accessibility, and rich interactions | 2018 | Regular search |
| 33 | François Portet, Sybille Caffiau, Fabien Ringeval, Michel Vacher, Nicolas Bonnefond, Solange Rossato, Benjamin Lecouteux, Thierry Desot | Context-Aware Voice-based Interaction in Smart Home-VocADom@A4H Corpus Collection and Empirical Assessment of its Usefulness | 2019 | Regular search |
| 34 | Björn Bittner, Ilhan Aslan, Chi Tai Dang, Elisabeth André | Of smarthomes, IoT plants, and implicit interaction design | 2019 | Regular search |
| 35 | George Margetis, Stavroula Ntoa, Margherita Antona, Constantine Stephanidis | Augmenting natural interaction with physical paper in ambient intelligence environments | 2019 | Regular search |
| 36 | David Sheppard, Nick Felker, John Schmalzel | Development of Voice Commands in Digital Signage for Improved Indoor Navigation Using Google Assistant SDK | 2019 | Regular search |
| 37 | Markus Rittenbruch, Jared Donovan | Direct end-user interaction with and through IoT devices | 2019 | Regular search |
| 38 | Tamino Huxohl, Marian Pohling, Birte Carlmeyer | Interaction guidelines for personal voice assistants in smart homes | 2019 | Regular search |
| 39 | Jaime Zabalza, Zixiang Fei, Cuebong Wong, Yijun Yan, Carmelo Mineo, Erfu Yang, Tony Rodden, Jorn Mehnen, Quang-Cuong Pham, Jinchang Ren | Smart Sensing and Adaptive Reasoning for Enabling Industrial Robots with Interactive Human-Robot Capabilities in Dynamic Environments—A Case Study | 2019 | Regular search |

## 9.9 Summary of the articles

Wang et al. [#1] describe a grammar for **gesture-based interaction and controls**, validated with users who repeated gestures in Kinect (a device composed of an RGB camera, a low-cost depth sensor, and a multi-array microphone) and on a mobile phone. Kinect enables full-body 3D motion capture, facial recognition, and voice recognition. The study focuses on exploring gesture-based interaction in smart spaces.

Gamecho et al. [#2] proposes and describes an architecture for **disabled people's interaction** with ubiquitous systems and implements two services as part of a case study. It describes interaction with ubiquitous services as cognitive (e.g., attention, concentration) or physical (e.g., precision, strength). The project’s main goal was to grant access to interactive machines to users with disabilities provided with their own mobile devices.

Kordts et al. [#3] describe an architecture for detecting and interpreting **movement interactions with cameras**. The architecture is validated by an experiment with participants who performed gestures in a realistic setup. The study intends to allow for recording movement scores and subscribing to events from live motion data streams. The task lists several types of motion gestures, which are explored in the experiment.

Lowens et al. [ #4] search the Internet for users' opinions regarding 11 **WWD (Wrist-Worn Devices)**. The opinions were analyzed and categorized. WWDs are mainly employed as fitness trackers and smartwatches. They have been successfully applied to support less conventional activities as well, including gesture recognition and authentication. The study collected feedback about existing wearable bands and analyzed the most critical issues currently faced in the interaction with such devices.

Corno et al. [#5 ] briefly describe a literature review on intelligent environments and, based on the results, define a set of guidelines in the field of User Experience for Reliable Intelligent Environments. The guidelines are the main contribution of the study, which also proposes to use the concept of user confidence to offer **user–system interactions** that create the perception of a reliable and understandable intelligent system.

Oakley et al. [#6] propose the idea and **design of beating gestures** that they define as rapid sequences of screen touches and releases made by the index and middle fingers of the user’s hand. The interaction options are more comprehensive with this simple mechanism. It is possible to increase the richness of available inputs via the touchscreen of tiny devices such as smartwatches.

In their paper, Vacher et al. [#7] experimented with seniors and people with **visual impairment interacting** in a multi-room smart home using voice commands. The system uses several mono-microphones set in the ceiling of a smart home equipped with home automation devices and networks. It employs decoding and voice commands, which is then analyzed by selecting adequate action. Hands-free interaction is ensured by constant keyword detection, so the user can command the environment without wearing a specific device for physical interaction.

Lovato et al. [#8] explore how young children, who do not know how to read yet, **interact with voice input systems**. The study is evaluated by an online survey and a content analysis of Youtube videos of children using Siri. The study highlights various opportunities and challenges voice input systems present for children and parents due to children’s speech differing significantly from that of adults, both in sound characteristics like pitch and in the patterns of stress and intonation or prosody.

The SmartDisability Framework is proposed by Whittington et al. [#9]. It considers mappings between **disability type, Range of Movement, and Interaction mediums** to produce technology and task recommendations to enhance user interaction. The framework was conceived through the knowledge obtained from state-of-the-art literature reviews of disability classification, Range of Movement, interaction mediums, ‘off-the-shelf’ technologies and tasks, and evolved by requirements elicitation results and a usability evaluation involving a simulation. The interaction medium was touch and head-based methods measured using the System Usability Scale (SUS) and NASA Task Load

Index (NASA TLX). As a result, it demonstrated that fingers were a more suitable interaction method, as head tracking required a full range of neck movement.

Grguric et al. [#10] explore designing user interfaces (UIs) for elderly and disabled users. Their focus is on **Enhanced Living Environment** (ELE) and performs a survey of state-of-the-art user interaction in these environments. It shows how a challenge of user interaction is approached in different Ambient Assisted Living platforms. The results of the survey assessments were classified according to Effectiveness, Efficiency, User Acceptance, Learnability, Human Errors, Satisfaction, and Sales.

As an emerging field, affective haptics focuses on analyzing, designing, and evaluating systems that can capture, process, or display emotions through the sense of touch. In their paper, Eid et al. [#11] present an overview of the field and how it integrates ideas from **affective computing, haptic technology, and user experience**. They argue about how a haptic medium can communicate affective information and present various applications in the area ranging from interhuman social interaction systems to human-robot interaction applications. From their conclusion, we can highlight that a haptic stimulation can be successfully used to achieve a higher level of emotional immersion during media consumption or emotional. The interpretation of the haptic stimulation by human beings is highly contextual.

Di Nuovo et al. [#12] present the user evaluation of an MMUI (Multi-Modal User Interface), which was designed to enhance the user experience in terms of usability and increase the acceptability of assistive robot systems by elderly users. The quantitative and qualitative analyses performed showed a positive evaluation by users and provided insights into how **multi-modal means of interaction (visual and speech**) can help make elderly-robot interaction more natural.

Shirehjini et al. [#13] describe the conceptualization, design, implementation, and evaluation of a **3D-based user interface for accessing IoT-based Smart Environments** (IoT-SE). Their model addresses the cognitive overload associated with manual device selection, loss of user control, and other issues. They argue that the 3D-based user interface makes it easier for users to observe smart environments, find and select devices for interaction purposes. The interaction model provides integrated control of the connected appliances and multimedia artifacts (such as files and videos), allowing more users flexibly. A subjective satisfaction evaluation of the two-handed mobile multi-touch 3Dbased environment controller using the ISONORM 9241/110 questionnaire was performed.

Mayer et al. [#14] compare three systems in terms of architecture and features, focusing the analysis on usability factors. It describes **Interactivity in terms of user interaction and user-friendliness** when users with different characteristics use a system. It also provides a generic framework for designing and deploying user interfaces tailored to particular user groups.

Kim et al. [#15] investigate **multi-modal interaction** with smart TVs. First, a literature review is executed with questions regarding previous user experience with multi-modal interaction, types of integration patterns affecting the interaction, and the selection of input modes. An experiment was designed with a comparison between two groups interacting with smart TVs in different patterns.

Zallio et al. [#16] collect data regarding **house design and interaction** from users in "worldwide," European and Italian scenarios, focusing on aging people. The study questions if houses and appliances have the potential to improve both autonomy and people's quality of life. It explores design characteristics such as fewer stairs, minimum doors, handles in toilets, among others.

Hensen et al. [#17] examine the current state of research of systems that focus on **touchless human-computer interaction** in operating rooms and interventional radiology suites. They present new ways of interaction with medical software under sterile conditions. The interactions described in this study are camera-based, eye tracking, inertial sensors, and voice commands.

Carreira et al. [#18] evaluate **older people's interaction with Kinect**, using in-air gestures, in which users move their bodies to play video games. In this case, the users' body acts as a video game controller. If senior users' interest in technology could be captivated, this would contribute to fight isolation and exclusion and allow older people to be more productive, independent, and have a more social and fulfilling life.

Shimon et al. [#19] discuss some **touch interaction** limitations on smartwatches due to their limited input space. To close this gap, they explore how to design non-touchscreen gestures to extend smartwatches' input space. From an elicitation study eliciting gestures for 31 smartwatch tasks, they mapped gestures and commands and defined a user gesture set. From that, they proposed taxonomy and heuristics for designing smartwatches.

Zhang et al. [#20] present TapSkin, an interaction technique that recognizes up to 11 **distinct tap gestures** on the skin around the watch using only the inertial sensors and microphone a commodity smartwatch. This research opens the discussion for introducing any further on-body instrumentation. They evaluated it with 12 participants that the system can provide classification accuracy from 90.69% to 97.32% in three gesture families – number pad, d-pad, and corner taps.

Yoon et al. [#21] focused **on wearables' Interactivity**, with the main contributions of design and recommendations of advanced wearable UI prototypes by proposing a framework for testing and evaluating wearable devices and executing a case study.

Pineal, a design tool for end-users, is proposed by Ledo et al. [#22]. With this tool, users can (1) modify 3D models to include a smartwatch or phone as its heart; (2) specify high-level interactive behaviors through visual programming; and (3) have the phone or watch act out such programmed behaviors. The goal is to enable users to create interactive, self-contained, smart objects and prototype smart objects with little knowledge of circuitry, software development, or 3D modeling. Based on user-defined interactive behaviors, the system automatically modifies an imported 3D model to fit the mobile device inside and expose the necessary input and outputs.

Andrade et al. [#23] discuss IoT and UbiComp by highlighting frameworks, middlewares, and other development artifacts from the UbiComp community and can be used for IoT applications. Part of the discussion is related to interaction evaluation. They argue that they can be more complicated in IoT than in UbiComp systems once we have two different perspectives: **Human-Thing and Thing-Thing interactions**. Three observation studies were performed: GreatTour, GreatMute, and GreatPrint, and the results are discussed in the light of particular quality characteristics (Context-awareness, Mobility, Transparency, Attention, and Calmness).

Davani et al. [#24] develop **an interaction model for a meta-UI** for **smarter systems**. The model is implemented and evaluated with a proof-of-concept and a user evaluation. It explores the interaction between a human (smart player) and a meta-system. Its main contribution is the analysis, design, implementation, and evaluation of a 3D based meta user interface for ambient intelligence.

Carvalho et al. [#25] identified, in previous studies, a set of quality characteristics for ubiquitous mobile applications that impact the **user's interaction quality**. This paper proposes measures for a subset of five of these quality characteristics (context-awareness, transparency, attention, calmness, and mobility), given that all the others were classified as generic for any system. The study proposes a suite of software measures to evaluate the quality of interaction in ubiquitous applications.

Van der Werff et al. [#26] discuss the **interaction of users with lighting systems**, investigate how different user interface characteristics influence the use of lighting systems, and evaluate this interaction with qualitative studies, in which three interfaces for a shared lighting system are evaluated by 17 people working in an open-plan office.

Marulli et al. [#27] propose a case study describing an IoT infrastructure supporting HCI models for art recreation, supporting holographic projections. The study describes the **interaction with the holographic interface**, which considers an ultra-high-definition resolution human and triggers events based on sensors deployed in the room.

Alanwar et al. [#28] propose a **pointing framework solution to interact with pointing devices** and implement hardware solutions to evaluate it. The study's main contributions are designing a smartwatch hardware prototype equipped with UWB radio and inertial sensors and developing machine-learning models related to hand gesture recognition.

How do users envision **interacting with public displays** (using mobile phones and gestures) to scrutinize urban planning material, i.e., 3D objects? This research question guided Du et al. [#29] to provide insights into how public display designers can engage a broader range of citizens to participate in urban planning projects more actively. They performed an elicitation study for determining the gestures made when interacting with 3D objects – focusing on hand and phone gestures. They identified that two gesture sets that participants produced. These sets were assessed regarding their consistency and user acceptance, curating a result that can help designers select suitable interaction modalities for citizen consultations via public displays.

In their Human-Computer-Context Interaction Framework, Van Hove et al. [#30] defines five interaction levels to be considered for a new product development process when regarding user experience. Their levels are defined as **user-object, user-user, user-content, user-platform, and user-context interactions**. For the authors, the interaction level is related to the type of user interaction that can be engaged with the product. This preliminary work is proposed for HCI researchers, designers, and developers to achieve a more holistic development approach.

Rosales et al. [ #31] propose a model **representing interaction levels using an intelligent hybrid system approach with agents**. For the authors, the interaction level is related to the interaction complexity. The higher the level, the more complex is the interaction description and implementation. The agents represent a high-level abstraction of the system, where communication occurs between the user, the system, and the environment. The proposal was evaluated, qualitatively and subjectively, in a museum exhibition module, bearing in mind uncertain results of exchanges of messages.

Rajanna et al. [#32] present an overview of a study on **gaze-assisted interaction** with systems, using the eyes' movements. The study develops solutions for situational impairments, accessibility, and for implementing a rich interaction paradigm.

Portet et al. [#33] summarize the effort developed in the VOCADOM project to collect a corpus of **users' interaction in a smart home** to be made available to the community. The VocADom@A4H corpus collection considered typical tasks undertaken as part of Smart Home systems, such as the automatic location of dwellers, Human Activity Recognition, Voice Activity Detection, speaker identification, Automatic Speech Recognition, Natural Language Understanding, context-aware decision, among others.

Bittner et al. [#34] conducted a series of design inquiries and a user study in the field with 24 participants, addressing two research questions. How can we use IoT technology to support inhabitants properly watering their plants without replacing the benefits inhabitants may have from interactions with their plants? Would inhabitants prefer related future **interactive technology solutions based on Augmented Reality** (AR) or purely embedded/physical technology in their homes? The study results highlight shortcomings of today's smartphone mediated augmented reality compared to physical interface alternatives, considering measurements of perceived attractiveness and expected effects on determinants of wellbeing, and discusses the potentials of combining both modalities for future solutions.

Margetis et al. [#35] propose a printed matter augmentation framework, exemplify it in three scenarios and validate it with an experiment using a real implemented system, given some hypothesis: H1. The system is easy to use with minimum guidance, H2. Touch-based interaction with the augmented paper is natural, H3. The overall user experience is positive, H4. The system will be more comfortable to use and more appealing to younger users, H5. Experienced computer users will find the system more attractive and more comfortable to use. As a result, "all the evaluations have highlighted that interaction with digitally augmented physical paper is useful, usable, and well-accepted."

Sheppard et al. [#36] propose an architecture for digital signage in buildings like malls. Digital signs are television-like devices that can display advertisements, maps, menus, and much more. It allows users to find information faster than a standard touchscreen-only modality. Therefore, their focus is on voice commands, so the user of a device can find what they're looking for with a single voice command instead of either scanning a static screen or navigating an unfamiliar user interface.

Rittenbruch et al. [#37] design and evaluate an **IoT device that interacts** with the ambient and gives feedback. The direct end-user evaluates it. Such devices can mediate the interpretation of sensed data by end-users and help collect crowd-sourced data directly related to sensed data. The results show how IoT devices can support end-user interaction by combining ambient and tangible interaction approaches.

Huxohl et al. [#38] describe an open online survey about **smart home interaction** and propose some resulting design guidelines for further developing PVAs. These guidelines are Authentication & Authorization, Activity-Based Interaction, Situated Dialogue, and Explainability & Transparency. The study explores voice assistance interaction.

Zabalza et al. [#39] present a robotic manipulator with vision and collision avoidance. The **smooth interaction** between this robot and the dynamic environment is evaluated in a case study. The movement and reaction change through independent thinking and reasoning with the reaction times below the average human reaction time. It demonstrates the effectiveness of **human-robot and robot-robot interactions** through sensing techniques, efficient planning algorithms, and systematic designs.

## 9.10 Tracking matrix

| Ref | Paper | WHAT | HOW | WHERE | WHO | WHEN | WHY |
|---|---|---|---|---|---|---|---|
| 1 | A gesture-based method for natural interaction in smart spaces | X | X | X | | | X |
| 2 | Automatic Generation of Tailored Accessible User Interfaces for Ubiquitous Services | X | X | X | X | | X |
| 3 | Capturing and Analysing Movement using Depth Sensors and Labanotation | X | X | | | | X |
| 4 | Design Recommendations to Improve the User Interaction with Wrist-Worn Devices | X | X | | | | X |
| 5 | Designing for user confidence in intelligent environments | X | X | X | X | X | X |
| 6 | Beats: Tapping Gestures for Smart Watches | X | X | | X | | X |
| 7 | Evaluation of a Context-Aware Voice Interface for Ambient Assisted Living: Qualitative User Study vs. Quantitative System Evaluation | X | X | X | X | | X |
| 8 | “Siri, is this you?”: Understanding young children’s interactions with voice input systems | X | X | X | X | X | X |
| 9 | A SmartDisability Framework: enhancing user interaction | X | X | | X | | |
| 10 | A survey on user interaction mechanisms for enhanced living environments | X | | | | | X |
| 11 | Affective Haptics - Current Research and Future Directions | X | X | | | | X |
| 12 | Experimental evaluation of a multi-modal user interface for a robotic service | X | X | X | X | | |
| 13 | Human Interaction with IoT-based smart environments | X | X | X | | | X |
| 14 | A comparative study of systems for the design of flexible user interfaces | | X | X | X | | X |
| 15 | Effective quality factors of multi-modal interaction in simple and complex tasks of using a smart television | X | X | X | | | X |
| 16 | Lifelong Housing Design: User Feedback Evaluation of smart objects and accessible houses for healthy aging | X | X | X | X | | X |
| 17 | Touchless interaction with the software in interventional radiology and surgery: a systematic literature review | X | X | X | X | X | X |
| 18 | Evaluation of in-air hand gestures interaction for older people | X | X | X | X | | X |
| 19 | Exploring Non-touchscreen Gestures for Smartwatches | X | | X | X | | X |
| 20 | TapSkin: Recognizing On-Skin Input for Smartwatches | X | X | | X | | X |
| 21 | A case study on iteratively assessing and enhancing wearable user interface prototypes | X | | | | X | |
| 22 | Pineal - Bringing passive objects to life with embedded mobile devices | X | X | | | | |
| 23 | What changes from ubiquitous computing to the Internet of things in interaction evaluation | X | X | | | | |
| 24 | Towards interacting with smarter systems | X | X | X | X | | X |
| 25 | AQUArIUM - A suite of software measures for HCI | X | X | | | | X |
| | quality evaluation of ubiquitous mobile applications | | | | | | |
| 26 | Evaluating interface characteristics for shared lighting systems in the office environment | X | X | X | | | X |
| 27 | Internet of Things for Driving Human-Like Interactions: A Case Study for Cultural Smart Environment | X | X | X | X | X | X |
| 28 | SeleCon: Scalable IoT Device Selection and Control Using Hand Gestures | X | X | | | | X |
| 29 | Gestural interaction with 3D objects shown on public displays - An elicitation study | X | X | | | | |
| 30 | Human-Computer Interaction to Human-Computer-Context Interaction: Towards a conceptual framework for conducting user studies for shifting interfaces | X | X | | | | |
| 31 | Modelling the interaction levels in HCI using an intelligent hybrid system with interactive agents A case study of an interactive museum exhibition module in Mexico | X | X | | | X | |
| 32 | A gaze gesture-based paradigm for situational impairments, accessibility, and rich interactions | X | X | | X | | X |
| 33 | Context-Aware Voice-based Interaction in Smart Home-VocADom@A4H Corpus Collection and Empirical Assessment of its Usefulness | X | X | X | X | X | |
| 34 | Of smarthomes, IoT plants, and implicit interaction design | X | X | X | X | | |
| 35 | Augmenting natural interaction with physical paper in ambient intelligence environments | X | X | X | X | | X |
| 36 | Development of Voice Commands in Digital Signage for Improved Indoor Navigation Using Google Assistant SDK | X | X | X | | | X |
| 37 | Direct end-user interaction with and through IoT devices | X | X | X | | | X |
| 38 | Interaction guidelines for personal voice assistants in smart homes | X | X | X | | | X |
| 39 | Smart Sensing and Adaptive Reasoning for Enabling Industrial Robots with Interactive Human-Robot Capabilities in Dynamic Environments—A Case Study | X | X | X | X | | X |

## 9.11 Summary of the Findings

**RQ1: WHAT is the understanding and management of Interactivity in IoT projects?**

Interactivity in IoT projects occurs in several ways due to the wide variety of actors, hardware, places, and purposes. Most of the identified interactions focus on users and describe how they interact with things or other systems, and then things among them. In our understanding, the following classification comprises generic types of interaction:

- **Human-Thing**: Interaction between users and things. [23]
    - **User-Object**: Interaction between a user and technology. [30]
    - **User-Content**: Interaction and perception of information. [30]
    - **User-Platform**: User's direct interaction (e.g. smartphone) or indirect (e.g. cloud) interaction with a platform. [30]
    - **User-Meta System**: Interaction with a system that governs multiple information systems to support various autonomous entities residing and acting within the same physical space. [24]
    - **User-Context**: The user’s interaction with the context, which cannot be treated as static information, results from the user’s internal and external characteristics. [30]
- **User-Thing-User**: Interaction between the user and another nearby or computer-mediated communicator, resulting in a user-user interaction intermediated by technology. [30]
    - **Direct end-user**: Interaction with other users through IoT devices, mediated by the interpretation of crowd-sourced sensed data. [37]
- **Thing-Thing**: Interaction between things themselves. [23]

The great majority of interaction scenarios described in literature explores **Human-Thing** interactions. Di Nuovo et al. [12] describes the robot’s automatic speech recognition, and Zabalza et al. [39] describe autonomous interactions of robots with the environment for industrial context. Concerning **Human-Thing** interactions, the prevailing type in literature, how users interact with things might change depending on the scenario, hardware, available sensor data, and context.

Bittner et al. [34] differ **implicit** and **explicit** interactions. Implicit interactions are those in which users can concentrate on their actual goals instead of computer systems, a definition somehow related to the characteristic quality Transparency presented by Carvalho et al. [23].

Wearables and smartwatches are recurring devices in literature [4] [19] [21] [22]. Some ways that users may interact with these devices are:

- **Capacitive**: Touch the device [22]
- **Acoustic**: Speech and audio recognition [22]
- **Motion**: Identification of context of use given gestures, with the assistance of accelerometers, gyroscopes, and magnetometers. [4] [21] [22]
- **Vision**: Interact by taking photos and reading barcodes. [22]

- **Outputs**: Visual (displays), audio (speakers), and vibration. [22]
- **Navigation and Actions**: Zoom, pan, set hour, view time. [19]
- **External Input**: a joystick module or a potentiometer as external input [21]
- **Other sensors**: NFC, Temperature, IR, Force. [22]

Many authors also refer to **Gestures** as standard input, a natural and effective communication method [28]. They may be read in different ways and depict different interpretations. They may be captured by **motion** and **vision** interactions, the same described for smartwatches, which are also present in other systems and devices. Several studies [1] [3] [8] explores **in-air gestures**, those that users accomplish in front of a camera or sensor, to use hands or even the full body as input. Kinect, for instance, is a device composed of an RGB camera, a depth sensor, and a multi-array microphone capable of motion capture, facial and voice recognition [1].

Other types of gesture interactions are described by Du et al. [29], who explore interaction with 3D objects using hand gestures; Alanwar et al. [28] explore interactions starting from pointing to a device to select it and then drawing hand gestures in the air to specify a control action. Margetis et al. [35] cite five types of gestures: (I) Action-time, in which the user pauses fingers for a short time. (II) Handshape, in which a hand pose is mapped to a specific action. (III) Drawing, in which users virtually draw simple shapes with a finger (e.g., a number 3). (IV) Based on Movement, in which a movement posture is mapped to a specific action. (V) Two Hands, which considers movements such as a pinch. Oakley et al. [6] add tapping to the list of gestures by detecting finger placement sequences and finger release timing from smartwatches.

Two studies mention **Affect Detection**, the ability of a computer to characterize aspects of the emotional status of its user, as a type of interaction, which may be collected from the user given text input mood, speech temperament, facial or gestural expressions of an avatar, among other methods [11]. Gamecho et al. [2] point that transmission and interpretation of emotions are part of human communication.

In the medical context, Mewes et al. [17] explore the touchless interaction with medical software under sterile conditions, to be used by surgeons.

In the tourism context, Rosales et al. [31] describe the interaction with interactive exhibitions, and Marulli & Vallifuoco [27] describes the interaction with holographic projections in exhibitions. In this case, the interaction must be attractive to engage visitors. Rosales et al. list parameters that might positively influence these interactions: presence, Interactivity, control, feedback, creativity, productivity, communication, and adaptation.

In the context of smart homes [8] [15] [16] [33] [36], additional types of interactions are described:

- **3D-based interaction**, in which objects in the environment are managed and manipulated by their position instead of their IP addresses. [13]
- **Voice-based interaction**, which permits to perform operations by distance [33] [36]. Lovato & Piper [8] study children’s voice interaction.
- **Multi-modal interaction** consists of manipulating multimedia content via two or more input modes in given usage contexts. [15]

Interaction in the context of disabilities is addressed in two studies. Whittington & Dogan [9] describes a system that couples a joystick to a wheelchair and turns them capable of autonomous control using a laser guidance system. Rajanna & Hammond [32] explores eye movements, called gaze-assisted interaction, and list several of these movements.

Numerous studies discuss quality characteristics present in systems to improve the interaction between users and things. Quality in ubiquitous systems is explored by Carvalho et al. [25], who presents a set of five context-specific quality characteristics: Context-awareness, Transparency, Attention, Calmness, and Mobility. Vacher et al. [7] address elements related to fears of older people when using ubiquitous systems: not being able to use them (Usability), being dependent on a course which may fail (Dependence), and wishing that the system do not interfere with daily activities (Intrusiveness). Finally, Grguric et al. [10] indicate effectiveness, efficiency, acceptance, learnability, human error avoidance, and satisfaction as quality in use characteristics of UIs.

The analyzed studies develop interaction to facilitate users' necessities or understand the characteristics of the users to deliver something valuable. Interactions such as gestures and gaze-assisted are convenient types of input which facilitate the use of IoT systems by human users. Systems with the previously described Affect Haptics ability try to understand user's emotions to increase immersion with the system while interacting. Wearables, smartwatches, and sensors try to understand the user and its surroundings to deliver important information, such as the user's state of health.

**RQ2: HOW do IoT projects deal with software technologies (techniques, technologies, methods, and solutions) and their operationalization regarding Interactivity?**

To deal with the types of Interactivity mentioned in RQ1, and to make them possible, studies discuss several strategies, guidelines, and characteristics. The literature cites different ways to make Interactivity possible.

Considering smart objects, Ledo et al. [22] couple mobile devices in target objects to use their abilities. Shirehjini & Semsar [13] describe an interaction model with which users can identify and select devices based on their position, orientation, and form in a smart environment instead of IP address parameters. Zallio & Casiddu [16] also explore Interactivity with smart objects by reorganizing these objects' colors and layout. Di Nuovo et al. [12] describe using a simple GUI with web technology to support interaction with robotic devices connected to the network.

Eid & Osman [11] explores interaction by interpreting emotions and proposes an internet pajama capable of capturing the force generated by a hug. Whittington & Dogan [9] develop an interactive wheelchair that captures touch, joystick input, head tracking, and transmit to be interpreted in a web service. Marulli & Vallifuoco [27] explore holographic interfaces by giving them a UHD human figure and event management by detecting visitors' movements and reactions. Mewes et al. [17] describe the context of touchless interaction for use in surgeries by using cameras, eye tracking, inertial sensors, and voice commands.

Regarding gestures, Lowens et al. [4], in the context of wrist-worn devices, point out the importance of characteristics such as sensitiveness, comfort, ease of use, accuracy, and context-sensitivity for gesture interfaces. Du et al. [29] describe the use of hand gestures to interact with 3D objects. Alanwar et al. [28] use a specific gesture as a trigger to save energy in IoT devices. Kordts et al. [3] list some types of gesture movements: swipe, wave, scroll, pinch, throw, never mind, hammer, and clap. Wang et al. [1] present a gesture vocabulary composed of: Up, Down, Left, Right, Forward, Backward, Clockwise Circle, Counterclockwise circle, and Spirals.

Margetis et al. [35] explore gestures when interacting with physical paper in smart environments. They describe actions such as holding the paper with both hands, collocate, collate, flipping, rubbing, stapling papers, and handwriting. Unlike the others, which only considers hand gestures, Rajanna & Hammond [32] consider the eyes' movement as input to support accessibility. Eye-tracking is also addressed by Mewes et al. [17]. Carreira et al. [18] add Swipe, Grab, Drag, Point and Hold, and Point and Push to the gestures list. Zhang et al. [20] and Oakley et al. [6] make interaction possible using different tapping patterns.

Regarding voice input, Sheppard et al. [36] point out that speech may be used to interact with a smart sign, to provide directions to a room. Kim et al. [15] indicate that multi-modal interaction with smart TVs occurs by voice, motion-based pointer modes, and combinations. Lovato & Piper [8], focused on children's voice interaction, list three ways to interact with a system: exploration, information seeking, and functional operating. Voice commands are used in the context of surgeries [17] when touchless interaction is mandatory. Vacher et al. [7] propose a system with audio processing, speech recognition, and voice command recognition.

Sensors are commonly used to make interactive possible in systems such as smart homes. Portet et al. [33] explore home automation by using plenty of sensors such as lighting, shutters, security systems, energy management, and heating to support voice-based interaction. Zabalza et al. [39] indicate that human-robot interaction is supported by interpreting sensing information of the robot's surroundings. Van der Werff et al. [26], which studies interaction with lighting, defines logic for color and intensity by detecting motion, light level, and temperature. Rittenbruch & Donovan et al. [37] quote that interactions between users through IoT devices may happen by interpreting temperature, lighting, humidity, and noise levels and identifying user's preferences concerning these factors.

Studies also analyze characteristics, parameters, requirements, guidelines, and quality characteristics of Interactivity. Several studies discuss the importance of planning them when developing systems such as ubiquitous [23], smart systems [31] [24], smart homes [38], and wrist-worn devices [4]. Quality characteristics are:

- **Context-awareness** [4] [14] [23] [25]: System's capability of monitoring contextual information regarding the user, the system, and the environment. Context is broken into three kinds by Van Hove et al. [30]:
  - **Socio-cultural context**: refers to the context on a societal level. [30]
  - **Situational context**: refers to the interpretation of situations. It is at the local level. [30]
  - **Interaction context**: refers to the micro-level of context around the interaction between the user and the product. [30]
- **Mobility** [23] [25]: Continuous or uninterrupted use of the systems while the user moves across several devices.
- **Transparency** [23] [25]: Ability to hide its computing infrastructure in the environment, so the user does not realize that it is interacting with a set of computational devices.
- **Attention** [23] [25]: The system's ability to keep the user's focus on real-world interactions rather than technology.

- **Calmness** [23] [25]: The system's capability to interact with the user at the right time and situation and presenting only relevant information.

Some studies report **problems** which are related to the systems that can **prevent them from reaching expected qualities**, such as:

- **Conflicts** [23]: multiple mobile devices when entering/exiting a room generate an inconsistency in the data presented in the application's interface to different users.
- **Delay** [23]: When users turned on a costly device, it affected the system's responsiveness.
- **Loss of Context** [23]: the state of things was not being displayed correctly in the interface when the user changed the focus to another thing.
- **Synchronicity** [23]: Two or more things may be uncoordinated.
- **Reliability** [23]: A command sent through the Internet can fail to arrive.
- **Battery** [23]: Some things need to be connected all the time and require much energy.
- **Installability** [23]: The difficulty of connecting things with other things.
- **Accuracy of data** [4]: The device should efficiently analyze user performance and activity level and make adjustments based on context.

Parameters of Interactivity described in the literature are:

- **Presence** [31]: do users have a constant presence?
- **Creativity** [31]: do users change the way they interact?
- **Productivity** [31]: do users propose something that changes their interaction?
- **Communication** [31]: do users communicate with the system?
- **Adaptation** [31] [14]: do users adapt their actions according to the content?

Huxohl et al. [38] propose guidelines for voice assistants in smart homes:

- **Authentication & Authorization** [38]: It should identify the person they interact with and verify their permissions.
- **Activity-Based interaction** [38]: Relates the context to their activities, such as reading or sleeping.
- **Situated Dialogue** [38]: processes the natural language.
- **Explainability** [38]: provides the user with info about possible commands.

Davani et al. [24] list requirements for interaction with smart systems:

- **Enabling end-user programming** [24]: Decide what the system should do when specific events occur.
- **Transparency** [24]: Enable users to explore the environment. The same requirement is cited by Carvalho et al. [23] [25] as a quality characteristic.
- **Learnability** [24]: Include a learnable visualization of the environment.
- **Predictability** [24]: Provide a history of interaction and past behaviors.
- **Controllability** [24]: Enable users to trigger or cancel system behaviors.

Corno et al. [17] propose a set of guidelines for intelligent environments, strongly related to previously listed characteristics: (i) Consider people as the driver and technology as the enabler, (ii) Design for all persons and cultures, (iii) Design in a simple and emotional way, (iv) Balance system autonomy with the user will and needs, (v) Design for positive behavior changes, (vi) Consider the world as the interface; Explore new interaction means, (vii) Do not forget personalization, security and privacy issues, (viii) Design for uncertainty and cope with complexity, (ix) Learn in and from the field,

(x) Consider social aspects, (xi) Consider the environment as a single intelligent entity, (xii) Explore "strange, new" environments.

The analyzed studies present strategies and techniques to make interactivity possible in IoT projects. Some studies are also concerned about the quality of the interaction, presenting different viewpoints of quality characteristics and lists of guidelines. Some projects take advantage of the characteristics of smart objects, such as sensor and visual outputs. Similarly, voice commands and gesture patterns were widely adopted as input in studies. The interaction may also be improved by applying simple algorithms of even machine learning to the collected data.

**RQ3: WHERE do IoT projects locate the activities regarding Interactivity?**

Interaction in the context of IoT projects may happen approximately everywhere, according to literature. Several papers do not specify a location, even when there is an implicit location for the system, assuming that the interaction may happen anywhere and will possibly only vary its context.

At all, 24 papers out of 39 specified a location for their specific case. Several studies limit the range of the described system to an office, house, smart home or a room [7] [8] [12] [13] [16] [18] [24] [26] [33] [34] [37] [38]. Kim et al. [15] describe the interaction with smart television and defines its location in a living room in front of the user. A few studies are not so generic for considering the range of the system as “anywhere” but calls the range of the system smart spaces, environment, or smart environment [1] [5] [14] [39]. Shimon et al. [19] refers to the location as “Environment” but breaks into home, work, or the public.

Although some types of interaction are so specific that it requires the establishment of a location. Marulli & Vallifuoco [27] defines an interaction which is specific to cultural environments and exhibitions. Sheppard et al. [36] describe the interaction with digital signs, which may happen in places like malls, fast food restaurants, or gas pumps.

Mewes et al. [17] refer to touchless interaction in the surgery context, so the location may only be in hospitals. Oakley et al. [6] and other studies regarding wearables, smartwatches, among others, do not specify a location because it may be interpreted as the user itself or the places the user visits.

Some studies have created scenarios to exemplify the use of a system or a type of interaction. Gamecho et al. [2] exemplify a scenario in a cafeteria and a bus station. Margetis et al. [35] define scenarios in a school, bus station, and library.

From the articles, we conclude that IoT has a wide range of locations since the activities in IoT projects may happen anywhere within the project's specified scope and context. However, at present, it seems that IoT projects tend to be set more inside rooms, houses, offices, or other limited areas such as art galleries. Given that the Human-Thing interaction was the most common type in the analyzed papers, it seems plausible for projects to be set in controlled places where interaction results may deliver value.

**RQ4: WHOM do IoT projects allocate to deal with Interactivity?**

According to the literature, interaction in the context of IoT projects may be used by approximately everyone, depending on the system's purpose. In RQ3, many studies imply that everyone may interact with the system and do not refer to people neither as "users."

Few studies refer to the users of a system broadly, as "users" [33], "anyone" [35], or even "humans" [5] [24]. Studies focusing on interaction with wearables refer to the users as "Wearable Users" [6] [19] [20]. Studies about smart homes refer to users as homeowners [34].

Some studies focus on users of specific ages. Some studies deal with Interactivity with older people in a variety of contexts [7] [12] [14] [16] [18], and Lovato & Piper [8] deals with voice interaction with children.

Regarding specific context interactions, Zabalza et al. [39] describe a case of human-robot interaction, in which both may be considered actors. Marulli & Vallifuoco [27], who represents a case study in cultural environments, names "visitors" the system's users. Mewes et al. [17] explores touchless interaction in the medical context and treats the surgeon as the user.

Finally, some studies address interaction in the context of accessibility of impairment, focusing on a specific disability type. Gamecho et al. [2] specify people with physical, sensory, and cognitive disabilities. Margetis et al. [32] define users in a situation where there is a situational impairment, such as the inability to work on a computer due to busy hands. Whittington & Dogan [9] deals with interaction regarding some types of impairments: limited movements of neck, shoulders, elbow, wrists, fingers, ankle, scoliosis, contractures, dyskinesia, atrophy, paraplegia, hemiparesis, visual impairments, and cataracts.

Analyzed studies, at large, do not address who will use IoT solutions. Those which explicit the user type indicates the usage by humans. However, some studies are not explicit about the users and treat them just as "everyone." We may imply that the users are also human, given the nature of the described IoT projects. Some studies focus their solutions, not for every human, but everyone inside segments such as kids, elderly, or people with disabilities. Users may also be given a role under certain circumstances, such as being a visitor while visiting an art gallery. In this context, employees of the art gallery are not users even though they are humans. Despite the possibility of IoT projects to deal with other living actors such as animals and plants, none of the analyzed studies detailed them. The absence may only suggest a higher proportion of studies focusing on systems for people.

**RQ5: WHEN do the effects of time, transformations, and states of Interactivity affect IoT projects?**

Interactivity in IoT projects is usually addressed as a facet that may happen at any time, depending on the users' necessities. The types of interaction described in the literature may trigger preset events; consequently, they may be used anytime. Portet et al. [33] specify that the system is real-time, given that it is voice-based and needs to be always operational. Studies about interaction with wearables, such as Yoon et al. [21], differ from within and outside the system. The system must always be available to capture and polish external interaction. Corno et al. [5], who defines user experience guidelines for reliable, intelligent environments, indicate that the interaction may happen only during daily activities. The system may not necessarily be available the whole time.

Systems specific to a context or location may also be limited to a particular time range. Marulli & Vallifuoco [27] and Rosales et al. [31], who describe systems to interact with visitors in exhibitions, might only be used when visitors are nearby and during working days. Mewes et al. [17] study touchless interaction made by surgeons in both real or test cases, and it is expected that it may occur only when surgery is taking place, or at least when a surgeon is in a working hour.

The availability of IoT systems may be either anytime or during a specific time range, in the context of analyzed studies. Most of the described projects are focused on human users, so it is expected that systems are available in moments when people are present, such as during daytime for art galleries and during working time for offices. However, a great majority of studies approach IoT projects within houses. These systems, or at least part of them, should be available all the time to fulfill their goals when users need them.

**RQ6: WHY do IoT projects motivate the implementation of Interactivity?**

Several motivations for Interactivity were spotted in IoT projects. Eid & Osman [11] study the detection and interpretation of user's emotions in a variety of ways by analyzing their gestures, facets, and other factors to find out if they are bored, frustrated, or angry, and so on. The results may be used to improve several systems, such as E-learning.

Zabalza et al. [39] want to demonstrate that human-robot and robot-robot interactions can be realized by integrating sensing techniques, planning algorithms, and systematic designs. Marulli & Vallifuoco [27] want to improve cultural interactions and make visitors in art galleries more interested by developing interaction with holographic projections.

Some studies are focused on improvements in user interfaces [14]. Davani et al. [24] present both the analysis and design of a meta-UI to support operations on a metasystem for ambient intelligence. Mewes et al. [17] examine literature searching for touchless human-computer interaction occurrences, motivated by assisting radiologists and surgeons.

Three studies are motivated by the specific interaction of older people. Grguric et al. [10] intend to improve the elderly's integration in smart environments by studying suitable interaction design for this age group. Carreira et al. [18] also focus on the elderly and intend to fight their isolation and exclusion, allowing them to be more productive, independent, and have a more social and fulfilling life. Vacher et al. [7] describe Ambient Assisted Living, with which the necessities of older people may be anticipated to respond to their needs.

Alanwar et al. [28] list four contributions of their studies regarding pointing gestures. (i) introduce a method for selecting and controlling IoT devices using pointing and hand gestures, (ii) prototype a smartwatch with ultra-wideband radio and inertial sensors, (iii) develop machine-learning models for hand gesture recognition, and (iv) develop a module for pointing event detection, only based on inertial sensors.

Several studies are interested in improving the quality of smart environments. Shirehjini & Semsar [13] expect to simplify people's interaction with tiny objects without their interfaces, which are only accessible through technical infrastructure such as IP addresses. The result may positively affect the implementation of IoT devices in smart environments. Zallio & Casiddu [16] desire to improve people's quality of life by designing better houses and appliances. Huxohl et al. [38] propose explicit guidelines to improve personal voice assistants' development in smart homes. Van der Werff et al. [26]

investigate how different users of the same surroundings behave when sharing the same interactive lighting system. Rittenbruch & Donovan [37] investigate shared sensors' interaction as well, but in IoT devices as a whole. Kim et al. [15] investigate multi-modal interaction with smart televisions.

Voice interaction was previously referred by Huxohl et al. [38] in smart homes. Still, the possibility of facilitating interaction with voice commands was also explored by Sheppard et al. [36], who used this type of input to present path directions in digital signs. It might be quicker than using a touchscreen modality. Lovato & Piper [8] are motivated by the challenges of studying children's voice interaction, who eventually have their voices wholly changed.

Corno et al. [5] are motivated explicitly by stimulating intelligent environments researchers to consider the study of User Confidence, which is defined as the property of a system to offer user–system interactions that create the perception of a reliable and understandable intelligent system, whose behavior is transparent and on whose actions users can build trust. The authors propose guidelines to facilitate these studies.

Regarding gesture interactions, Kordts et al. [3] study full-body gestures and movements and describe the architecture, implementation, and evaluation of an ambient movement analysis engine. Margetis et al. [35] present a framework for gesture-based interaction with physical paper. Wang et al. [1] propose a vocabulary of gestures to improve gesture-based interaction in smart spaces. Oakley et al. [6] specifically study tapping gestures for smartwatches, and Lowens et al. [4] collect feedback about existing wearable bands to understand and analyze interaction issues. Shimon et al. [19] are motivated by smartwatches' limited touch space and study different gestures to overcome this limitation. Zhang et al. [20] explore the different types of gesture interactions between smartphones and smartwatches.

Gamecho et al. [2] are concerned with integrating disabled users. Their main goal was to grant access to ubiquitous interactive machines by using their mobile devices. Carvalho et al. [25] are motivated by improving the interaction with ubiquitous applications and propose a suite of software measures to evaluate the quality of interaction. Rajanna & Hammond et al. [32] are interested in developing solutions for situational impairments, such as not using both hands for a short period.

Overall, studies are motivated by the possibility of improving interaction techniques themselves or by delivering some value to users. Some are explicitly motivated by improving the life of a group such as the elderly. Several studies simply explore a way of interaction to understand them and prove the possibility of using them in other IoT projects, such as voice commands, gestures, patterns, multimodal interaction, or understanding users' emotions.

## 9.12 Final Considerations

Based on this study on the Interactivity facet, it was possible to identify some weaknesses, gaps, and topics not defined. Considering the definitions of Interactivity, we may observe that most of the studies delve into the interaction between people and things. Just a few cases explore the relation between things themselves without involving a human user. Besides, it is possible to find clusters of voice-based interaction, gestures, wristbands and wearables, multimodal interaction, and ubiquitous systems, possibly indicating that these are essential contemporary subjects.

The questions about the type of user involved in the interaction, its location, and when it happens were downsides of the research. The actors involved in the Interactivity were usually dealt with generically, like "everyone," except in specific cases in which interaction occurs in places such as a hospital, an art gallery, or when the focus was some characteristics such as age (kids or elderly) and disabilities. Similarly, the location was frequently treated as "everywhere," except when the study focused on a specific site such as a smart home, a hospital, or an art gallery. Some studies softened this downside by illustrating scenarios in schools, cafeteria, and bus stations. The quantity of studies about smartwatches and wearables affected this matter, given that it is difficult to specify the location of these devices. Identifying when interaction occurs was a problem as well because most of the presented systems were real-time.

Finally, regarding the motivation of studies, several kinds of research engaged in exploring new technologies, such as Whittington & Dogan [9] who describe an interactive wheelchair, Marulli & Vallifuoco [27] who propose holographic interactions in museums, and Eid & Osman [11] who interpret emotions to improve interactions with the environment. Many studies aimed to improve smart homes' interaction characteristics, smartwatches, gestures, and voice commands.

With these results, we confirmed the importance of Interactivity in IoT projects and verified how much researchers are committed to understanding and improving it. Despite the commitment, we noticed that the studies do not explicitly define what Interactivity is, being up to us interpret how it was addressed in each study to spot similarities. Our brief arrangement of knowledge previously dispersed in a diversity of other studies permits them to use them as guides or insight for authors eventually interested in starting IoT projects focused on Interactivity aspects.

By bringing up the difficulty in obtaining answers regarding who uses the systems, where interaction takes place, and when it happens, we might influence future projects and researches to plan these dimensions when developing their projects explicitly. Besides contributing to knowledge regarding the tendencies of interactivity in IoT projects, the quality characteristics extracted from several studies may help researchers ponder their projects' quality. Concerns such as Mobility, Context-Awareness, and Transparency may help them model their new projects' interactivity since the requirements phase.

## 9.13 References

**Final Set:**

[1] Wang, Xian, et al. “A gesture-based method for natural interaction in smart spaces.” Journal of Ambient Intelligence and Smart Environments 7.4 (2015): 535-562.

[2] Gamecho, Borja, et al. “Automatic generation of tailored accessible user interfaces for ubiquitous services.” IEEE Transactions on Human-Machine Systems 45.5 (2015): 612-623.

[3] Kordts, Börge, Bashar Altakrouri, and Andreas Schrader. “Capturing and analyzing movement using depth sensors and Labanotation.” Proceedings of the 7th ACM SIGCHI Symposium on Engineering Interactive Computing Systems. 2015.

[4] Lowens, Byron, Vivian Motti, and Kelly Caine. “Design recommendations to improve user interaction with wrist-worn devices.” 2015 IEEE international conference on pervasive computing and communication workshops (PerCom Workshops). IEEE, 2015.

[5] Corno, Fulvio, et al. “Designing for user confidence in intelligent environments.” Journal of Reliable Intelligent Environments 1.1 (2015): 11-21.

[6] Oakley, Ian, et al. “Beats: Tapping gestures for smartwatches.” Proceedings of the 33rd Annual ACM Conference on Human Factors in Computing Systems. 2015.

[7] Vacher, Michel, et al. “Evaluation of a context-aware voice interface for ambient assisted living: qualitative user study vs. quantitative system evaluation.” ACM Transactions on Accessible Computing (TACCESS) 7.2 (2015): 1-36.

[8] Lovato, Silvia, and Anne Marie Piper. "Siri, is this you?” Understanding young children’s interactions with voice input systems. Proceedings of the 14th International Conference on Interaction Design and Children. 2015.

[9] Whittington, Paul, and Huseyin Dogan. “A SmartDisability Framework: enhancing user interaction.” (2016).

[10] Grguric, Andrej, et al. “A survey on user interaction mechanisms for enhanced living environments.” International Conference on ICT Innovations. Springer, Cham, 2015.

[11] Eid, Mohamad A., and Hussein Al Osman. “Affective haptics: Current research and future directions.” IEEE Access 4 (2015): 26-40.

[12] Di Nuovo, Alessandro, et al. “Experimental evaluation of a multi-modal user interface for a robotic service.” Annual Conference Towards Autonomous Robotic Systems. Springer, Cham, 2016.

[13] Shirehjini, Ali Asghar Nazari, and Azin Semsar. “Human interaction with IoT-based smart environments.” Multimedia Tools and Applications 76.11 (2017): 13343-13365.

[14] Mayer, Christopher, et al. “A comparative study of systems for the design of flexible user interfaces.” Journal of Ambient Intelligence and Smart Environments 8.2 (2016): 125-148.

[15] Kim, Seong M., Eui S. Jung, and Jaekyu Park. “Effective quality factors of multi-modal interaction in simple and complex tasks of using a smart television.” Multimedia Tools and Applications 76.5 (2017): 6447-6471.

[16] Zallio, Matteo, and Niccolò Casiddu. “Lifelong Housing Design: User Feedback Evaluation of smart objects and accessible houses for healthy aging.” Proceedings of the 9th ACM International Conference on PErvasive Technologies Related to Assistive Environments. 2016.

[17] Mewes, Andre, et al. “Touchless interaction with the software in interventional radiology and surgery: a systematic literature review.” International journal of computer-assisted radiology and surgery 12.2 (2017): 291-305.

[18] Carreira, Micael, et al. “Evaluation of in-air hand gestures interaction for older people.” Universal Access in the Information Society 16.3 (2017): 561-580.

[19] Arefin Shimon, Shaikh Shawon, et al. “Exploring non-touchscreen gestures for smartwatches.” Proceedings of the 2016 CHI Conference on Human Factors in Computing Systems. 2016.

[20] Zhang, Cheng, et al. “Tapskin: Recognizing on-skin input for smartwatches.” Proceedings of the 2016 ACM International Conference on Interactive Surfaces and Spaces. 2016.

[21] Yoon, Hyoseok, et al. “A case study on iteratively assessing and enhancing wearable user interface prototypes.” Symmetry 9.7 (2017): 114.

[22] Ledo, David, et al. “Pineal: Bringing Passive Objects to Life with Embedded Mobile Devices.” Proceedings of the 2017 CHI Conference on Human Factors in Computing Systems. 2017.

[23] Andrade, Rossana MC, et al. “What changes from ubiquitous computing to internet of things in interaction evaluation?.” International Conference on Distributed, Ambient, and Pervasive Interactions. Springer, Cham, 2017.

[24] Davani, Aida Mostafazadeh, Ali Asghar Nazari Shirehjini, and Sara Daraei. “Towards interacting with smarter systems.” Journal of Ambient Intelligence and Humanized Computing 9.1 (2018): 187-209.

[25] Carvalho, Rainara Maia, Rossana Maria de Castro Andrade, and Káthia Marçal de Oliveira. “AQUArIUM-A suite of software measures for HCI quality evaluation of ubiquitous mobile applications.” Journal of Systems and Software 136 (2018): 101-136.

[26] van der Werff, Thomas, et al. “Evaluating interface characteristics for shared lighting systems in the office environment.” Proceedings of the 2017 Conference on Designing Interactive Systems. 2017.

[27] Marulli, Fiammetta, and Luca Vallifuoco. “Internet of things for driving human-like interactions: a case study for a smart cultural environment.” Proceedings of the Second International Conference on the Internet of things, Data, and Cloud Computing. 2017.

[28] Alanwar, Amr, et al. “Selecon: Scalable IoT device selection and control using hand gestures.” Proceedings of the Second International Conference on Internet-of-Things Design and Implementation. 2017.

[29] Du, Guiying, et al. “Gestural interaction with 3D objects shown on public displays.” Interaction Design and Architecture (s) 2018.38 (2018): 184-202.

[30] Van Hove, Stephanie, et al. “Human-Computer Interaction to Human-Computer-Context Interaction: Towards a conceptual framework for conducting user studies for shifting interfaces.” International Conference of Design, User Experience, and Usability. Springer, Cham, 2018.

[31] Rosales, Ricardo, et al. “Modelling the interaction levels in HCI using an intelligent hybrid system with interactive agents: a case study of an interactive museum exhibition module in Mexico.” Applied Sciences 8.3 (2018): 446.

[32] Rajanna, Vijay, and Tracy Hammond. “A gaze gesture-based paradigm for situational impairments, accessibility, and rich interactions.” Proceedings of the 2018 ACM Symposium on Eye Tracking Research & Applications. 2018.

[33] Portet, François, et al. “Context-aware voice-based interaction in smart home-vocadom@ a4h corpus collection and empirical assessment of its usefulness.” 2019 IEEE Intl Conf on Dependable, Autonomic and Secure Computing, Intl Conf on Pervasive Intelligence and Computing, Intl Conf on Cloud and Big Data Computing, Intl Conf on Cyber Science and Technology Congress (DASC/PiCom/CBDCom/CyberSciTech). IEEE, 2019.

[34] Bittner, Björn, et al. “Of Smarthomes, IoT Plants, and Implicit Interaction Design.” Tangible and Embedded Interaction. 2019.

[35] Margetis, George, et al. “Augmenting natural interaction with physical paper in ambient intelligence environments.” Multimedia Tools and Applications 78.10 (2019): 13387-13433.

[36] Sheppard, David, Nick Felker, and John Schmalzel. “Development of Voice Commands in Digital Signage for Improved Indoor Navigation Using Google Assistant SDK.” 2019 IEEE Sensors Applications Symposium (SAS). IEEE, 2019.

[37] Rittenbruch, Markus, and Jared Donovan. “Direct end-user interaction with and through IoT devices.” Social Internet of Things. Springer, Cham, 2019. 143-165.

[38] Huxohl, Tamino, et al. “Interaction Guidelines for Personal Voice Assistants in Smart Homes.” 2019 International Conference on Speech Technology and Human-Computer Dialogue (SpeD). IEEE, 2019.

[39] Zabalza, Jaime, et al. “Smart sensing and adaptive reasoning for enabling industrial robots with interactive human-robot capabilities in dynamic environments—A case study.” Sensors 19.6 (2019): 1354.

**Additional References:**

C. Tricco et al. A scoping review of rapid review methods. BMC Medicine, 2015.

B. Cartaxo et al.: The Role of Rapid Reviews in Supporting Decision -Making in Software Engineering Practice. EASE 2018.

B. Cartaxo et al. Evidence briefings: Towards a medium to transfer knowledge from systematic reviews to practitioners. ESEM, 2016.

# DEVELOPING IOT SOFTWARE SYSTEMS? TAKE ENVIRONMENT INTO ACCOUNT

## CONTEXT

IoT allows composing systems from *things* equipped with identifying, sensing, or actuation behaviors and processing capabilities that can communicate and cooperate to reach a goal. For so, different facets should be taken into account, such as environment.

## ENVIRONMENT

It is the place holding *things*, actions, events, and people. IoT systems provide smart services to adapt to users' needs and behavior according to the context of a given environment.

## SUMMARY

Different types of technological solutions and techniques are employed in several types of ambient to help users on their activities, provide comfort and reduce costs and effort. To engineer IoT software systems is necessary to understand the aspects, capabilities, techniques and solutions for support the environment infrastructure and context of the solution.

## OBJECTIVE

This briefing aims to enlighten project planning and to provide guidance when dealing with Environment in IoT. The information presented in this briefing seeks to answer: "What taken into account when considering environment for IoT?".

## MOTIVATION

IoT offers challenges for their construction, with concerns that rather than just developing, we need to engineer software systems. This motivated us to a more comprehensive investigation to address these concerns and to provide evidence-based support for IoT engineering.

## ADDITIONAL INFORMATION

**Who are the briefing's clients?** Software developers and practitioners who want to make decisions about how to deal with *environment* in IoT, considering scientific evidence.

**Where do the findings come?** All findings of this briefing were extracted from scientific studies about *environment* through a Rapid Review[1]. The Technical Report[2] containing all the findings is available for further information.

**What is included in this briefing?** Technologies, challenges, and strategies to deal with *environment* in IoT projects.

**What are the Challenges and Opportunities?** It involves a lot of devices generating a large amount of data that ca lead to problems related with: connectivity and interoperability, efficient, reliable and high-speed communication, energy efficiency, and others. These high complex environments require system integration, security, privacy, data access and data reliability, that are important aspects to engineer IoT software systems. There is also a necessity for trustful and legal regulation as well as sustainability when considering environment in IoT.

**Does environment represent a concern in the Engineering of Internet of Things Software Systems?** Yes. Environment should be considered thought the development of IoT being present in the architecture, design, and technologies that will be employed on these systems.

## HIGHLIGHTS

❖ *Several technologies and solutions have been offered but there* ***isn't a complete solution***

❖ ***Security and privacy*** *represent important characteristics of these systems but are still open problems*

## WHAT
### is the the understanding of environment in CSS?

- The environment where the solution is deployed is a multi-dimensional contextual space with **different levels of importance**, that can change over time.
- To translate the environment into computing technologies when considering context, it is necessary to **state the contextual** variables to be used.
- IoT can **adapt their behavior** according to the information they receive about the environment or the user's, and this information is the context the systems should be aware.

## HOW
### do CSS projects deal with the operacionalization of environment?

- The environments are enhanced with sensors and actuators to sense and change a state of the ambient. Technologies like cloud and edge computing, Wireless Sensor Networks, Vehicular Ad-hoc Networks, artificial intelligence, and others can be employed on these systems.
- It is necessary to **define relevant environment information**, how it will be collected and shared.
- The interplay between environment and solution can have an effect on each other that can alter the desired outcome, therefor is necessary to verify **the environmental impact** in the solution and vice-versa.

## WHERE
### CSS projects locate the activities regarding environment?

- The activities' location is basically the own environment and depends on the domain that are employed. Based on the literature found environment can be places like city, home, ambient assisted living, university campus, building, industry, transportation, street, bike station, parking space and others.

## WHO
### do CSS projects allocate to deal with environment?

- In software engineering the phases that allocate environment activities allocate developers, system designers, domain experts, technical professionals, end-users and stakeholders to build the ambient solution.

## WHEN
### do the effects of time, and states of environment affect CSS projects?

- Concerning the solutions presented, the majority deals with software activities related to analysis, design, and implementation phases on activities like system architecture definition, software design, requirement specification and software implementation.

## WHY
### do CSS projects implement environment?

- Adapt ambient to users' needs and behavior. Provides comfort, quality of life, benefit daily lives, accessibility, high productivity, reduce costs and effort, save time, use resources efficiently and give autonomy to users.
- Helps on health diseases, pollution management, traffic efficiency, deterioration and management of infrastructure, criminality, climate change, cyber-security and economic development. Provides easy and sustainable user-centric quality services.

[1] More on Rapid Reviews: https://dl.acm.org/citation.cfm?id=3210462
[2] Full Report: https://goo.gl/no7jtR

# 10 RAPID REVIEW ON ENVIRONMENT

# Rapid Reviews Meta-Protocol:

## Engineering of Internet of Things Software Systems

**Danyllo Silva, Rebeca C. Motta, Guilherme H. Travassos**

# Environment

In the investigation regarding Internet of Things Software Systems (IoT), it has been observed that these modern software systems offer challenges for their construction since they are calling into question our traditional form of software development. Usually, they rely on different technologies and devices that can interact-capture-exchange information, act, and make decisions. It leads to concerns that, rather than just developing software, we need to engineer software systems embracing multidisciplinarity, integrating different areas. From our initial research, we analyzed the concerns related to this area. We categorized them into a set of facets - Connectivity, Things, Behavior, Smartness, Interactivity, Environment, and Security - representing such projects' multidisciplinarity, in the sense of finding a set of parts composing this engineering challenge.

Since these facets can bring additional perspectives to the software system project planning and management, acquiring evidence regarding such facets is of great importance to provide an evidence-based framework to support software engineers' decision-making tasks. Therefore, the following question should be answered:

***"Does Environment represent a concern in the engineering of***

***Internet of Things software systems?"***

This Rapid Review (RR) aims to analyze the environment to characterize it in the IoT field, regarding *what, how, where, when, and why* is used in the context of IoT projects, verifying the existence of published studies supporting the previous results. The 5W1H aims to give the observational perspective on which information is required for the understanding and management of the facet in a system (what); to the software technologies (techniques, technologies, methods, and solutions) defining their operationalization (how); the activities location being geographically distributed or something external to the software system (where); the roles involved to deal with the facet development (who); the effects of time over the facet, describing its transformations and states (when); and to translate the motivation, goals, and strategies going to what is implemented in the facet (why), in respect of IoT projects.

## 10.1 Research Questions

- **RQ1:** What is the understanding and management of the Environment in IoT projects?

- **RQ2:** How do IoT projects deal with software technologies (techniques, technologies, methods, and solutions) and their operationalization regarding the Environment?
- **RQ3:** Where do IoT projects locate the activities regarding the Environment?
- **RQ4:** Whom do IoT projects allocate to deal with the Environment?
- **RQ5:** When do the effects of time, transformations, and states of Environment affect IoT projects?
- **RQ6:** Why do IoT projects implement the Environment?

## 10.2 Search Strategy

The Scopus[11] search engine and the following search string support this RR, built using PICOC with five levels of filtering:

**P**opulation - Internet of Things software systems

Synonymous:

"ambient intelligence" OR "assisted living" OR "multiagent systems" OR "systems of systems" OR "internet of things" OR "Cyber-Physical Systems" OR "Industry 4" OR "fourth industrial revolution" OR "web of things" OR "Internet of Everything" OR "contemporary software systems" OR "smart manufacturing" OR digitalization OR digitization OR "digital transformation" OR "smart cit*" OR "smart building" OR "smart health" OR "smart environment"

**I**ntervention - environment

“use* context” OR “surrounding environment” OR “smart space” OR “smart environment” OR “contextual environment” OR “use* environment” OR “physical environment” OR “system ambient” OR “software ambient” OR “system surrounding” OR “system context” OR “software context” OR “emergent environment” OR “social environment” OR “social context” OR "smart context" OR "smart ambient"

**C**omparison – no

**O**utcome - Synonymous:

understanding OR management OR technique OR “technolog*” OR method OR location OR place OR setting OR actor OR role OR team OR time OR transformation OR state OR reason OR motivation OR aim OR objective

**C**ontext –

engineering or development or project or planning OR management OR building OR construction OR maintenance

Limited to articles from 2015 to 2018

Limited to Computer Science and Engineering

11 https://www.scopus.com

LIMIT-TO (SUBJAREA, "COMP" ) OR LIMIT-TO (SUBJAREA, "ENGI" ) ) AND ( LIMIT-TO (PUBYEAR, 2018 ) OR LIMIT-TO (PUBYEAR, 2017 ) OR LIMIT-TO (PUBYEAR, 2016 ) OR LIMIT-TO (PUBYEAR, 2015)

| |
|---|
| TITLE-ABS-KEY (("ambient intelligence" OR "assisted living" OR "multiagent systems" OR "systems of systems" OR "internet of things" OR "Cyber Physical Systems" OR "Industry 4" OR "fourth industrial revolution" OR "web of things" OR "Internet of Everything" OR "contemporary software systems" OR "smart manufacturing" OR digitalization OR digitization OR "digital transformation" OR "smart cit*" OR "smart building" OR "smart health" OR "smart environment" ) AND ("use* context" OR "surrounding environment" OR "smart space" OR "smart environment" OR "contextual environment" OR "use* environment" OR "physical environment" OR "system ambient" OR "software ambient" OR "system surrounding" OR "system context" OR "software context" OR “emergent environment” OR “social environment” OR “social context” OR "smart context" OR "smart ambient") AND (understanding OR management OR technique OR "technolog*" OR method OR location OR place OR setting OR actor OR role OR team OR time OR transformation OR state OR reason OR motivation OR aim OR objective) AND (engineering or development or project or planning OR management OR building OR construction OR maintenance) ) AND ( LIMIT-TO ( PUBYEAR,2018) OR LIMIT-TO ( PUBYEAR,2017) OR LIMIT-TO ( PUBYEAR,2016) OR LIMIT-TO ( PUBYEAR,2015) ) AND ( LIMIT-TO ( SUBJAREA,"COMP" ) OR LIMIT-TO ( SUBJAREA,"ENGI" ) ) |

## 10.3 Selection procedure

One researcher performs the following selection procedure:

1. Run the search string;
2. Apply the inclusion criteria based on the paper Title;
3. Apply the inclusion criteria based on the paper Abstract;
4. Apply the inclusion criteria based on the paper Full Text, and;

After finishing the selection from Scopus, use the included papers set to:

5. Execute snowballing backward (one level) and forward:
    a. Apply the inclusion criteria based on the paper Title;
    b. Apply the inclusion criteria based on the paper Abstract;
    c. Apply the inclusion criteria based on the paper Full Text.

The JabRef Tool[12] must be used to manage and support the selection procedure.

## 10.4 Inclusion criteria

- The paper must be in the context of **software engineering**; and
- The paper must be in the context of the **Internet of Things software systems**; and
- The paper must report a **primary or a secondary study**; and
- The paper must report an **evidence-based study** grounded in empirical methods (e.g., interviews, surveys, case studies, formal experiment, etc.); and
- The paper must provide data to **answering** at least one of the RR **research questions**.
- The paper must be written in the **English language**.

[12] http://www.jabref.org/

## 10.5 Extraction procedure

The extraction procedure is performed by one researcher, using the following form:

| <paper_id>:<paper_reference> | |
|---|---|
| Abstract | <Abstract> |
| Description | <A brief description of the study objectives and personal understanding> |
| Study type | <Identify the type of study reported by paper (e.g., survey, formal experiment)> |
| RQ1: WHAT<br>information required to understand<br>and manage the environment in IoT | - < A1_1><br>- < A1_2><br>- ... |
| RQ2: HOW<br>software technologies (techniques, technologies, methods and solutions) and their operationalization | - < A2_1><br>- < A2_2><br>- ... |
| RQ3: WHERE<br>activities location or something external to the IoT | - < A3_1><br>- < A3_2><br>- ... |
| RQ4: WHO<br>roles involved to deal with the environment development in IoT | -< A4_1><br>-<A4_2><br>- … |
| RQ5: WHEN<br>effects of time over environment, describing its transformations and states in IoT | - < A5_1><br>- < A5_2><br>- ... |
| RQ6: WHY<br>motivation, goals, and strategies regarding environment in IoT | - < A6_1><br>- < A6_2><br>- ... |
| Additional Information and Comments | - <Important information not necessarily related to research questions><br>- <Personal comments> |

## 10.6 Synthesis Procedure

In this RR, the extraction form provides a synthesized way to represent extracted data. Thus, we do not perform any synthesis procedure.

However, the synthesis is usually performed through a narrative summary or a Thematic Analysis when the number of selected papers is not high.

## 10.7 Report

An Evidence Briefing [2] reports the findings to ease the communication with practitioners. It was presented as the cover for this chapter.

## 10.8 Results

**Execution**

| Activity | Execution date | Result | Number of papers |
|---|---|---|---|
| First execution | 11/07/2018 - 08:55 | 925 documents added | 925 |
| Remove conferences/workshops/books | 11/07/2018 | 78 documents withdrawn | 847 |
| Included by Title analysis | 14/07/2018 | 677 documents withdrawn | 170 |
| Included by Abstract analysis | 15/07/2018 | 111 documents withdrawn | 59 |
| Papers not found | 15/07/2018 | Two documents were withdrawn | 57 |
| Articles for reading | 15/07/2018 | 57 documents | 57 |
| Removed after a full reading | 16/07/2018 – 23/07/2018 | 40 documents withdrawn | 17 |
| Snowballing | 23/07/2018 | 12 documents added | 29 |
| Snowballing after reading | 23/07/2018 – 29/07/2018 | 7 documents withdrawn | 22 |
| Total included | 29/07/2018 | 22 documents | 22 |
| Papers extracted | 29/07/2018 – 07/08/2018 | 22 documents | 22 |

**Final Set**

| Reference | Author | Title | Year | Source |
|---|---|---|---|---|
| [1] | Corno and Razzak | Real-time monitoring of high-level states in smart environments | 2015 | Regular search |
| [2] | Rahnama et al. | Synthesizing social context for making the Internet of Things environments more immersive | 2015 | Regular search |
| [3] | Lhotská et al. | Challenges and trends in Ambient Assisted Living and intelligent tools for disabled and elderly people | 2015 | Regular search |
| [4] | Jaen et al. | Customizing smart environments: A tabletop approach. | 2015 | Regular search |
| [5] | Connolly et al. | Enabling Factors for Smart Cities: A Case Study | 2015 | Regular search |
| [6] | Vinci et al. | Edge enabled development of smart cyber-physical environments | 2015 | Regular search |
| [7] | Sadik et al. | A literature review on Smart Cities: Paradigms, opportunities and open problems | 2015 | Regular search |
| [8] | Felemban et al. | Requirement engineering technique for smart spaces | 2015 | Regular search |
| [9] | Rodriguez et al. | An Experience of Engineering of MAS for Smart Environments: Extension of ASPECS | 2016 | Regular search |
| [10] | Wittenberg, C. | Human-CPS Interaction-requirements and human-machine interaction methods for the Industry 4.0 | 2016 | Regular search |
| [11] | Matera et al. | Empowering end users to customize their smart environments: model, composition paradigms, and domain-specific tools. | 2017 | Regular search |
| [12] | Semsar et al. | Human interaction with IoT-based smart environments. Multimedia Tools and Applications | 2017 | Regular search |
| [13] | Skeie et al. | A Review of Smart House Analysis Methods for Assisting Older People Living Alone | 2017 | Regular search |
| [14] | Al-Fayez et al. | A Survey on the Internet of Thing Enabled Smart Campus Applications | 2017 | Regular search |
| [15] | Silva, J et al. | Smart mobility: A survey | 2017 | Regular search |
| [16] | Kandasamy et al. | Smart Garbage Bin Systems–A Comprehensive Survey | 2017 | Regular search |
| [17] | Vinci et al. | A Metamodel Framework for Edge-Based Smart Environments | 2018 | Regular search |
| [18] | Bouchachia et al. | A review of smart homes in healthcare | 2015 | Snowballing |
| [19] | Dangelico et al. | Smart cities: Definitions, dimensions, performance, and initiatives. | 2015 | Snowballing |
| [20] | Murphy et al. | A communications-oriented perspective on traffic management systems for smart cities: Challenges and innovative approaches | 2015 | Snowballing |
| [21] | Vinci et al. | Metamodeling of smart environments: from design to implementation. | 2017 | Snowballing |
| [22] | Rjab, A. B., & Mellouli, S. | Smart cities in the era of artificial intelligence and the internet of things: a literature review from 1990 to 2017 | 2017 | Snowballing |

## 10.9 Summary of the articles

This section describes some studies about the environment in the context of the Internet of Things software systems. In the literature, we have some domains in which this kind of software system is built-in (e.g., smart homes, smart mobility, smart campuses, ambient assisted living and industry). To solve some problems of different ambient solutions were proposed to help developers, designers, and users in this context, from which we can highlight: metamodels to represent services and devices embedded in the environment, approaches to give autonomy to users on developing their ambient intelligence, architecture to monitor and interpret complex smart environments.

Several authors have proposed many definitions of Smart City, but still, there is no universal agreement about this term [19]. Sadik et al. [7] describe smart cities, defining it by a composition of multiple dimensions - technology, people, and community - and six characteristics: Smart Human Learning (Smart HL), Smart Governance, Smart Economy, Smart Mobility, Smart Environment, and Smart Living. At last, the authors discuss the challenges and opportunities. We can highlight the necessity for sustainable growth with harmonizing technologies, consumption of resources, and impact on the environment. The use of Artificial Intelligence and a Smart Grid for smart energy management is advised to solve these problems.

As open problems, we can mention many devices that produce a significant amount of data and turn the task of analysis and storage more challenging. Security and privacy are other significant challenges in Smart Cities. Interoperability is required to enable communication of heterogeneous devices [7].

Dangelico et al. [19] present a literature review of definitions, components, and performance measures of cities. Also, this work shows some projects of Smart City around the world. In general, the cities present some characteristics that we can attribute to an interconnected infrastructure, emphasis on business-led urban development, activities for promoting urban growth, social inclusion, social capital, and the natural environment as a strategic component for the future. To achieve these characteristics, Smart Cities uses information and communication technology (ICT) to improve the way subsystems operate to enhance the quality of life [19][22].

Smart City is a solution for current society problems (pollution, traffic congestion, poverty, deterioration of infrastructure, criminality) by adopting technological tools (connected objects, self-driven cars or self-driving trucks, drones, chatbots, robots, practical quantum computers, Botnets of Things, etc.) to promote services to increase economic competitiveness, quality of life and urban development. This article mentions some application areas in Smart Cities: Smart Home, Smart Transport, Smart Healthcare, Urban environment, Industry, Smart Offices, Smart Learning, Smart Government, Decision-making, Security, Human-Machine Interaction, and E-service [22].

In an extensive literature review, Rjab and Mellouli [22] mention the Internet of Things (IoT), Artificial Intelligence (AI), and Cloud computing as the most applied technological solutions to provide an intelligent infrastructure to solve urban problems. The Internet of Things can provide connectivity, real-time data processing by improving the quality and accessibility of services and increasing security. Meanwhile, Artificial Intelligence helps with intelligent monitoring, behavioral modeling, intelligent network, natural resources treatment, interaction with citizens, industrial automation, and analyzing and dealing with Big Data. Besides that, the authors introduce ethical, social, and technical problems existing in Smart Cities: data processing, privacy, security, freedom control, authority, independence, automatism, adaptability, integrity, and autonomy of machines (responsibility of actions, unemployment problems).

Connolly et al. [5] present preliminary findings from a case study undertaken in Dublin that confirm a literature review's findings. The authors propose an ensemble of enabling factors for Smart Cities: Technology, Social Infrastructure, Governance, Triple Helix Partnerships, and Information Services. This study exhibits four dimensions to Smart Cities: Infrastructure (skills exchange, connected thinking, spatial issues, multi-operator telecommunications), Approach (competition, stakeholder engagement, local authority as enabler and leadership), Goals (branding, vision, demonstration showcase, demonstrate long-term thinking, overall definition of the area and identification of assets, storytelling as a success, employer location of choice, boots innovation) and Principles (citizen-centric approach to strategy, citizen-engagement through the design process, creative finance, transparency, technology as secondary/enabler and learning as outcome).

Kandasamy et al. [16] provide a survey in the area of smart monitoring of waste, smart bin collection, and route optimization for garbage collection. With the increase of population in urban areas, the volume of solid waste is increasing enormously. The management system we have today will not be able to deal with this large volume of waste generated by cities. The management problem provokes garbage accumulation and garbage degradation in public areas; this can cause diseases and grow bacteria and viruses.

Smart Cities offer several smart services to ensure the quality of life for their citizens. This term can be identified through six dimensions: Smart Economy (Innovation and Competitiveness), Smart Mobility (Infrastructure and Transport), Smart Environment (Resources and Sustainability), Smart People (Creativity and Social Capital), Smart Living (Culture and Quality of Life) and Smart Governance (Participation and Empowerment). The authors in [16] propose a smart Garbage Management System (GSM) framework that benefits the citizens' health. To achieve GSM, available technologies such as Pervasive Computing technologies, the Internet of Things (IoT), Cloud Computing, Big Data, and Wireless Sensor Networks (WSN) can be used.

Smart Cities can provide Smart Campus to their citizens [14]. The Smart Campus provides an interactive and creative environment for students and faculty, offering services promptly, and reducing effort and costs. Smart campus means that the institution will adopt advanced technologies to automatically control and monitor campus facilities and provide high-quality services by using IoT technology. Some Smart Campus applications were mentioned like Smart Buildings, Renewables, and Smart Grid, Smart Learning, and Waste and Water Management. Finally, they identify the main challenges to deliver a Smart Campus applying IoT: interoperability, information processing, system integration, and efficient/reliable/high-speed communication.

Smart Mobility [15] is a dimension in Smart Cities that enables smart urban traffic services using Information and Communication Technology. This dimension has six main areas: Driving Safety, Smart Lightning Systems, Sharing, Urban Mobility, Electric Mobility, Green Mobility, and Smart Payment Systems. The authors focused on Intelligent Transportation Systems, Smart Parking Systems, and Smart Traffic Lights Systems.

To achieve these services, we can use sensors, and the Vehicular Ad-hoc Network (VANET) enables Vehicle-to-Vehicle and Vehicles-to-Infrastructure communication [15]. Besides that, VANET provides high mobility, high computational ability, rapid changes in network topology, and variable network density. Smart Mobility brings accessibility, a better quality of life, reduces costs and gridlocks, efficient energy usage, lower accident rates, improves air quality, and helps achieve a sustainable city.

Murphy et al. [20] discuss Traffic Management Systems (TMS) for Smart Cities. TMSs offer services like Traffic Prediction, Routing Planning, and Parking Management Systems and services to reduce road traffic congestion, improve response time to incidents, and ensure a better travel experience for commuters.

These systems are composed of four phases: Data Sensing and Gathering, Data Fusion, Processing and Aggregation, Data Exploitation, and Service Delivery [20]. The leading technologies for Data Sensing and Gathering are Wireless Sensor Networks, M2M Communication, Mobile Sensing, Social Media, and VANETs. The authors cite initiatives in architecture, safety, efficiency, sustainability and energy awareness, reliability and security, and innovative services. At last, the main challenges mentioned are security, routing, challenging to address vehicles, real-time response, managing large amounts of data, etc.

Some researchers [13,18] discuss the Smart Home. Skeie et al. [13] offer an overview of Smart Homes, including a definition, technology devices, method analysis, and current challenges. Smart Homes can learn the user’s activities and behaviors to find patterns and then use these patterns to “predict” the user's future behavior. Therefore, Smart Houses projects can help older adults remain at home as long as possible, reduce bills, or just improve comfort for the occupant.

The authors discussed the first sensor technology devices like motion sensors, humidity sensors, acoustic sensors, temperature sensors, water flow sensors, positioning sensors, optical sensors, sensors on doors, pressure sensors, ultrasonic and infrared sensors. Considering method analysis, the ones mentioned are computer vision, pattern recognition, image processing, artificial intelligence, artificial neural networks (ANN), Bayesian networks, Markov models, fuzzy logic, LeZi algorithm, and Support Vector Machine (SVM). The main challenges of Smart Homes are data integration, privacy, data access, security, the need for standardization of devices and communication patterns, limitations in analysis methods, and legal regulations.

Bouchachia et al. [18] applied Smart Homes as ambient assisted living technologies to help residents with daily activities improve their quality of life and ensure the elderly live comfortably and independently. These systems are equipped with sensors, actuators, or cameras. Therefore, Smart Homes can sense and control the environment, predict actions, and track the residents' health condition.

This survey [18] provides a logical structure for smart home systems. This structure consists of three layers: sensing, processing, interacting. It provides the main techniques and technologies used, like sensing and networking technologies, data processing and knowledge engineering, human interfaces. Some smart home applications are energy saving, security and safety, fall detection, light management, smoke, and fire detection. Finally, the authors mention the limitation of current approaches like security, privacy, reliability, and activity recognition.

Lhotská et al. [3] provide a study of Ambient Assisted Living (AAL). These systems help disabled and older people in daily activities, giving them autonomy and improving the quality of life. The authors offer a set of solutions that act on the prevention and intervention of various scenarios in which patients have some diseases like Parkinson's disease and cancer.

Information and communication technologies (ICT) can be employed to achieve an Ambient Assisted Living. The authors mention wireless sensors networks, the internet of things, telemedicine applications, smart home, distributed systems, data mining, hardware and software tools (orientation and location, recognition of gestures, intelligent systems, nutrition therapy, and rehabilitation program), and many other approaches involving artificial intelligence and engineering, with medical applications [3].

Wittenberg [10] discusses the effects of Industry 4.0 and analyzes the use of mobile applications supporting technicians on Smart Factories. This paper gives an overview of industry history until Smart Factories and the lifecycle of automated production systems. The need to reduce time and effort strengthens automatization, but it brings some complexity to the systems. The authors also bring a study of mobile devices as tools to help technicians because of their robustness and the power of the rechargeable battery usability.

In software engineering, aspects related to the development and requirement engineering have to be addressed. Some researchers [8, 9, 17, 21] provides an approach for modeling a smart environment considering concepts like sensors, actuators, and services. Authors in [4] and [11] provides an approach for a comfortable user development environment.

Felemban et al. [8] propose a requirement specification technique for smart spaces based on software engineering using the Use Case concept. This solution proposes a common approach for requirement capture and specification, simplifying the process and aids developers and designers' understanding and communication. The authors adapted from the Unified Modeling Language (UML) the Use Case concept and Use Case scenarios and description.

This work [8] presents five steps for the specification of Use Case: identify the possible actions, describe the functionality provided by each use case, identify the participants (actors), develop the models of the smart space, and provide the use case description. This technique was applied to developing indoor smart space in Masjid Al-Haram. An advantage of this approach is that once requirements captured and documented in standard form can be reused in developing any other related project.

Rodriguez et al. [9] present a methodological approach to engineering systems of Smart Environments. This approach is based on existing MAS methodology, namely ASPECS. The Smart Environments Systems are difficult to analyze and design, as they are composed of numerous interacting entities immersed in a dynamic and partially predictable environment.

This methodology [9] provides identification of goals hierarchy, detailed requirements, and associations of goals to sensors/effectors, different levels of ontologies to describe problem conceptualization, and the several involved proficiency that results in a set of metamodels produced in an organizational structure. In the end, the authors present a Case Study in the eHealth project named ECare that monitor patients with heart failure disease.

Vinci et al. [6][17][21] introduce an approach to lead with Smart Environments. In [6], the authors present a methodology for designing and implementing Smart Cyber-Physical Environments providing reactivity, scalability, extensibility, and fault tolerance using the isapiens platform. Isapiens is a platform employed on the Internet of Things that applies the concept of edge computing by exploiting the agent metaphor. The methodology efficiency has been proved by the realization of a Smart Office Case Study. At last, the preliminary results of the Smart Office prototype have been shown.

The work presented in [21] provides a Smart Environment Metamodel (SEM) framework suited for designing Smart Environments (SEs). These metamodels provide a high-level abstraction and a common vocabulary to model specific Smart Environment concepts, attributes of concepts, and relations between concepts. Smart Environment Metamodel framework, called SEM, offers two metamodels: the functional metamodel and the data metamodel. This approach helps through the phases of analysis, design, and implementation of a specific SE. Also, it is offered a system engineering process using SEM. In the end, the authors offer a Case Study applying their approach to a Smart Office environment.

In [17] provides an extension of the existing Smart Environment Metamodel (SEM) framework suited for designing Smart Environments (SEs), focusing on the concept of edge computing. The authors defend the usage of edge computing to move the computation close to the smart environment to improve the system reactivity and avoiding bottlenecks related to the transmission of a significant amount of data. Finally, the authors apply the approach in a Smart Office and analyze the results and the behavior of the workers.

Jaen et al. [4] provide a rule editing tool for interactive tabletops to specify behavior in reactive smart environments. The authors provide an interactive tabletop to interact with users that can specify the rules. The approach is composed of Event-Condition-Action (ECA) rules and a domain-independent language. Additionally, an experimental study was conducted, and the system was evaluated. This method helps users with configuration and personalization of the environment by naturally programming their own rules.

Matera et al. [11] present three systems prototypes based on task-automation across creating rules for smart object composition performed by end-users. The approach consists of rule creation based on cause-effect allowing end users to customize their smart spaces easily. It permits users to adapt to the environment to accommodate their necessities. Finally, the authors present the platform organization and an analysis of the usability of the prototypes implemented.

Semsar et al. [12] present a mobile 3D UI application for interaction with IoT-based Smart Environments. Because of the complexity and the lack of infrastructure integration, it is necessary to help the user interact with the environment. The authors proposed a solution composed of 3D-based user interfaces, an interaction model, and physical objects. The authors implemented a smart meeting room and showed the architecture and technologies applied like Unity3D, IoT, Wireless Sensor Network gateway, Cloud, etc. Finally, they applied a questionnaire about the quality of the user interface and analyzed the results comparing experienced users and beginners.

Corno and Razzak [1] provide a high-level approach based on the concept of Domotic Effects for monitoring and interpreting complex smart environments achieving easy management of smart environments. This approach consists of an abstract model to design user goals or intentions using the Domotic Effects (DE) framework. In this method, the users can program their environments, and the Ambient Intelligence designers have an abstraction layer that enables the definition of generic goals inside the environment**.** At last, it shows the experimental results of the architecture to evaluate domotic effects.

Finally, Rahnama [2] discusses the social behavior of smart objects. The authors propose a solution named CANthings framework that can provide social network services. In this solution, the objects can create relationships and provide or consume services. The approach enables users to easily make their own rules based on the concept of if-this-then-that. In the end, a Case Study is presented using the proposed solution.

## 10.10 Tracking Matrix

| Ref | Paper | WHAT | HOW | WHERE | WHO | WHEN | WHY |
|---|---|---|---|---|---|---|---|
| [1] | Real-time monitoring of high-level states in smart environments | X | X | X | X | X | X |
| [2] | Synthesizing social context for making the Internet of Things environments more immersive | X | X | X | X | | X |
| [3] | Challenges and trends in Ambient Assisted Living and intelligent tools for disabled and elderly people | X | X | X | X | | X |
| [4] | Customizing smart environments: A tabletop approach. | X | X | X | X | X | X |
| [5] | Enabling Factors for Smart Cities: A Case Study | X | | X | X | | X |
| [6] | Edge enabled development of smart cyber-physical environments | X | X | X | X | | X |
| [7] | A literature review on Smart Cities: Paradigms, opportunities and open problems | X | | X | X | | X |
| [8] | Requirement engineering technique for smart spaces | X | X | X | X | X | X |
| [9] | An Experience of Engineering of MAS for Smart Environments: Extension of ASPECS | X | X | X | X | X | X |
| [10] | Human-CPS Interaction-requirements and human-machine interaction methods for the Industry 4.0 | X | X | X | X | | X |
| [11] | Empowering end users to customize their smart environments: model, composition paradigms, and domain-specific tools. | | X | X | X | X | X |
| [12] | Human interaction with IoT-based smart environments. Multimedia Tools and Applications | X | X | X | X | | X |
| [13] | A Review of Smart House Analysis Methods for Assisting Older People Living Alone | X | X | X | X | | X |
| [14] | A Survey on the Internet of Thing Enabled Smart Campus Applications | X | X | X | X | | X |
| [15] | Smart mobility: A survey | X | X | X | X | | X |
| [16] | Smart Garbage Bin Systems–A Comprehensive Survey | X | X | X | X | | X |
| [17] | A Metamodel Framework for Edge-Based Smart Environments | X | X | X | X | X | X |
| [18] | A review of smart homes in healthcare | X | X | X | X | | X |
| [19] | Smart cities: Definitions, dimensions, performance, and initiatives. | X | | X | X | | X |
| [20] | A communications-oriented perspective on traffic management systems for smart cities: Challenges and innovative approaches | X | X | X | X | | X |
| [21] | Metamodeling of smart environments: from design to implementation | X | X | X | X | X | X |
| [22] | Smart cities in the era of artificial intelligence and the internet of things: a literature review from 1990 to 2017 | X | X | X | X | | X |

## 10.11 Summary of the Findings

**RQ1: WHAT is the understanding and management of the environment in IoT projects?**

The environment is the place where things are, actions happen, events occur, and people are. Smart Environments (SE) or Smart Spaces provide intelligent services by acquiring knowledge about itself and its inhabitants to adapt to users' needs and behavior [8][17]. Contemporary Software Systems apply various technological solutions to meet specific requirements that differ according to the project [13][16].

These systems have a set of (smart) devices that can sense, reason, collaborate and act upon ambiance [17]. Besides having low costs, the devices have low storage capacity and are primarily employed. These pieces of collected information are processed, analyzed, and transformed into knowledge and employed to provide quality of life by predicting the users' behavior and adapting the ambient to their needs. An essential characteristic of this ambient is the user-centric thinking approach in which all of the systems have to be developed to attend to the users in the first place.

This collected data needs to be transferred, processed, analyzed, and stored. Therefore, there is a necessity to connect and integrate devices, systems, real-world data, and cloud computing. This data have to be delivered to the users, systems, or objects and respond faster as possible, giving them real-time characteristics. Some projects have to provide secure management and customize abstract concepts to offer easy control for the users.

Furthermore, a large number of devices can be of various manufacturers. It brings heterogeneity, scalability, extensibility, and fault tolerance capacity for the Internet of Things systems. The ambient has to provide safety and privacy to their users. Some domains have sensitive data, and the systems have to protect them from illegal access or manipulation.

Nowadays, exists a significant effort related to equalizing sustainability and technology. The systems have to optimize resources, being energy efficient, and do wise management of natural resources. These characteristics compose a complex environment and show what the Internet of Things systems have to worry about.

The environment can involve a lot of devices composed of sensors, actuators, and other objects generating a significant amount of data, causing problems related to connectivity and interoperability, data format, data aggregation, data analysis, data processing, data storage, efficient, reliable and high-speed communication, communication bandwidth, energy efficiency, etc. [13][20].

These high complex ambients bring a necessity for system integration [14]. Security, privacy, data access, and data reliability [13][20] are essential aspects of building IoT software systems. There is a necessity for trustful and legal regulation [13]. Sustainability is a crucial concept in these systems, as well [14].

**RQ2: HOW do IoT projects deal with software technologies (techniques, technologies, methods, and solutions) and their operationalization regarding the environment?**

To solve the daily problems and provides a quality of life for users, information, and communication technologies (ICT) are employed by many authors [3][15][22]. In general, smart environments have sensors and actuators to sense and change the ambient [6][8]. The authors frequently mention some technologies. A combination with the Internet of Things (IoT), cloud, and smart objects with sensors and actuators deal with majority problems and provides a general solution for the ambients. To lead with the high number of devices, an exciting approach uses Middleware that enables communication with diverse devices. Middlewares solve the problem of heterogeneity of devices but includes a layer between communication, then exists a tradeoff applying middlewares on the environment.

To enable communications and interactions with smart objects, a network solution is required. Some approaches mention Wireless Sensor Networks (WSN) and Vehicular Ad-hoc Network VANETs using by the systems. WSN is a wireless network composed of distributed autonomous devices and designed to exchange data through the network. A VANETs are an essential component of Intelligent Transportation Systems, enabling communication between vehicles and roads [15][20].

A common problem on Smart Environment is the real-time response necessity. The amount of data exchanging through the network causes an overhead of packages and makes the network's communication difficulties. To solve the problem of transmission and systems, response edge computing can be employed. This approach moves the computation close to the smart environment location and devices, enabling faster communication and reactivity with the user's systems.

To achieve some smartness level, concepts like artificial intelligence, machine learning, and data mining can be an optimal solution. These technologies can model and predict the user's behavior by improving the system's efficiency and quality.

The devices can communicate with each other. This social behavior of smart objects social gives new opportunities for services and new experiences for the users. This autonomy is increasing the quality of systems [2].

Furthermore, the activity of smart design systems is different from conventional systems. The objects and services demand another kind of treatment. Then a requirement engineering technique extending Use Cases can be part of the solution [8]. It provides a standard on the smart environment's specification and allows a common There is no general response to this question. The activities' location is the own environment and depends on the domain that is employed.

Based on the literature found, the environment can be anywhere. We can mention cities, homes (healthcare, ambient assisted living (AAL)), campus, offices, industry, buildings, transportation, streets, roads, bikes stations, parking spaces, etc.understanding to stakeholders.

Smart Environment Metamodels (SEM) [17][21] permit design and analysis using UML class diagrams. This approach enables modeling functional and data requirements (functionalities, objects, location, and data) employed on the application domain using a set of UML-compliant abstractions. A methodological process can be used to orientate the design phase of these systems.

Solutions giving some autonomy for users have been offered using task rules. It is an easy way to personalize functionalities. The end-users can program their services by specifying rules based on if-this-then-that [2][4][11]. Another approach is using the concept of Domotic Effects that provides monitoring and interpreting complex smart environments achieving easy management of smart environments [1].

User interaction is an essential requirement in a Smart Environment. A graphic interface facilitates the understanding and control for the users. A tabletop approach to program the rules is an easy way for users to interact with the systems [4]. Another approach uses a 3D interface to show and control the ambient with Unity3D, IoT, Wireless Sensor Network gateway, Cloud, and others [12].

**RQ3: WHERE do IoT projects locate the activities regarding the environment?**

There is no general response to this question. The activities' location is the own environment and depends on the domain that is employed.

Based on the literature found, the environment can be anywhere. The authors build Internet of Things software systems in places like city [19], home [18], ambient assisted living (AAL) [3], campus [14], office [17][21], industry [10], building [14], transportation [20], street, road, bike station, parking space [15] and others.

**RQ4: WHOM do IoT projects allocate to deal with the environment?**

The environment is projected to serve users' needs. Depending on the domain they can be: students, educators [14], citizens [22], politicians [7], drivers, pedestrians, cyclists [15], patients [9], residents [18], healthcare professionals [13], medical specialist [9], disabled people [18], elderly people [3], family members, doctors, researchers [18], governments, stakeholders, city admins [5], occupants [13], craft groups, operators, technicians [10] and workers [17].

In software engineering, the phases that allocate environment activities allocate developers [2], system designers [8], domain experts [11], technical professionals[8], end-users [4], and stakeholders [11] to built the ambient solution. Multi-stakeholders manage the smart environment. This brings the necessity of engagement [5] to build and manage these ambients.

**RQ5: WHEN do the effects of time, transformations, and states of the environment affect IoT projects?**

Concerning the solutions presented, the majority deals with software activities related to analysis [21], design [8], and implementation [6] phases of a specific Smart Environment. Activities like system architecture definition [6], software design [21], requirement specification [8], and software implementation [6] are performed on these phases.

**RQ6: WHY do IoT projects implement the environment?**

The environment is a result of the combination of other facets. In the Internet of Things systems, several systems the environment is capable of acquiring knowledge about itself and its inhabitants to adapt to users' needs and behavior [17].

This facet is intrinsic to any ambient and provides comfort, quality of life, benefit daily lives, accessibility, high productivity, reduces costs and effort, save time, use resources efficiently and give autonomy to users [3][14][15]. The target of this system benefits the users on your activities by using cutting-edge technologies [4].

There are many scenarios to apply smart environments in our society. This kind of ambient can helps with health diseases, pollution management, traffic efficiency, deterioration and management of infrastructure, criminality, climate change, cyber-security, and economic development [18][20][22]. When we talk about smart ambients, a keyword is sustainability; here is needed to provide user-centric quality services in a comfortable and sustainable way [19].

## 10.12 Final Considerations

Nowadays, the Internet of Things software systems is employed in various domains like smart homes, smart offices, smart cities, Ambient Assisted Living (AAL), etc. These smart environments help users with their daily activities improving quality of life, building sustainable environments, and reducing costs and time. To build these ambients, software engineer activities must be employed to build a model and architecture of these systems that happen on analysis, design, and implementation phases. Use Cases and Metamodels have been used by some approaches to describe and design these complex systems. These solutions permit the reuse of these models built to other systems with the same domains and requirements. An important point is the complexity of these models that must be improved and simplified to provide easy understanding and utilization.

Information and communication technologies (ICT) can be employed to collect, transfer, process, analyze, and store data provided by the ambient. Approaches can combine technologies like smart objects composed of sensors and actuators, Internet of Things (IoT), cloud, middlewares, Wireless Sensor Networks (WSN), Vehicular Ad-hoc

Networks (VANETs), edge computing, artificial intelligence, machine learning, data mining, and others. The combination of ICTs helps to instance ambients depending on the users' necessities. It brings some questions, such as if the existing technologies are sufficient to build the environments and how we can combine existing technologies based on the requirements.

The minimum requirements to be an environment in IoT are the capability to collect, transfer, process, transform, store, analyze data, and act through the ambient applied for most of the found projects. Besides that, connect and integrate smart devices are required for systems, people, and objects to use and manage the data collected and interact with these devices. Large-scale systems or specific systems can demand heterogeneity, interoperability, scalability, extensibility, fault tolerance, real-time response, sustainability, and energy efficiency. For all systems, natural human interfaces are required people to manage and interact with the environment.

In general, the solutions proved by some authors leads to some technological problems. Most of them implement a type of architecture composed of a sensor system, network communication, data analysis and store, and user interface. We can abstract this architecture by four layers: the physical layer, data processing layer, communication layer, and interface layer like Bouchachia et al. [18] suggest in their solution.

Some approaches include layers like cloud services and edge computing to lead with storage, analysis, network, bandwidth, real-time response, and others. To solve problems with the heterogeneity of hardware and software, middleware can be applied. The utilization of these approaches brings a tradeoff; it consists of creating more one layer that intermediates the communication. It brings a system dependency of middleware that must be working full time and leads to network problems like bandwidth and disponibility.

Data is one of the most important things in an environment, and actions to protect data must be provided. Some authors mentioned and highlight security and privacy as problems related to the environment. All of these authors state problems but don't offer a solution for them. There is difficulty implementing security on these ambients because the security must be on all systems' levels. Other difficulties are that devices can interact with systems; these complex interactions bring challenges to identify, grant permissions, and associated devices and services.

User interaction is another problem in the environment. Some approaches have been offered to provide a more straightforward interface and management by task-automation, creating rules relating cause and effect to program actions performed by smart objects in a specific environment. Besides that, a 3D and interactive tabletop interface has been proposed to provide easy management but don't resolve all user interaction problems in smart environments that can't depend on user device or tabletop to interact with these systems.

The system's operationalization is more complex and embracing than their own system. Some solutions reuse the existing services like communication infrastructure (e.g., wi-fi network) provided previously, reducing the complexity. On the other hand, other solutions must establish the systems' ambiance with all required infrastructure services to deploy the application. IoT projects can reduce costs, time, and effort by reusing required services provided in the environment. Therefore, an easy way to build these systems is to provide essential infrastructure services. However, it requires high initial investment and effort demanding previous engagement, for example, by the government or stakeholders. Security and privacy are vital aspects in this shared infrastructure to provide data access control and data privacy enabling secure data access and protecting users and applications.

Finally, we can conclude there isn't a complete technological solution for an environment that attends to all the problems and necessities. Most of them are complicated and limited and have to be adapted by including approaches to specific system problems.

## 10.13 References

**Final Set:**

- Corno, F., & Razzak, F. (2015). Real-time monitoring of high-level states in smart environments. Journal of Ambient Intelligence and Smart Environments, 7(2), 133-153.
- Davoudpour, M., Sadeghian, A., & Rahnama, H. (2015, September). Synthesizing social context for making the Internet of Things environments more immersive. In Network of the Future (NOF), 2015 6th International Conference on the (pp. 1-5). IEEE.
- Geman, O., Sanei, S., Costin, H. N., Eftaxias, K., Vyšata, O., Procházka, A., & Lhotská, L. (2015, October). Challenges and trends in Ambient Assisted Living and intelligent tools for disabled and elderly people. In Computational Intelligence for Multimedia Understanding (IWCIM), 2015 International Workshop on (pp. 1-5). IEEE.
- Pons, P., Catala, A., & Jaen, J. (2015). Customizing smart environments: A tabletop approach. Journal of Ambient Intelligence and Smart Environments, 7(4), 511-533.
- Petercsak, R., Maccani, G., Donellan, B., Helfert, M., & Connolly, N. (2016). Enabling Factors for Smart Cities: A Case Study.
- Cicirelli, F., Fortino, G., Guerrieri, A., Spezzano, G., & Vinci, A. (2016, October). Edge enabled development of smart cyber-physical environments. In Systems, Man, and Cybernetics (SMC), 2016 IEEE International Conference on (pp. 003463-003468). IEEE.
- Arroub, A., Zahi, B., Sabir, E., & Sadik, M. (2016, October). A literature review on Smart Cities: Paradigms, opportunities, and open problems. In Wireless Networks and Mobile Communications (WINCOM), 2016 International Conference on (pp. 180-186). IEEE.
- Aziz, M. W., Sheikh, A. A., & Felemban, E. A. (2016, March). Requirement engineering technique for smart spaces. In Proceedings of the International Conference on the Internet of things and Cloud Computing (p. 54). ACM.
- Descamps, P., Hilaire, V., Lamotte, O., & Rodriguez, S. (2016). An Experience of Engineering of MAS for Smart Environments: Extension of ASPECS. In Intelligent Interactive Multimedia Systems and Services 2016 (pp. 649-658). Springer, Cham.

- Wittenberg, C. (2016). Human-CPS Interaction-requirements and human-machine interaction methods for Industry 4.0. IFAC-PapersOnLine, 49(19), 420-425.
- Desolda, G., Ardito, C., & Matera, M. (2017). Empowering end users to customize their smart environments: model, composition paradigms, and domain-specific tools. ACM Transactions on Computer-Human Interaction (TOCHI), 24(2), 12.
- Shirehjini, A. A. N., & Semsar, A. (2017). Human interaction with IoT-based smart environments. Multimedia Tools and Applications, 76(11), 13343-13365.
- Sanchez, V. G., Pfeiffer, C. F., & Skeie, N. O. (2017). A Review of Smart House Analysis Methods for Assisting Older People Living Alone. Journal of Sensor and Actuator Networks, 6(3), 11.
- Abuarqoub, A., Abusaimeh, H., Hammoudeh, M., Uliyan, D., Abu-Hashem, M. A., Murad, S., ... & Al-Fayez, F. (2017, July). A Survey on the Internet of Thing Enabled Smart Campus Applications. In Proceedings of the International Conference on Future Networks and Distributed Systems (p. 38). ACM.
- Faria, R., Brito, L., Baras, K., & Silva, J. (2017, July). Smart mobility: A survey. In the Internet of Things for the Global Community (IoTGC), 2017 International Conference on (pp. 1-8). IEEE.
- Soni, G., & Kandasamy, S. (2017, December). Smart Garbage Bin Systems–A Comprehensive Survey. In International Conference on Intelligent Information Technologies (pp. 194-206). Springer, Singapore.
- Cicirelli, F., Fortino, G., Guerrieri, A., Mercuri, A., Spezzano, G., & Vinci, A. (2018, April). A Metamodel Framework for Edge-Based Smart Environments. In Cloud Engineering (IC2E), 2018 IEEE International Conference on (pp. 286-291). IEEE.
- Amiribesheli, M., Benmansour, A., & Bouchachia, A. (2015). A review of smart homes in healthcare. Journal of Ambient Intelligence and Humanized Computing, 6(4), 495-517.
- Albino, V., Berardi, U., & Dangelico, R. M. (2015). Smart cities: Definitions, dimensions, performance, and initiatives. Journal of Urban Technology, 22(1), 3-21.
- Djahel, S., Doolan, R., Muntean, G. M., & Murphy, J. (2015). A communications-oriented perspective on traffic management systems for smart cities: Challenges and innovative approaches. IEEE Communications Surveys & Tutorials, 17.
- Cicirelli, F., Fortino, G., Guerrieri, A., Spezzano, G., & Vinci, A. (2017). Metamodeling of smart environments: from design to implementation. Advanced Engineering Informatics, 33, 274-284.
- Rjab, A. B., & Mellouli, S. (2018, May). Smart cities in the era of artificial intelligence and the internet of things: a literature review from 1990 to 2017. In Proceedings of the 19th Annual International Conference on Digital Government Research: Governance in the Data Age (p. 81). ACM.

**Additional References:**

C. Tricco et al. A scoping review of rapid review methods. BMC Medicine, 2015.
B. Cartaxo et al.: The Role of Rapid Reviews in Supporting Decision -Making in Software Engineering Practice. EASE 2018.
B. Cartaxo et al. Evidence briefings: Towards a medium to transfer knowledge from systematic reviews to practitioners. ESEM, 2016

# 11 CONCLUSION

From the data recovered in our research, we realize that IoT concepts and properties can change according to the context and actors involved. For this reason, the representation should be as comprehensive as possible to represent all aspects involved, motivating this facets proposition. The software is only one of the latest technologies since further development is necessary for requirements representation, data infrastructure, network configuration, etc. (Tang, Jun Han, and Pin Chen, 2014).

Our aim is regarding the recovered data's conceptual organization that should take into account the requirements of different stakeholders and the activities in the different facets. Having such a conceptual structure, we do not aim to guide the software system development but rather to organize the concepts more explicitly and support the decision making when engineering IoT. We want to organize our findings according to Figure 1.

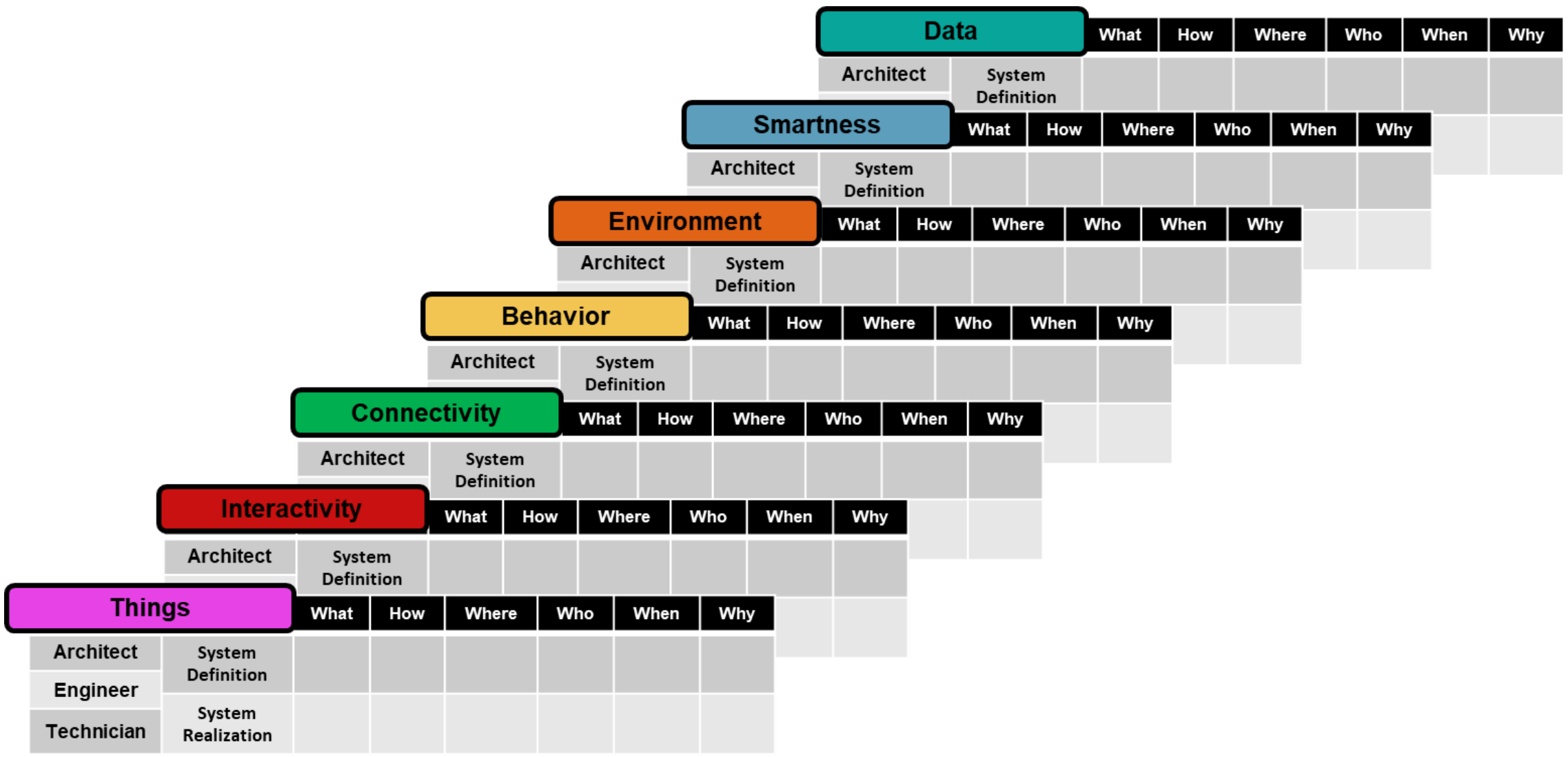

**Figure 2. IoT Body of Knowledge**

The activities conducted to date confirm the adequacy of the facets in IoT and give us input to instantiate the body of knowledge. The next activities include investigating possible relationships between the facets and conducting a more in-depth analysis of the findings. Therefore, the report is still under development.

We aim to keep the information up to date; therefore, if any adjustment is necessary, we sought to make available these reports and the protocols available at Delfos[13]

[13] http://146.164.35.157/

(Observatory of Contemporary Software Engineering) to facilitate access, dissemination, re-execution, and evolution of the findings to keep the body of knowledge updated.